\documentclass[%
 aip,
 amsmath,amssymb,
 reprint,%
]{revtex4-1}

\pdfoutput=1
\usepackage[usenames,dvipsnames]{xcolor}
\usepackage{hyperref}
\usepackage{soul}
\usepackage{tablefootnote}
\usepackage{mathtools}
\usepackage{graphicx}
\usepackage{amssymb}
\usepackage[english]{babel}
\usepackage{amsmath}
\usepackage{epstopdf}
\usepackage{diagbox}
\usepackage{bbm}
\usepackage{dsfont}
\usepackage{bbold}
\usepackage{color,soul}
\usepackage{dcolumn}
\usepackage{latexsym}
\usepackage{bm}
\usepackage{upgreek}
\usepackage{ulem}
\usepackage{listings}
\usepackage{cases}
\usepackage{enumerate}
\usepackage[version=3]{mhchem} 
\usepackage{csquotes}
\usepackage{subcaption}

\definecolor{lg}{gray}{0.93}

\DeclareRobustCommand*{\citen}[1]{%
  \begingroup
    \romannumeral-`\x 
    \setcitestyle{numbers}%
    \cite{#1}%
  \endgroup
}

\renewcommand\thefigure{\arabic{figure}}
\begin{document}
\normalem

\title{Magnetic Interactions between Nanoscale Domains in Correlated Liquids}

\author{Mohammadhasan  Dinpajooh}
\email{hadi.dinpajooh@pnnl.gov}
\affiliation
{Physical and Computational Sciences Directorate, Pacific Northwest National Laboratory, Richland WA 99352, USA}
\author{Giovanna Ricchiuti}
\affiliation
{Physical and Computational Sciences Directorate, Pacific Northwest National Laboratory, Richland WA 99352, USA}
\author{Andrew J. Ritchhart}
\affiliation
{Physical and Computational Sciences Directorate, Pacific Northwest National Laboratory, Richland WA 99352, USA}
\author{Tao E. Li}
\affiliation
{University of Delaware, Newark, DE 19716 USA}
\author{Elias Nakouzi}
\affiliation
{Physical and Computational Sciences Directorate, Pacific Northwest National Laboratory, Richland WA 99352, USA}
\author{Sebastian T. Mergelsberg}
\affiliation
{Physical and Computational Sciences Directorate, Pacific Northwest National Laboratory, Richland WA 99352, USA}
\author{Venkateshkumar Prabhakaran}
\affiliation
{Physical and Computational Sciences Directorate, Pacific Northwest National Laboratory, Richland WA 99352, USA}
\affiliation
{Voiland School of Chemical Engineering and Bioengineering, Washington State University, Pullman, WA 99164, USA}
\author{Jaehun Chun}
\affiliation
{Physical and Computational Sciences Directorate, Pacific Northwest National Laboratory, Richland WA 99352, USA}
\affiliation
{Department of Chemical Engineering, CUNY City College of New York, New York, New York 10031, USA}
\author{Maria L. Sushko}
\affiliation
{Physical and Computational Sciences Directorate, Pacific Northwest National Laboratory, Richland WA 99352, USA}

\date{\today}

\begin{abstract}
The formation of nanoscale domains (NDs) in correlated liquids and the emerging collective magnetic properties have been suggested as key mechanisms governing ion transport under external magnetic fields (eMFs). However, the molecular-level understanding of these magnetic field-driven phenomena and the interaction between these domains remain elusive. To this end, we introduce a simplified model of a solvated nanoparticle (NP) that consists of localized magnetic domains at their surfaces to represent groups of paramagnetic ions, forming NDs, whose effective magnetic dipole moments are at least one order of magnitude greater than the individual ions. We use classical density functional theory (cDFT) to estimate the effective interactions between these localized magnetic NPs (LMNPs). Our findings indicate that, unlike individual ions, magnetic dipole interactions of NDs in the LMNP model can indeed compete with the electrostatic, van der Waals, and hydration interactions. Depending on the direction of eMF, the cDFT effective interactions between two LMNPs turn out to become more attractive or repulsive, which may play a critical role in ion separation and nucleation processes. This indicates that the cDFT interaction barrier heights can be significantly affected by the magnetic dipole interactions and the barrier heights tend to increase as the size of LMNPs increases.
\end{abstract}

\maketitle

Complex phenomena in electrolyte solutions cannot be explained by electric fields alone and require consideration of magnetic fields.
Analogous to the interaction between electric fields and electric charges, magnetic fields interact with magnetic dipoles. 
The migration of ions in inhomogeneous external magnetic fields (eMFs) is an intriguing phenomenon, suggested by experimental interpretations to occur within micro-size domains of highly correlated ions and water molecules.\cite{Fujiwara-2001,Fujiwara-2006,Pulko-2014,Rodrigues-2017,Rodrigues-2018,Rodrigues-2019}
The collective movement of ion groups under eMFs occurs on time and length scales of seconds and meters, respectively, and has been attributed to Kelvin forces arising from eMF gradients.\cite{Butcher-2023,Rassolov-2024,Weston-2010} To this date, it remains unclear how the eMFs interact with the ionic species in electrolyte solutions at the molecular scale to cause such collective motions at the macroscopic scale.

At the nanoscale (e.g., $\sim 20-100$ nm), one may conceive the formation of suspended domains in solutions in the presence of eMFs, with given magnetic properties, where the Kelvin force can act significantly. The mechanism of formation of such domains\cite{Anovitz-2024} and their magnetic properties is an ongoing area of research. 
However, recent research in electrolyte solutions in the absence of eMFs has already discussed the existence of suspended domains\cite{Gibbs-2024,Omta-2003,Turton-2008} and hydrated clusters\cite{Georgalis-2000,Kim-2014,Kathmann-2019,Jin-2024,Dinpajooh-2024,Dinpajooh-2024-2} consisting of cations and anions with an overall charge that is nearly neutral.\cite{Hartel-2023,Komori-2023,Safran-2023,Consta-2002}
These clusters are amorphous, transient, fluctuating in and out of existence and can form domains within a liquid\cite{Shi-2012,Kang-2016,Dunne-2020,Lei-2021,Shuwen-2023} and exhibit unique structural, dynamical, and magnetic properties compared to the surrounding medium. These domains are significantly different from the forming permanent structures like crystals or colloids, which represent distinct phases from the host medium.\cite{Lee-2023,Jaehun-2023,Butreddy-2024}
Specific interactions between such domains may lead to further growth and emergent phenomena over different spatiotemporal scales, which would be responsible for the net movements of groups of ions in the presence of eMFs. 
However, the responses of groups of ions in the nanoscale domains (NDs) under eMFs in aqueous solutions and the effective interactions, $W$s, between NDs at the molecular scale are less understood. It is worth mentioning that $W$s between two NDs at $1-2$ nm are not normally affected by the Kelvin forces (gradient of eMFs)\cite{Butcher-2023,Rassolov-2024,Weston-2010} because the spatial variations of experimental eMFs remain almost constant at these length scales. 
The role of the Lorentz forces on $W$s between NDs can also be studied, but recent classical molecular dynamics studies have concluded they have negligible effects.\cite{Panczyk-2021,Anovitz-2024,Spreiter-1999} In this work we focus on $W$s only due to the interaction of eMFs with the magnetic dipoles of NDs.

\begin{figure*}[tbh]
\centering
\includegraphics[width=\textwidth]{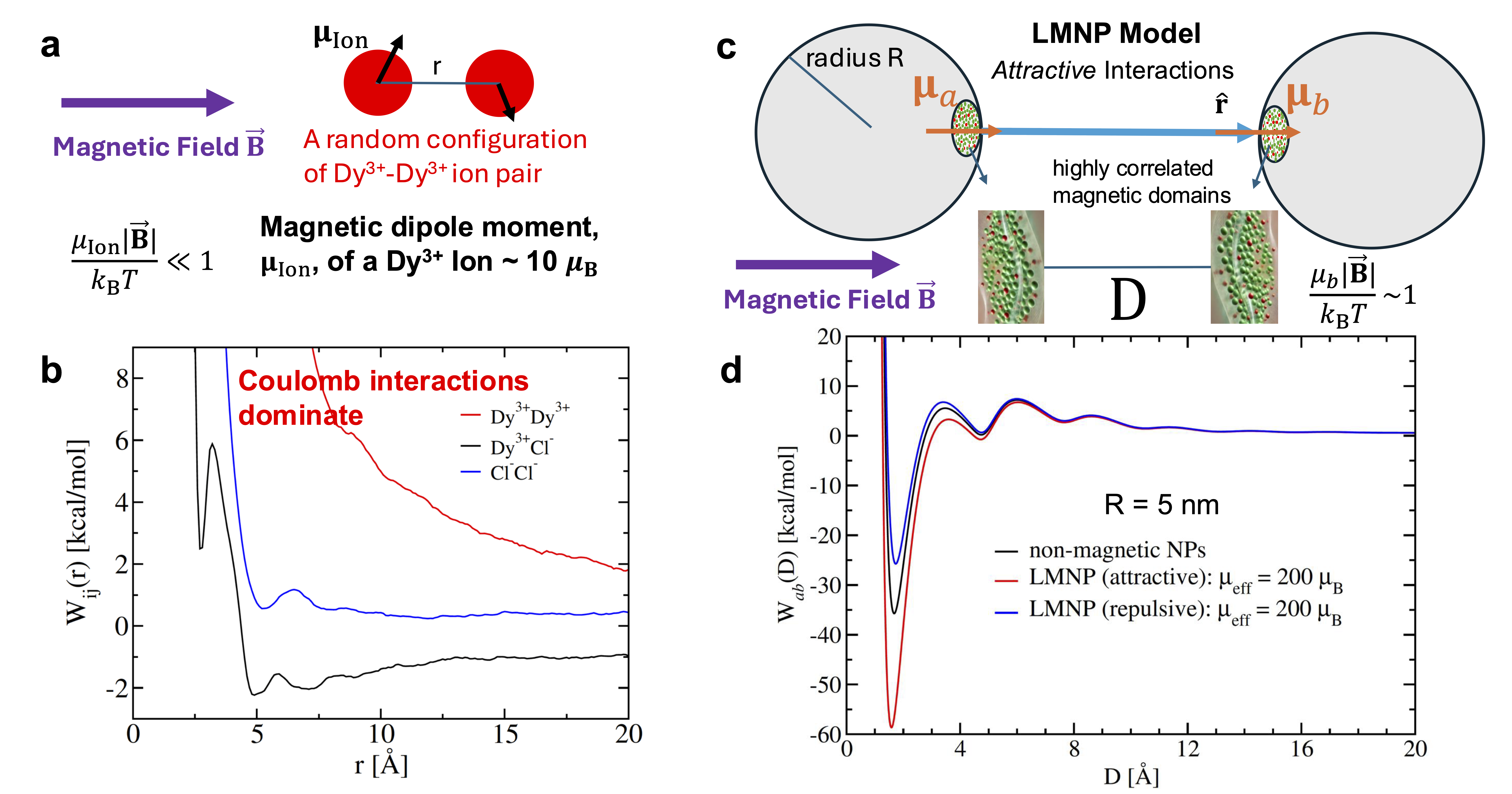}
\caption{({\bf a}) Illustration of a \ce{Dy^{3+}-Dy^{3+}} ion pair and the magnetic dipoles of Dy$^{3+}$ ions in the presence of external magnetic field (eMF) {\bf B}. Because the Zeeman energy is orders of magnitude smaller than the thermal energy, the magnetic dipole moments of ions on average do not align to eMF. Here, the magnetic dipole moment of each Dy$^{3+}$ ion is approximated to be $\sim 10 \mu_{\rm B}$, where $\mu_{\rm{B}}$ represents one Bohr magneton. ({\bf b}) Effective interactions, $W$s, between various ion pairs in water obtained from classical molecular simulations in the absence of eMF. The interactions due to the magnetic dipoles are negligible when compared to the electrostatics interactions that dominate the \ce{Dy^{3+}-Dy^{3+}}  $W$s at relevant distances.
({\bf c}) The localized magnetic NP (LMNP) model consists of a highly correlated magnetic domain close to the surface of NP $a$ consisting groups of ions whose $\mu_{\rm eff}$ is shown by $\mu_a$. In the presence of eMF {\bf B}, because the magnitudes of thermal energies and Zeeman are comparable (at  $B>\sim 0.1$ T), the effective magnetic domains tend to align to the direction of eMF resulting in an attractive magnetic interaction between NPs $a$ and $b$ in the setup shown in this panel. For illustrative purposes, the interacting magnetic nanoscale domains in the outer-shell layers of NPs are represented as ellipsoids. The magnetic interactions are obtained using Eq. \ref{eq:u_NPmag}. ({\bf d}) The classical density functional theory $W$s between NPs (see Eq. \ref{np-np-pmf-corr}) with radii of $5$ nm obtained from the classical density functional theory (see the main text). The black line shows $W$s between two NPs with no localized magnetic domains. The red line shows $W$ between two NPs when eMF aligns them as shown in panel {\bf c} resulting in attractive magnetic interaction. The blue line shows $W$ between two NP when the magnetic dipoles are aligned such that the magnetic interactions are repulsive (when eMF {\bf B} is applied from bottom to top in panel {\bf c}).}
\label{fig:overview}
\end{figure*}

Given the complex properties of NDs in electrolyte solutions, we introduce a simplified model of a liquid-like nanoparticle (NP), where groups of paramagnetic ions are located at its surface and represented as NDs. By ``liquid-like NPs'', we refer to NPs that are solvated in water, appearing externally as liquid droplets while maintaining stable configurations of ions and water molecules in their interiors.\cite{Sun-2014,Cao-2019,Kathmann-2019,Hartel-2023,Komori-2023,Safran-2023,Dinpajooh-2024,Jin-2024,Liu-Xubo-2021,Liu-2019,Wu-2021} 
This model allows one to address the significance of magnetic interactions in $W$s between two NPs while comparing them with other types of interactions; namely, van der Waals (vdW), electrostatics, and hydration interactions. Importantly, the effective magnetic dipole moment, $\mu_{\rm eff}$, of a given ND representing groups of paramagnetic ions are assumed to be at least one order of magnitude larger than the individual ions. The enhancement of the magnetic dipole moment of a given ND is likely attributable to the collective nature arising from strongly correlated interactions among paramagnetic ions within the domain and has been suggested in previous models.\cite{Luttinger-1946,Bergman-2006,Majetich-2006,Doyle-2007,Liu-Xubo-2021,Liu-2019,Wu-2021,Schaller-2010,Muratov-2002} A rigorous investigation of the origin of such an enhancement and the related coarse-graining is an interesting future study. In this study, we focus on the response of NDs consisting of \ce{Dy^{3+}} ions to eMFs and address how they affect $W$s.

A simple way to estimate the magnetic response is to compare the Zeeman energies of isolated ions and NDs with given magnetic dipoles in solutions. In the Appendix, we present the details of Bartington MS2 magnetic susceptibility meter measurements to obtain the bulk magnetic susceptibility of electrolyte solutions at low concentrations. The bulk magnetic susceptibilities have been suggested to estimate the magnitudes of the magnetic dipole moments of individual ions\cite{Dunne-2020} including \ce{Dy^{3+}} and \ce{Nd^{3+}} paramagnetic ions.  Here we estimate that Zeeman energy of a separated ion is much smaller than its thermal energy (see panel {\bf a} in Fig. \ref{fig:overview}) indicating that only a small fraction of paramagnetic ions tend to align to the direction of eMF. 
In future work, we will present a statistical mechanical approach to explain the enhancement of magnetic properties of NDs by considering their spatial heterogeneity. Here, we estimate the Zeeman energy of a given ND by assuming its effective magnetic dipole moment is at least one order of magnitude greater than individual ions (see below). This indicates that in the presence of a sufficiently strong eMF ($B>\sim 0.1$ T) its magnitude is comparable to the thermal energy (see panel {\bf b} in Fig. \ref{fig:overview}).

Next, we consider the magnetic dipole-dipole ($dd$) interactions between separate ions and compare them with the magnetic $dd$ interactions between NDs.
The magnetic $dd$ interaction can be considered as a special case of anisotropic spin-spin coupling, which is always present in addition to the exchange interaction, a quantum mechanical effect that arises due to the overlap of electronic wavefunctions and dominates at short distances.\cite{Luttinger-1946,Majetich-2006,Furrer-2013} 
While in solids the exchange interactions are significant to enhance or suppress their magnetic properties, in aqueous solutions, shielding effects and limited proximity render exchange interactions negligible, making magnetic $dd$ interactions the most plausible consideration.
The magnetic $dd$ interaction between separate ions $i$ and $j$ at a fixed orientation is given by

\begin{equation}
\begin{split}
W_{dd,\:\rm{ions}} (r,\theta_{ij}) =  \frac{\mu_0}{4 \pi r^3} \Big( \bm{\mu}_i \cdot \bm{\mu}_j - 3 (\bm{\mu}_i \cdot \hat{\bf{r}})  (\bm{\mu}_j \cdot \hat{\bf{r}})  \Big) 
\end{split}
\label{eq:u_mag}
\end{equation}
where $\bm{\mu}_i$ is the magnetic dipole moment of ionic species $i$, and $r$ is the distance between interaction ionic species $i$ and $j$, and $\hat{\bf{r}}$ is the unit vector parallel to the line joining the centers of two magnetic dipoles. Here $\mu_0$ is the magnetic permeability of the medium and $\theta_{ij}$ shows that the magnetic $dd$ interaction depends on the angle between magnetic dipoles $i$ and $j$. 
Simple estimations of magnetic $dd$ interactions between paramagnetic ions at relatively short distances show that their magnitudes are negligibly small when compared to the effective interaction energy between solvated ions in water, as obtained from molecular simulations in the absence of eMFs (see panel {\bf b} in Fig. \ref{fig:overview}). This disparity suggests the question:
how do magnetic $dd$ interactions between NDs with Zeeman energies comparable to the thermal energy differ from magnetic $dd$ interactions between separate ions in aqueous solutions?

To address the above question, we introduce a reduced model for a solvated NP to investigate the magnetic interactions between groups of ions. The NP model assumes the existence of localized magnetic domains close to its surface with effective magnetic dipoles, and the resulting $W$s between two NPs are investigated in the presence of eMFs. 
Inside the NP, stable configurations of cations are surrounded by anions to form charge-neutral small clusters. These small clusters are modeled by Lennard-Jones (LJ) particles and are assumed to be uniformly distributed within the NP. At the surface of NP (its outer-shell layer), we assume groups of paramagnetic ions form magnetic NDs whose $\mu_{\rm eff}$ are at least one order of magnitude larger than the individual ions. We call this model, the localized magnetic NP (LMNP) (see panel {\bf c} in Fig. \ref{fig:overview}).
In particular, the NDs can practically consist of a relatively small number of ions compared to the total number of particles in the liquid-like NP. For instance, with a packing efficiency of $55\%$, one can pack $\sim 2000$  Dy$^{3+}$ ions in a spherical NP with a radius of $5$ nm, resulting in a density that is about half of the crystalline density of \ce{DyCl3} ($3.67$ g/cm$^3$). If one assumes that a few percentages of NP ions ($\sim 20-80$ Dy$^{3+}$ ions) in a given ND, here labeled $a$, are correlated to result $\mu_{\rm eff}$ with magnitudes of $\mu_a\sim150-400\mu_{\rm B}$, the Zeeman energy of ND is comparable to the thermal energy and the magnitudes of magnetic $dd$ interactions between two LMNPs (see panel {\bf c} in Fig. \ref{fig:overview}) at $D=5$ \AA\ can be $\sim 0.5-4 k_{\rm B}T$.
In what follows, we consider magnetic NDs with relatively high packing efficiency and evaluate the interaction energies ($W$) between them by fixing their relative orientations, a reasonable assumption given their comparatively large Zeeman energies.

The cDFT $W$s are obtained by considering vdW, electrostatic, and magnetic interactions as well as the correlational (hydration) potential due to the solvent between NPs.
A rigorous way to treat the vdW interactions can be formulated using a Lifshitz-Hamaker approach.\cite{Hamaker1937,Evans-1999,Priye-2013,Lee-2023} Here, for the spherical reduced models of liquid-like NPs, we determine the Hamaker constant and the resulting vdW interactions by analytically integrating over the interacting LJ particles within the two NPs,\cite{Evans-1999,Henderson-1997,Rabani-2002,Everaers2003,Grest2008,Cheng-2012} which results in 

\begin{equation}
    \upsilon_{ab}^{\mathrm{LJ}}(r_{ab}) = \upsilon^{\rm{A}}_{ab}(r_{ab}) + \upsilon^{\rm{R}}_{ab}(r_{ab})
    \label{Eq:vdW}
\end{equation}
where $r_{ab}$ is the distance between the centers of NPs, $\upsilon^{\rm{A}}_{ab}(r_{ab})$ and $\upsilon^{\rm{R}}_{ab}(r_{ab})$ are the attractive (long-range) and repulsive (short-range) parts of the interaction between NPs $a$ and $b$, respectively whose expressions are given in the Appendix. This vdW formulation can result in strongly repulsive behavior at short separations as discussed before.\cite{Grest2008,Cheng-2012,Chuev-2022} 
The electrostatic interactions between LMNPs can be calculated using Coulomb's law when NPs carry net charges. Considering the short-range nature of magnetic interactions, in the LMNP model, only NDs at the nearest surface areas of two NPs contribute to the magnetic interactions. Therefore, the magnetic $dd$ interaction is given by 

\begin{equation}
\begin{split}
W_{\rm LMNP} (D,\theta_{ab}) =  \frac{\mu_0}{4 \pi D^3} \Big( \bm{\mu}_a \cdot \bm{\mu}_b - 3 (\bm{\mu}_a \cdot \hat{\bf{r}})  (\bm{\mu}_b \cdot \hat{\bf{r}})  \Big) 
\end{split}
\label{eq:u_NPmag}
\end{equation}
where $D$ is the edge-to-edge separation distance between LMNPs (see panel {\bf c} in Fig. \ref{fig:overview}). Here $\theta_{ab}$ is the angle between the $\bm{\mu}_a$ and $\bm{\mu}_b$.
In the presence of an eMF {\bf B}, the magnetic $dd$ interactions can become attractive or repulsive depending on the alignments of dipoles. An attractive setup is shown in panel {\bf c} of Fig. \ref{fig:overview}. 
Noting that calculating $W$s between relatively large NPs ($R>1$ nm) by molecular simulations of NPs in explicit solvent become computationally intractable, we use a classical density functional theory (cDFT) approach.

The cDFT approach calculates the solvent density profiles around the NPs, the details of which are presented in Ref. \citen{Chuev-2022}. At a fixed orientation, $W$ between two NPs is given by

\begin{equation}
    W_{ab,\:\mathrm{tot}} (r) = W_{ab,\:\mathrm{bare}} (r) + W_{ab,\:\mathrm{corr}} (r)
    \label{np-np-pmf-corr} 
\end{equation}
where $W_{ab,\:\mathrm{bare}}$ is the bare potential between NPs $a$ and $b$ in the absence of solvent, which
includes vdW, electrostatics, and magnetic $dd$ interactions between NPs (see Eqs. \ref{Eq:vdW} and \ref{eq:u_NPmag}).
The $W_{ab,\:\mathrm{corr}}$ term refers to the correlational potential due to the presence of solvent, and in $Q$-space is given by

\begin{equation}
    W_{ab,\:\mathrm{corr}} (Q) = - \frac{k_{\mathrm{B}}T}{\rho} \Delta \bm{\rho}_a(\mathbf{Q}) \cdot \mathbf{S^{-1}(\mathbf{Q})} \cdot \Delta \bm{\rho}_b (\mathbf{Q})  
    \label{eq:corr-k} 
\end{equation}
where $\rho$ is the bulk solvent density, $\Delta \bm{\rho}_a$ and $\Delta \bm{\rho}_b$ are the solvent density fluctuations caused by the isolated NPs $a$ and $b$, respectively and $\mathbf{S(Q)}$ is the homogeneous solvent structure factor. Note that $W_{ab,\:\mathrm{corr}}$ is closely related to the hydration interaction recognized in colloidal science as the interaction that modifies the ordering of liquid near the surfaces when two such surfaces approach.\cite{Butreddy-2024,Nakouzi-2023,Israelachvili-1987,Israelachvili-1988,Israelachvili-1996,Israelachvili2011,Evans-1999} 
In this work, we obtain $W_{ab,\:\mathrm{corr}}$ in Eq. \ref{eq:corr-k} by making use of the structure factor of pure solvent, $\mathbf{S(Q)}$ (dilute limit approximation).\cite{Chuev-2022} In the presence of eMF, one needs to consider its effect on $\mathbf{S(Q)}$. However, in the Appendix we show that both Monte Carlo simulation and X-ray scattering experimental results indicate that $\mathbf{S(Q)}$ remains unaffected at the eMF strengths commonly used in experiments ($0-2$ T).

Panel {\bf d} of Fig. \ref{fig:overview} shows differences in the resulting $W$s between the NPs with radii of $5$ nm for three cases: NPs with non-magnetic NDs (black line), LMNPs with attractive magnetic NDs (red line), and LMNPs with repulsive magnetic NDs (blue line).
Depending on the direction of eMF, the LMNP $W$s can become noticeably more attractive or repulsive due to magnetic $dd$ interactions as reflected in heights of several extreme points in $W$s at relatively short $D$ separations. Importantly, at short $D$, the LMNP model can capture the hydration layers that result in the oscillatory behaviors in $W$s and magnetic interactions can compete with the hydration interactions (see below).  
Nevertheless, the positions of the minima and maxima in $W$s appear at almost the same $D$. If two LMNPs approach each other from infinity to closer distances, they all need to pass a barrier height of $\sim 7$ kcal/mol at $D\sim 6$ \AA\ below which the barrier heights are significantly affected by the magnetic $dd$ interactions. 

As a comparison to the LMNP model, one may argue that each NP can consist of magnetic NDs at regions far from the surface of NP and ask about the significance of magnetic $dd$ interactions between such domains in two NPs. 
In a simple case, $\mu_{\rm eff}$ of spherical NPs have been previously modeled by placing them at the centers of NPs,\cite{Dormann-1988,Hansen-1998,Schaller-2010,Ahrentorp-2015,Bender-2017} which we call the centralized magnetic NP (CMNP) model. The magnetic $dd$ interaction between NPs $a$ and $b$ is then given by 

\begin{equation}
\begin{split}
& W_{\rm CMNP} (r_{ab},\theta_{ab}) =  \frac{\mu_0}{4 \pi r_{ab}^3} \times \\
& \Big( \bm{\mu}^a_{\rm core} \cdot \bm{\mu}^b_{\rm core} - 3 (\bm{\mu}^a_{\rm core} \cdot \hat{\bf{r}})  (\bm{\mu}^b_{\rm core} \cdot \hat{\bf{r}})  \Big) 
\end{split}
\label{eq:CMNP_NPmag}
\end{equation}
where $r_{ab}$ is the separation distance between the centers of NPs $a$ and $b$, and $\bm{\mu}^a_{\rm core}$ is $\mu_{\rm eff}$ at the center of NP $a$. Similarly, more detailed models such as core-shell models can be proposed,\cite{Biasi-2002} which include center-center and surface-surface magnetic $dd$ interactions as well as two center-surface magnetic $dd$ interactions one of which is given by 

\begin{equation}
\begin{split}
& W_{\rm core-shell} (p_{ab},\theta_{ab}) =  \frac{\mu_0}{4 \pi p_{ab}^3} \times \\ 
& \Big( \bm{\mu}^a_{\rm core} \cdot \bm{\mu}^b_{\rm shell} - 3 (\bm{\mu}^a_{\rm core} \cdot \hat{\bf{r}})  (\bm{\mu}^b_{\rm shell} \cdot \hat{\bf{r}})  \Big) 
\end{split}
\label{eq:core-shell}
\end{equation}
where $p_{ab}$ is the distance between the surface of NP $a$ and the center of NP $b$, and $\bm{\mu}^b_{\rm shell}$ is $\mu_{\rm eff}$ at  surface of NP $b$. Therefore, the magnetic $dd$ interactions in the core-shell NP model consists of all terms presented in Eqs. \ref{eq:u_NPmag}, \ref{eq:CMNP_NPmag}, and  \ref{eq:core-shell}.  

\begin{figure}[tbh]
\centering
\includegraphics[width=\columnwidth]{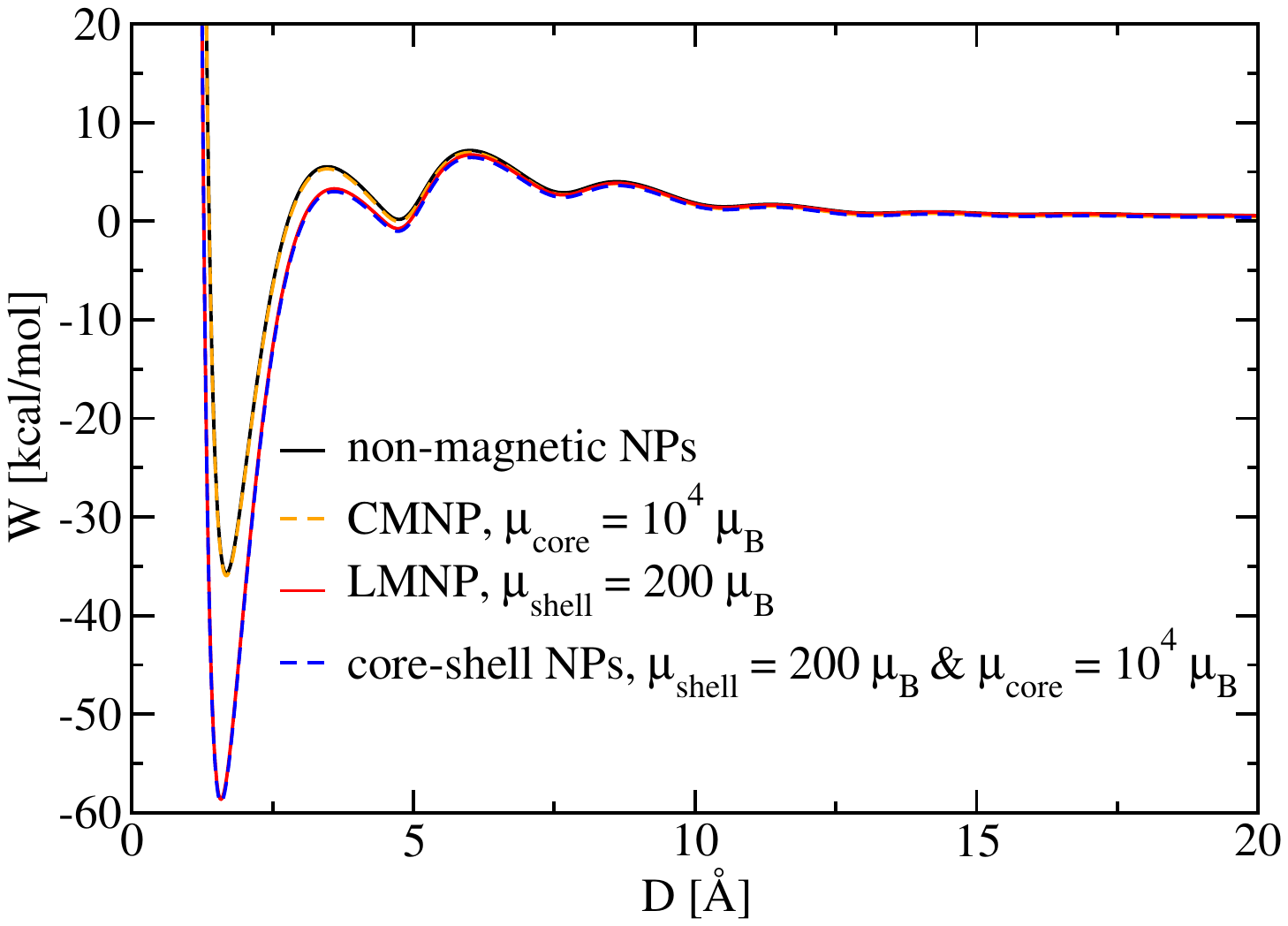}
\caption{Comparisons of $W$s between different models of NPs with radii of $5$ nm and various effective magnetic dipole moments, $\mu_{\rm eff}$. They include non-magnetic NPs, NPs whose effective magnetic dipoles with $\mu_{\rm core}=10^4\mu_{\rm B}$ are located at their centers, and the NPs whose effective magnetic interactions are located at their surfaces (the LMNP model) with magnitudes of $\mu_{\rm eff}=\mu_{\rm shell}=200\mu_{\rm B}$. All the magnetic dipoles  are aligned with the eMF similar to setup shown in panel {\bf c} of Fig. \ref{fig:overview}.
}
\label{fig:CMNP}
\end{figure}

Figure \ref{fig:CMNP} compares $W$s for non-magnetic NPs, CMNPs, LMNPs, and core-shell NPs for NPs with radii of $5$ nm whose magnetic interactions are only attractive (panel {\bf c} in Fig. \ref{fig:overview}). Here $\mu_{\rm eff}$s at the centers of NPs are assumed to be $50$ times greater than $\mu_{\rm eff}$s at the surfaces of the NPs.
The resulting $W$s for the LMNP model are in excellent agreement with the core-shell NP model suggesting that the LMNP model can capture the essential features of magnetic $dd$ interactions. 
This argument is further supported by the remarkable agreement observed between $W$s of non-magnetic NPs and CMNPs.
Therefore, for relatively large NPs, the magnetic $dd$ interactions in the LMNP model are in general much more significant than the ones for the CMNP model even if the effective magnetic dipole of the CMNP model is assumed to be about $50$ times greater than $\mu_{\rm eff}$ at the surface of the NP in the LMNP model. This can be attributed to the fast decays of the magnetic $dd$ interactions with respect to the center-center or center-surface distances (decays with the cube of the distance) which may be compensated with the large magnitudes of magnetic dipoles moments. One can also conceive cases where the fast decays of magnetic $dd$ interactions for center-center distances do not fully overcome the relatively large magnitudes of the dipole moments (see Section VII in Appendix). However, in normal cases, the LMNP model is a satisfactory model as shown in Fig. \ref{fig:CMNP}.

\begin{figure}[tbh]
\centering
\includegraphics[width=\columnwidth]{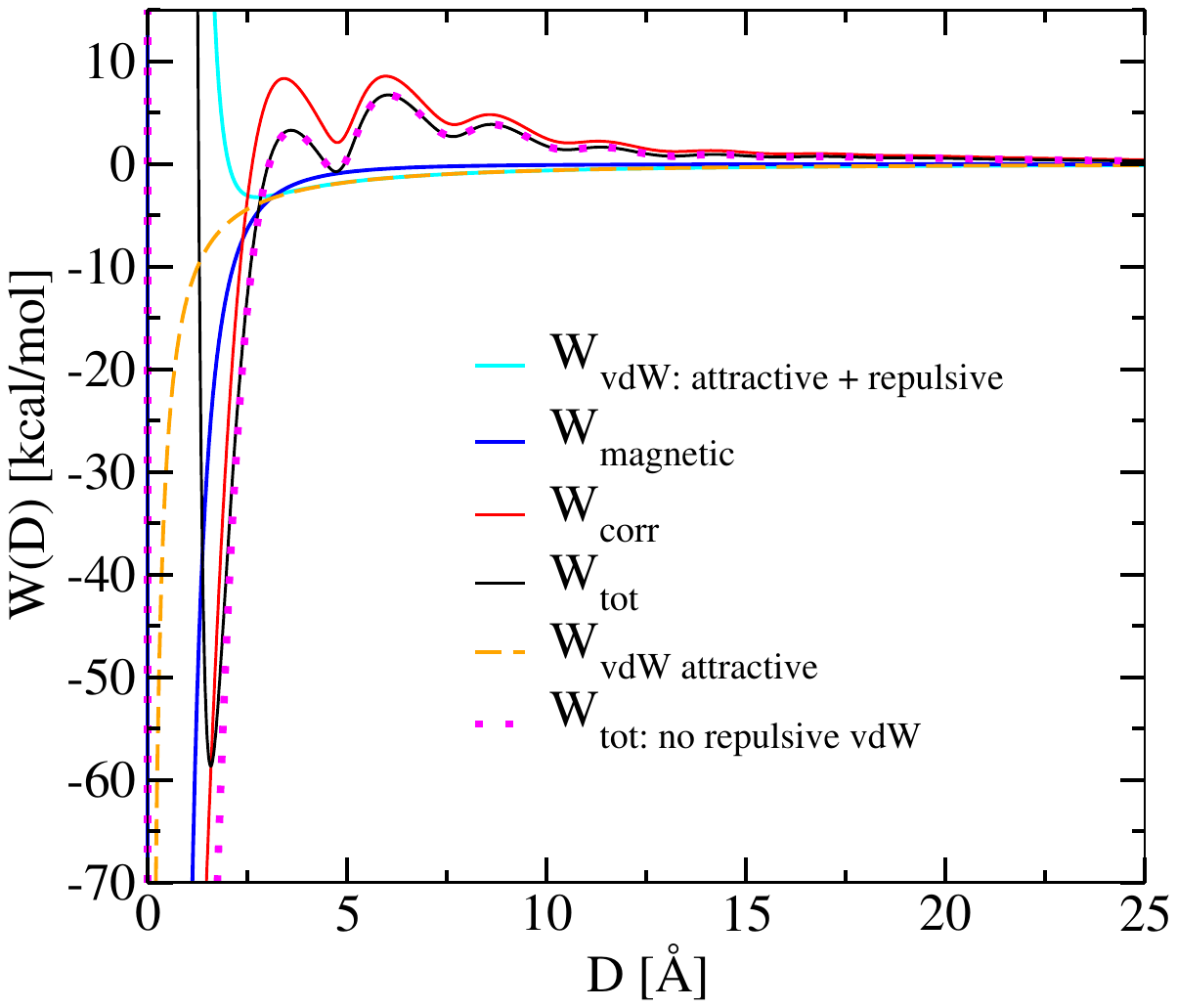}
\caption{ 
Decomposition of cDFT NP-NP $W$s into van der Waals (vdW: cyan), magnetic (blue), and correlational potentials (corr: red) showing that the oscillations in the cDFT $W$s originate from the correlational interactions. The vdW $W$, in the current LMNP model, consists of both attractive and repulsive contributions, which result in strongly repulsive $W$ at short separations. For comparison, the vdW $W$ and total $W$ are also shown (orange and magenta dashed lines, respectively) when the repulsive contributions are neglected in the LMNP model. The LMNP model is used for NPs with the radii of $5$ nm and the magnetic NDs with magnitudes of $200\mu_{\rm B}$. The orientations of magnetic NDs are similar to the panel {\bf c} in Fig. \ref{fig:overview}.}
\label{fig:contribution-np-np}
\end{figure}

The cDFT $W$ is decomposed into vdW, magnetic, and correlational potential in Fig. \ref{fig:contribution-np-np} for the LMNPs. The effects of electrostatic interactions are discussed below. 
As shown in Eq. \ref{Eq:vdW}, consistent with previous studies\cite{Grest2008,Cheng-2012} the vdW interactions in the current LMNP model consists of both attractive and repulsive contributions, which result in strongly repulsive $W$ at short separations. While the Lifshitz-Hamaker approach predicts the attractive part of vdW interactions, the repulsive part of vdW is less understood. If one simply neglects the repulsive vdW contributions, Fig. \ref{fig:contribution-np-np} shows that the LMNPs can have contact and aggregation consistent with previous findings in colloidal science.\cite{Lee-2023} It remains to future studies to further examine the vdW interactions for the liquid-like LMNP model studied in this work.    
Notably, the correlational potential exhibits strong attractive behavior at relatively short separations, transitioning to a repulsive regime for $D > 2.5$~\AA. It also displays oscillations with a periodicity approximately matching the size of a water molecule at shorter separations. Interestingly, the oscillation wavelengths increase at large distances and they completely vanish at separations greater than $20$ \AA. Therefore, the many-body solvent interactions lead to the correlational potentials with oscillations that vanish at separations greater than NP diameter. At long distances, the bare potential almost cancels the correlational potential contributions.
Importantly, the interactions due to Kelvin forces\cite{Butcher-2023,Rassolov-2024,Weston-2010} are not considered here because experimentally eMFs remain almost constant at $D$ separations studied.

\begin{figure}[tbh]
\centering
\includegraphics[width=\columnwidth]{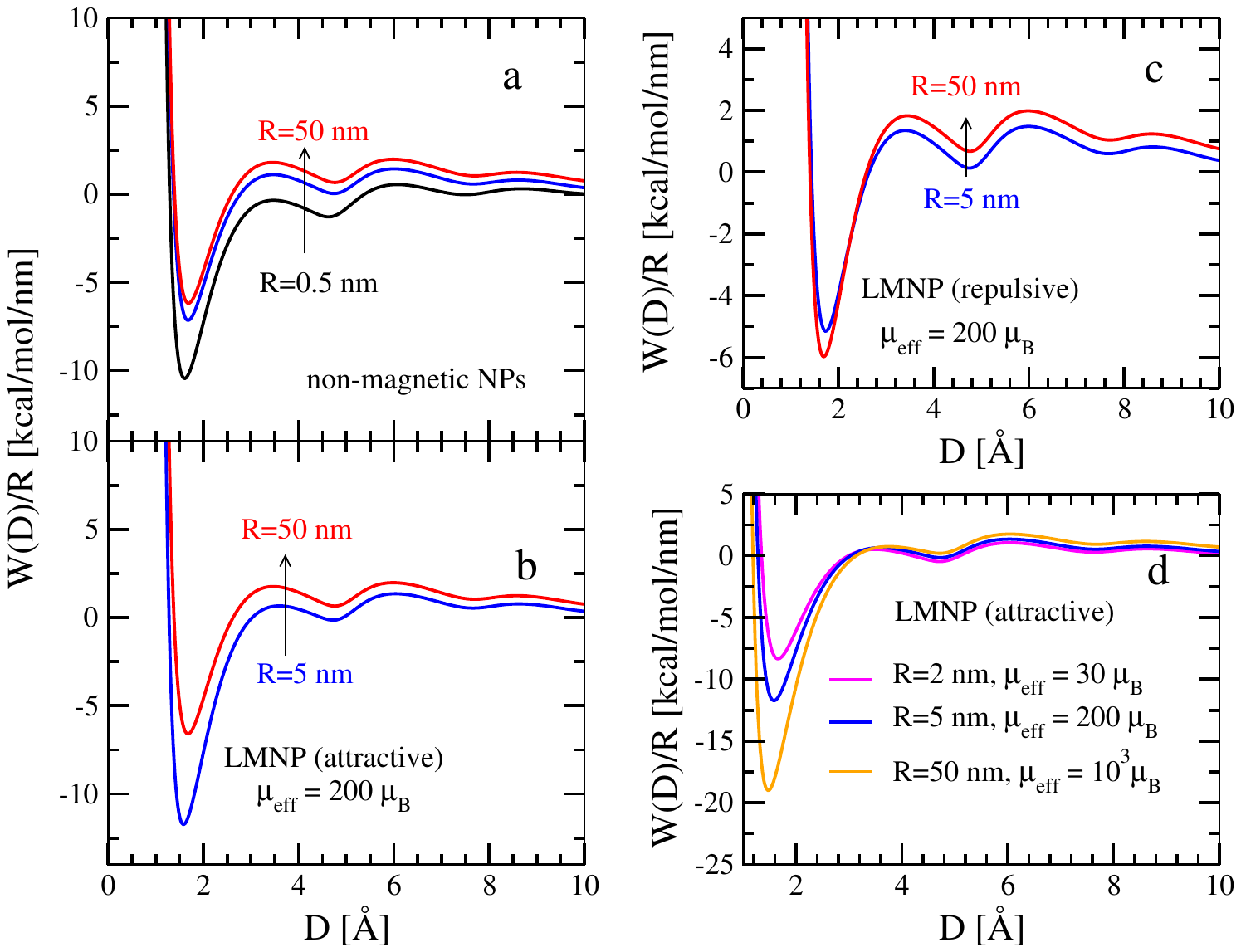}
\caption{Size dependencies of $W$s between NPs. a) The scaled $W$s between non-magnetic NPs with sizes of $0.5$ nm (black line), $5$ nm (blue line), $50$ nm (red line). b) The scaled $W$s between LMNPs of various sizes with $\mu_{\rm eff}=200\mu_{\rm B}$ with $\theta_{ab}=0$. c) The scaled $W$s between LMNPs of various sizes with $\mu_{\rm eff}=200\mu_{\rm B}$ with $\theta_{ab}=\pi/2$. d) The scaled $W$s between LMNPs with sizes of $2$ nm \& $\mu_{\rm eff}=30\mu_{\rm B}$ (magenta line), $5$ nm \& $\mu_{\rm eff}=200\mu_{\rm B}$  (blue line), $50$ nm \& $\mu_{\rm eff}=1000\mu_{\rm B}$  (orange line).}
\label{fig:pmf-size}
\end{figure}

Next, we discuss $W$s between NPs of different sizes, charges, and orientations. Figure \ref{fig:pmf-size} shows that $W$s between NPs are in general much stronger for larger size NPs, which can be attributed to the increased effective interaction area between the two NPs.\cite{Derjaguin1934,Wennerstroem2020} Here we explore the trends in $W$s between NPs when scaled by their radii. 
Panel {\bf a} of Fig. \ref{fig:pmf-size} illustrates this for non-magnetic NPs, where an upward shift is observed in the height of the first maximum for the NPs with radii of $50$ nm ($\sim2.1$ kcal/mol/nm). A similar upward shift is observed for the LMNPs with attractive magnetic interactions (panel {\bf b}) while their sizes increase at constant $\mu_{\rm eff}$. Due to attractive magnetic $dd$ interactions in panel {\bf b}, the scaled $W$s can show larger depth magnitudes as compared with the non-magnetic NPs.
On the other hand, the repulsive magnetic $dd$ interactions in magnetic NDs of NPs of various sizes (panel {\bf c}) with the same $\mu_{\rm eff}$, result in smaller depth magnitudes of scaled $W$s for smaller NPs. Importantly, panel {\bf d} shows the sensitivity of attractive scaled $W$s between LMNPs when their $\mu_{\rm eff}$s change as the NP size increases, and the scaled long-range $W$s for various sizes of NPs almost collapse to each other. An examination of the scaled short-range $W$s clearly shows that the depth magnitudes of the first minima are largest for LMNPs with larger NP sizes with larger $\mu_{\rm eff}$. As expected, a comparison of panels {\bf b} and {\bf d} shows that larger $\mu_{\rm eff}$ for the same NP sizes dominates the depth magnitudes of scaled $W$s at short-range distances.      

Interestingly, panel {\bf d} of Fig.\ \ref{fig:pmf-size} illustrates that the maximum energy barrier opposing the approach of two LMNPs with radii of $5$~nm is approximately $87$~kcal/mol, indicating a strong resistance to aggregation for large-size LMNPs. However, for LMNPs with radii of $2$ nm, the cDFT predicts maximum barrier height around $2$ kcal/mol ($3.4 k_{\rm B}T$), which shows their tendencies to approach each other. Therefore, panel {\bf d} supports the hypothesis that NDs can form and be stable, noting that their formations are limited by their barrier heights, which tend to increase for larger LMNPs showing the self-limiting growth of these LMNPs.

\begin{figure}[tbh]
\centering
\includegraphics[width=\columnwidth]{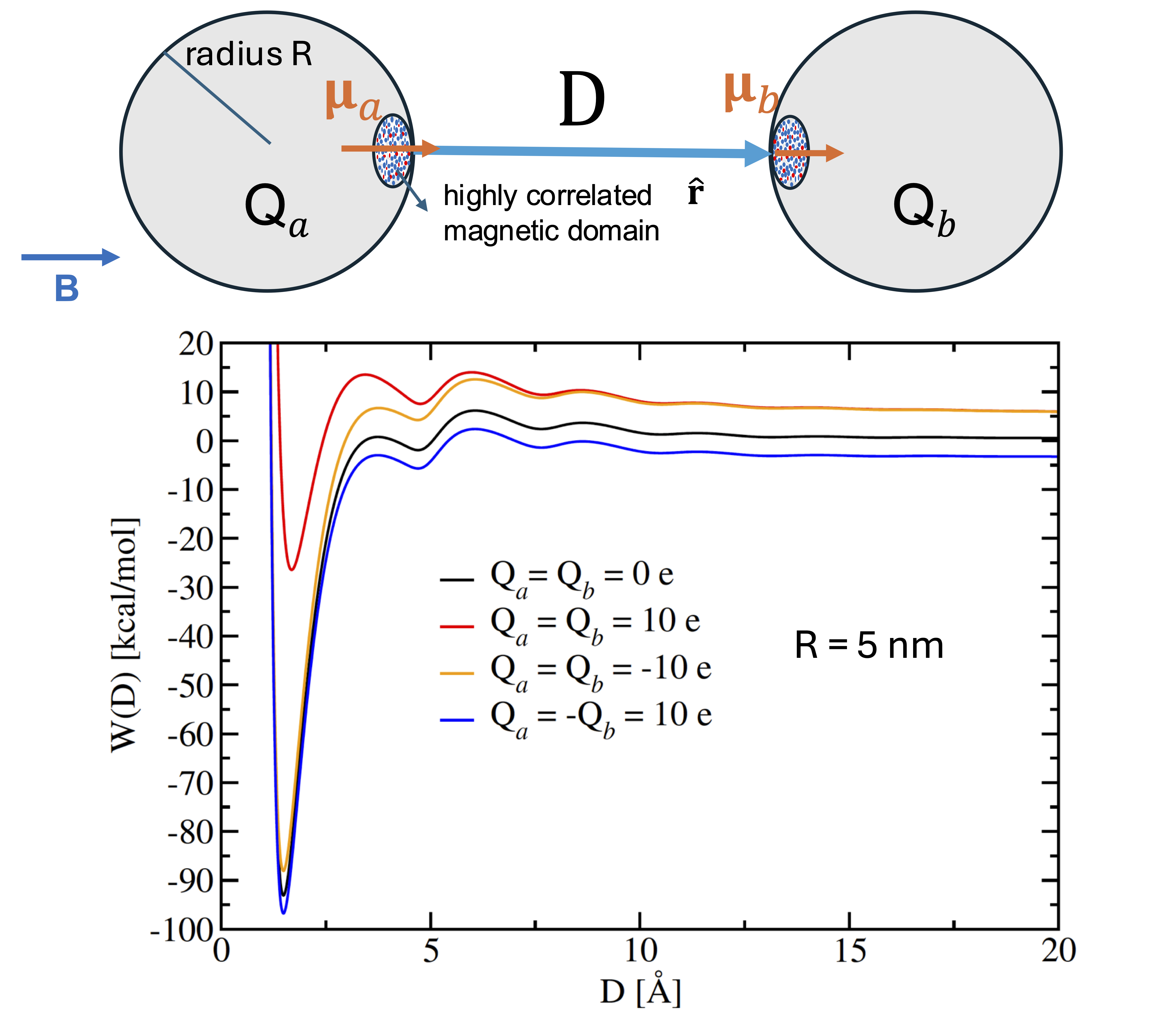}
\caption{Charge dependencies of $W$s between neutral and charged LMNPs in a setup shown at the top panel, where magnitudes of magnetic domains are set to $300$ $\mu_{\rm B}$.}
\label{fig:pmf-chrg}
\end{figure}

When NDs are not completely neutral, one can discuss the significance of electrostatics interactions in cDFT $W$s. We assume that the NPs can carry charges uniformly distributed over their surfaces (see Sect. III in Appendix) and compare $W$s between the neutral and charged LMNPs with the same $\mu_{\rm eff}$ and radii (of $5$ nm). Figure \ref{fig:pmf-chrg} shows that the first minima of $W$s for all studied NPs appear around $D\sim1.5$ \AA\ with the depth magnitudes largest for the oppositely charged NPs ($Q_a=-Q_b=10$ e) and the lowest one for the positively charged NPs ($Q_a=Q_b=10$ e).   
Interestingly, for the oppositely charged NPs, $W$ between NPs at $D>\sim 7$ \AA\ is attractive and a repulsive barrier height of $2.4$ kcal/mol appears at $D\sim6$ \AA. On the other hand, $W$s between other NPs show repulsive behaviors at $D>\sim10$ \AA\ noting that the positively and negatively charged NPs show almost identical repulsive behaviors mainly due to the significant reduction of the correlational potentials (see Fig. \ref{fig:contribution-np-np}).
The positively and negatively charged NPs show significantly different behaviors at $D<\sim10$ \AA, which can be attributed to the asymmetry of local densities of water around opposite charges. \cite{Remsing2014,Dinpajooh2015-born} In particular, $W$ between the negatively charged NPs is about $62$ kcal/mol more stable than the one for the positively charged NPs at $D\sim1.5$ \AA\ and the barrier height of the negatively charged NPs at $D\sim3.5$ \AA\ is about $7$ kcal/mol less than the one for the positively charged NPs. As expected, the long-range $W$s for the studied charged LMNPs almost completely overlap with the long-range behaviors of Coulomb interactions, $Q_aQ_b/(\epsilon r_{ab})$ with $\epsilon$ as the dielectric constant of water, around $35$ nm (results not shown).

\begin{figure}[tbh]
\centering
\includegraphics[width=\columnwidth]{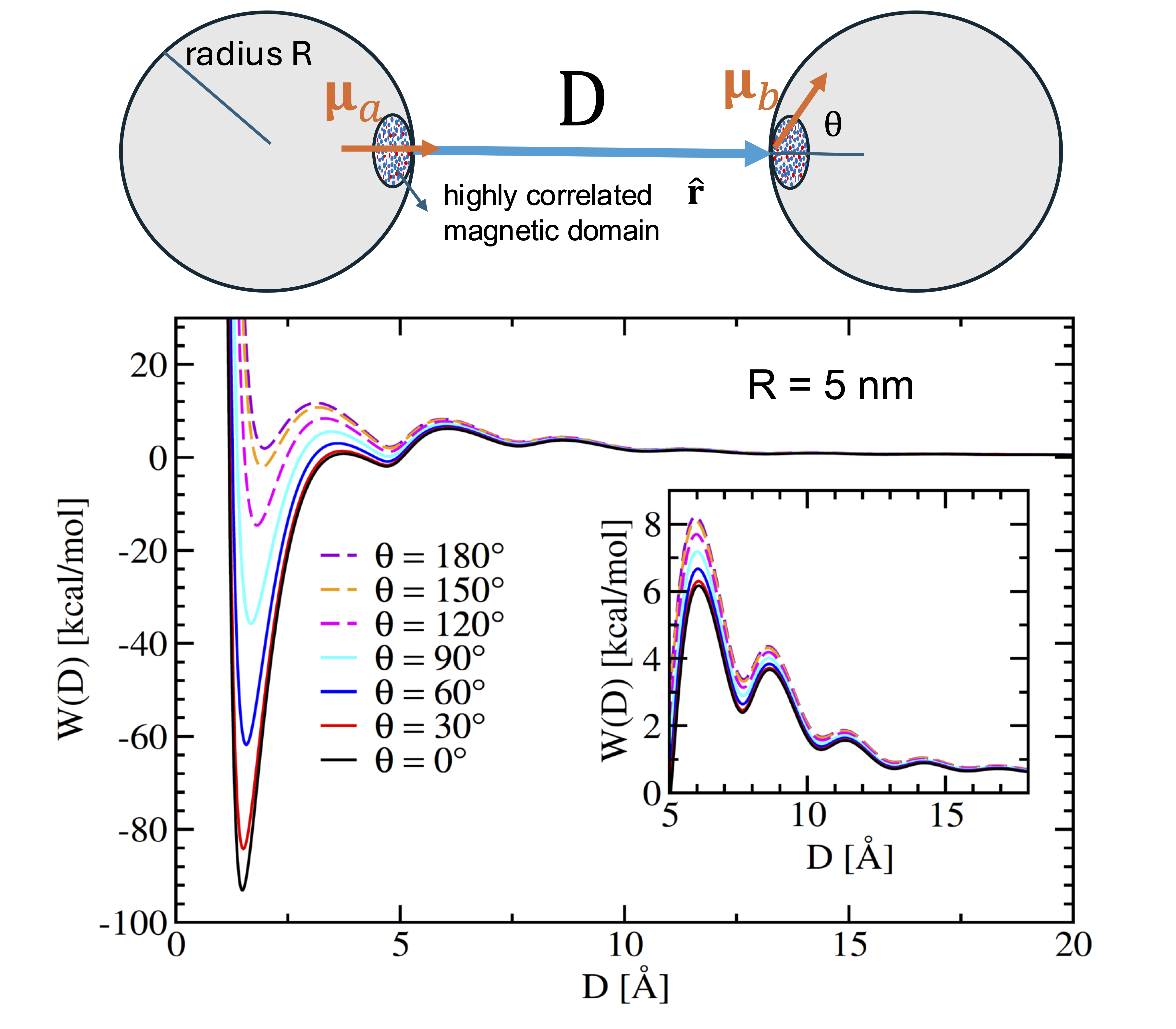}
\caption{Orientation dependencies of $W$s between NPs whose active magnetic domains with magnitudes of $300$ $\mu_{\rm B}$ are located at the surfaces of NPs. Here $\theta$ denotes the angle between the magnetic NDs in the LMNP model.}
\label{fig:pmf-orient}
\end{figure}

When the Zeeman energies of NDs are not comparable to thermal energy ($B\sim<0.1$ T or relatively small $\mu_{\rm eff}$), their relative orientation can change considerably at room temperature. To provide more insights for simplicity, we fix the directions of magnetic NDs to various orientations and report $W$s between the LMNPs in Fig. \ref{fig:pmf-orient}. Interestingly, $W$s between LMNPs with $\mu_{\rm eff}=300\mu_{\rm B}$ at all orientations are repulsive at $D>5$ \AA\  and several barrier heights exist between $W$s of two LMNPs. In particular, the barrier heights at $D\sim16.8$ \AA\ range from $0.72-0.81$ kcal/mol with the lowest barrier heights for the attractive LMNPs that are completely aligned ($\theta=0$ in Fig. \ref{fig:pmf-orient}). The range of barrier heights at $D\sim14.2$ \AA\ is around $0.85-1.05$ kcal/mol and this range increases to barrier heights of $1.5-1.9$ kcal/mol and $3.7-4.4$ kcal/mol at $D\sim11.4$ \AA\ and $D\sim8.6$ \AA\, respectively. Therefore, the probability of crossing the barrier height at  $D\sim8.6$ \AA\ is about $0.1\%$ at room temperature. The barrier heights at $D\sim6.0$ \AA\ turn out to be $6.0-8.2$ kcal/mol, which make the close-approach of such LMNPs relatively unlikely. However, in the case of rare events, the closest barrier heights to the surface of the LMNPs are about $0.7$ kcal/mol (with the peak position $\sim3.7$ \AA) and $11.8$ kcal/mol (with the peak position $\sim3.0$ \AA) for $\theta$ of $0$ and $180$ degrees, respectively, showing the significance of orientations on the magnetic interactions.

In summary, we have presented simple models to improve our understanding of magnetic interactions between NDs in correlated liquids, where each domain represents groups of ions.\cite{Kathmann-2019,Hartel-2023,Komori-2023,Safran-2023,Dinpajooh-2024} 
The LMNP model involves localized magnetic NDs at the surfaces of NPs solvated in water. A cDFT approach is developed to calculate $W$s between LMNPs in water. The results show that unlike the magnetic $dd$ interactions between two ions in the aqueous solutions, the magnetic $dd$ interactions between two LMNPs can compete with electrostatic, vdW, and hydration interactions. The limitations of the LMNP model as well as the sensitives of $W$s between NPs with respect to the sizes, charges, and orientations of NPs are discussed. 
In a future work, we will address the movements of the NDs due to various forces in the presence of eMFs\cite{Rassolov-2024,Bazant-2024,Guo-2024} and the implications of this theoretical work to the related phenomena such as self-assembly\cite{Swan-2014,Sherman-2016,Kachman-2017} and 
analyses of experimental observations in atomic force microscopy\cite{Higashitani-1997,Koga1999,Tikhomirov2008,Harada2010,Ballone2012,Li2020,Li-2020-2,Nakouzi-2021,Nakouzi-2023,Butreddy-2024,Liu-2024} and Mach-Zehnder Interferometry.\cite{Yuan-2013,Lei-2017,Huang-2019,Ricchiuti-2022,Rodrigues-2017,Rodrigues-2018,Rodrigues-2019}
This requires incorporating a consistent formalism for estimating the forces.\cite{Ayansiji-2020,Leong-2020,Weston-2010,Butcher-2023,Franczak-2016,Rassolov-2024} When combined with cDFT for time-dependent analysis,\cite{Vrugt-2023,Brader-2024} such an approach holds the potential to bridge the gap between experimental observations and theoretical predictions of magnetophoretic separation and pave the way for innovative applications in biomedical, environmental, and industrial contexts.\cite{Dunne-2020}

\hfill
\section*{Appendix}
The Appendix provides detailed descriptions of the Bartington MS2 Magnetic Susceptibility Meter measurements and related analyses. Additionally, it discusses the Small-Angle X-ray measurements conducted in the presence of eMF, as well as the molecular dynamics simulations, Monte Carlo simulations, and classical density functional theory calculations.

\section*{Acknowledgements}
The authors would like to thank Herman M. Cho, Grant E. Johnson, Shawn M. Kathmann, Christopher J. Mundy, and Gregory K. Schenter for useful discussions. This study was supported by a collaborative effort funded by the Laboratory Directed Research and Development (LDRD) program at PNNL, under the Non-Equilibrium Transport Driven Separations Initiative (NETS). T.E.L. acknowledges the start-up funds from the University of Delaware Department of Physics and Astronomy.

\section*{Author Declarations}
\subsection*{Conflict of interest}
The authors have no conflicts to disclose.

\section*{Data Availability}
The data that supports the findings of this study are available within the article. Source code implementations are also available from the corresponding author upon reasonable request.

\appendix

\setcounter{equation}{0}
\setcounter{figure}{0}
\setcounter{table}{0}
\setcounter{section}{0}
\makeatletter
\renewcommand{\thefigure}{A\arabic{figure}}

\section{Experimental methods for Magnetic susceptibility measurements}

\subsection{Chemicals}
Lanthanum(III) chloride hydrate (99.9\%, 211605-500G), Neodymium(III) chloride hexahydrate (99.9\%, 289183-25G), Dysprosium(III) chloride hexahydrate (99.9\%, 289272-25G), Hydrochloric acid (37\%, ACS reagent, 258148-500ML) were purchased from Sigma-Aldrich (St. Louis, MO) and used as-received without further purification. The deionized water (ASTM Type II, LC267505) was purchased from Labchem Inc. (Zelienople, PA) and used as received to prepare the solutions.  

\subsection{Details}
The bulk volume magnetic susceptibility, $\chi_{\rm vol}$, of LaCl$_3$, NdCl$_3$, DyCl$_3$ electrolyte solutions were measured using Bartington MS2 Magnetic Susceptibility Meter, with MS2G single frequency sensor (San Carlos, CA). The MS2G detects changes in inductance within an inductor when the permeability of its core that contains the sample is altered. The $\chi_{\rm vol}$ is directly linked to the relative permeability, $\mu_r$, of the sample through the relationship $\chi_{\rm vol}=\mu_r-1$, where $\mu_r$ represents the ratio of the permeability of the sample to the permeability of a vacuum. The lanthanide solutions at different concentrations (0.5, 0.25, 0.1, 0.075, 0.05, 0.01, and 0.005 M) were prepared by dissolving salts in 0.1 M HCl to minimize the carbonate formation due to CO$_2$ absorption from the atmosphere. 1 mL of the sample (lanthanide salts or solution) is filled in the sample holder vial and placed in the rectangular block containing the measurement cavity core of the meter. The MS2G sensor applies a non-saturating field to the sample, enabling the measurement of initial susceptibility without affecting the magnetic remanence of the sample. Before placing the sample vial, the zero-reference state, which is the permeability of a vacuum, is determined when the measurement cavity contains only air, and the permeability of the air is approximated to that of a vacuum. The permeability of the sample is then determined with a sample vial in the measurement cavity and the $\mu_r$ is determined. For each sample, the acquisition time was set to 10 seconds, and a set of 5 repeated measurements was collected. The reference state is zeroed before each of these measurements to minimize the thermal or environmental drifts. The data points from the instrument are collected by Bartsoft software and the $\chi_{\rm vol}$ is directly obtained from the software in CGS units. All measurements are carried out at a room temperature of 25 $^\circ$C.

\begin{table}[tbh]
\centering
\caption{The dimensionless magnetic susceptibility values (in units of CGS) for three Lanthanide electrolyte solutions measured by Bartington MS2 Magnetic Susceptibility Meter, with MS2G single frequency sensor at various concentrations and a pH of $\sim1$ ($c_{\ce{HCl}}=0.1$ M). The dimensionless magnetic susceptibility of the acidic solution of \ce{HCl}  with no metal ions is measured as $-1.24\times 10^{-6}$. The magnetic dipole moments are not reported when their magnitudes cannot be determined via $\chi_{\rm solution}$ to within the uncertainties (NA). }
\begin{tabular}{ccc}
      c [M] & $\chi_{\rm solution}$  & $\mu_{\rm eff}/\mu_{\rm B}$  \\
     \hline
     \multicolumn{3}{c}{ \ce{DyCl3}} \\
      $0.01$ & $-7.78\times10^{-7}$ & $10.49$ \\
      $0.05$ & $3.16\times10^{-7}$ &  $8.61$ \\
      $0.075$ & $9.37\times10^{-7}$ & $8.31$ \\
      $0.10$ & $1.76\times10^{-6}$  &  $8.45$ \\
      $0.10$ & $2.57\times10^{-6}$ &  $9.53$ \\
      $0.25$ & $8.73\times10^{-6}$ &   $9.75$ \\
      $0.50$ & $1.93\times10^{-5}$ &   $9.89$ \\
     \hline
     \multicolumn{3}{c}{ \ce{NdCl3}} \\
      $0.01$ & $-1.25\times10^{-6}$ & NA \\
      $0.05$ & $-1.07\times10^{-7}$ &  $2.85$ \\
      $0.075$ & $-6.85\times10^{-7}$ & $4.20$ \\
      $0.10$ & $-8.10\times10^{-7}$  &  $3.20$ \\
      $0.10$ & $-7.90\times10^{-7}$ &  $3.30$ \\
      $0.25$ & $-1.80\times10^{-7}$ &   $3.20$ \\
      $0.50$ & $9.36\times10^{-7}$ &   $3.20$ \\
      \hline
\end{tabular}
\label{Chi_Giovanna}
\end{table}

Table \ref{Chi_Giovanna} reports the dimensionless magnetic susceptibilities  ($\chi_{\rm vol} = \chi_{\rm solution}$) in CGS units for the studied \ce{DyCl3} and \ce{NdCl3} aqueous electrolyte solutions. The $\chi_{\rm solution}$ of \ce{LaCl3} are not reported in Table \ref{Chi_Giovanna} and their values range $-1.24~-1.22 \times 10^{-7}$.

\begin{figure}[tbh]
\centering
\includegraphics[width=\columnwidth]{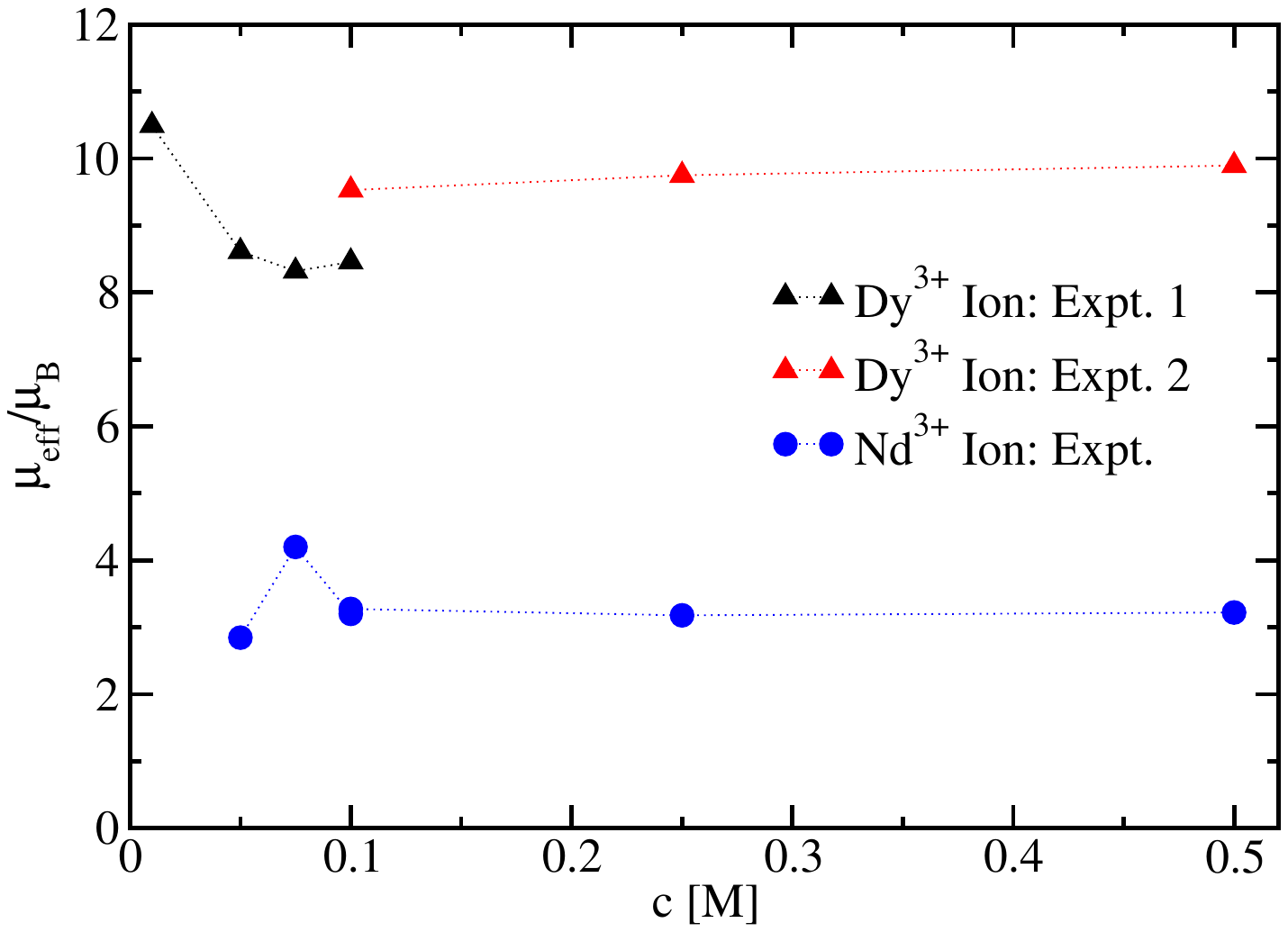}
\caption{Estimations of magnetic dipole moments of paramagnetic ions in Lanthanide aqueous solutions of \ce{DyCl3} and \ce{NdCl3} in the Bohr magneton unit, $\mu_{\rm B}$, from the Bartington MS2 magnetic susceptibility meter measurements.
Expt. 1 and 2 refer to two independent experimental measurements.}
\label{fig:dipole}
\end{figure}

The effective magnetic dipole moments of the ions can be determined using Curie's law and a model of an aqueous electrolyte solution.
Let us first consider a dilute limit model of an aqueous electrolyte solution in which the individual ions are homogeneously dispersed and calculate the magnetic properties of individual ions.
The paramagnetic ions such as Dy$^{3+}$ and Nd$^{3+}$ have intrinsic magnetic dipoles whose magnitudes are generally much larger than their induced magnetic dipoles originated from the response of ions to the external magnetic fields.
On the other hand, water and the diamagnetic ions such as La$^{3+}$ only have induced magnetic dipoles. Therefore, it is easy to notice that in an electrolyte solution consisting of paramagnetic ions, the magnetic $dd$ interactions are dominated by the paramagnetic ions.
Noting that Zeeman energy of an individual ion is much smaller than thermal energy, at equilibrium one can use Curie's law to extract the magnitudes of the intrinsic magnetic dipole moments of ions from the magnetic susceptibility measurements. In the dilute limit, the dimensionless magnetic susceptibility of an electrolyte solution of $\rm{MCl_3}(aq)$ may well be approximated as\cite{Kuchel-2003,Dunne-2020}

\begin{equation} 
\chi_{\text{solution}} = \chi_{\rm H2O} + [c_{\rm M^{3+}}] \chi_{\rm M^{3+}} + [c_{\rm Cl^{-}}] \chi_{\rm Cl^{-}}  
\label{chi_ion}
\end{equation}
where $ \chi_{\rm H2O}$ is the dimensionless magnetic susceptibility of pure or acidic water and $c_\alpha$ represents the molar concentration of dissolved ions $\alpha$.\cite{Dunne-2020} Since Cl$^{-}$ is a diamagnetic ion and its concentration is much less than water, for simplicity we treat $\chi_{\rm{Cl^{-}}}\approx 0$.
The effective magnetic dipole moments of homogeneously dispersed individual $\mathrm{M}^{3+}$ ions in aqueous solutions, $\mu_{\rm eff}$, can then be determined using Curie's law, $\chi_{\rm m, M^{3+}}=C/T$ with $C = \mu_0 \mu_{\rm B}^2 N_A \mu_{\rm eff}^2  /(3 k_{\rm B})$, where $\chi_{\rm m, M^{3+}}$ is the molar magnetic susceptibility, $\mu_0$ is the magnetic permeability of the medium, $\mu_{\rm{B}}$ is the Bohr magneton, $N_A$ is the Avogadro number, and $k_{\rm B}$ is the Boltzmann constant.\cite{Dunne-2020} Figure \ref{fig:dipole} shows the resulting magnetic dipole moments of  Dy$^{3+}$ and Nd$^{3+}$ paramagnetic ions in Bohr magneton predicted from experimental measurements at various concentrations using the Bartington MS2 magnetic susceptibility meter. Notably, the magnetic dipole moment values are almost independent of the salt concentration, which is consistent with previous studies.\cite{Ikeda-1968,Ikeda-1970} Considering the uncertainties in the experiments, they are in good agreement with the estimations of magnetic dipole moments from Hund's law ($10.6\mu_{\rm B}$ and $3.5\mu_{\rm B}$ for Dy$^{3+}$ and Nd$^{3+}$ ions, respectively).\cite{Dunne-2020} 
Therefore, at equilibrium a simple homogeneous model of an electrolyte solution can indeed predict the magnetic dipoles consistent with Hund's law predictions. In what follows, we demonstrate how to use these calculated ion magnetic dipole moments in the dilute limit to estimate the effective magnetic dipole moments of nanoscale domains.  
It is worth noting that the magnetic dipoles of La$^{3+}$ ions cannot be determined to within the uncertainties of experiments (similar to the first data point for \ce{NdCl3} at $0.01$ M shown by NA in Table \ref{Chi_Giovanna}).

Let us now consider a more realistic model of electrolyte solutions, which contains both individual ions and hydrated clusters consisting of ions. While the formation and experimental characterization of these clusters are beyond the scope of this study, recent research has demonstrated the presence of clusters consisting of cations and their counterions in aqueous electrolyte solutions, resulting in an overall charge that is nearly neutral.\cite{Kathmann-2019,Hartel-2023,Komori-2023,Safran-2023} In principle, if the cluster-size distributions are known, one can calculate the equilibrium $\chi_{\rm solution}$ enabling a comparison with experimental values. In the dilute limit, it is reasonable to assume that the proportion of such clusters remains small and Eq. \ref{chi_ion} holds to extract the magnetic dipole moments of individual ions.

\section{Structure factor of water in the presence of external magnetic field}

\subsection{X-ray Scattering Measurements of Water in the Presence of an External Inhomogeneous Magnetic Field}

Small Angle X-ray Scattering measurements were performed using a Xenocs Xuess $3.0$ instrument equipped with a Cu K$\alpha$ source and Dectris Eiger2 1M detector. Scattering profiles were acquired with 30 m acquisition time at a distance of $90$ mm across a $Q$ range of $0.0016$ to $3.75$ \AA$^{-1}$. Measurements were performed on $18$ M$\Omega$ purity water using a $1.0$ mm quartz capillary. A solution flow configuration was used allowing for exact background subtraction of quartz. Calibration was performed using a LaB$_6$ standard. Magnetic measurements were performed using a 1x1x4 cm, $0.5$ T permanent neodymium magnet placed $0.7$ mm directly below the capillary, aligned with the magnetic pole. Azimuthal integration was performed with linear binning using the Xenocs XSACT software package.

\begin{figure}[tbh]
\centering
\includegraphics[width=\columnwidth]{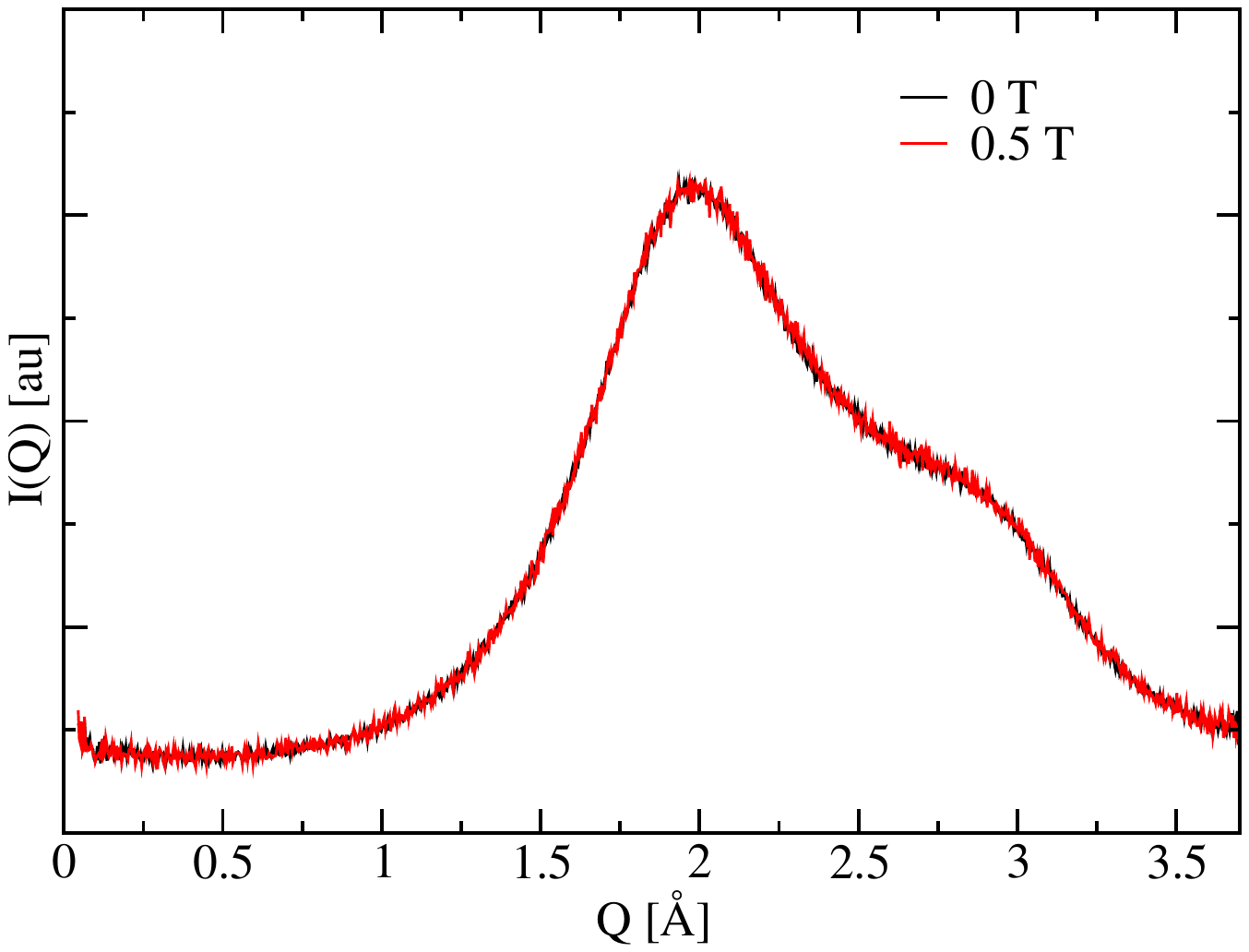}
\caption{A Comparison of experimental small-angle X-ray scattering signals for pure water in the absence of an external magnetic field and an external magnetic field of $0.5$ T.
}
\label{fig:WAXS}
\end{figure}

Figure \ref{fig:WAXS} shows the small-angle X-ray (SAXS) experiments for pure water in the absence and presence of external magnetic fields. As can be seen the differences between the SAXS signals are negligible suggesting that the structure factor of pure water does not change significantly at experimental external magnetic fields. In the next section, we confirm this observation by all-atom molecular simulations.

\subsection{Monte Carlo Simulations of Water in the Presence of an External Homogeneous Magnetic Field}

In the presence of homogeneous external magnetic field $\bf{B}$, the water interaction potential in the Monte Carlo simulations is given by
\begin{equation}
\begin{split}
U & =  
 \sum_i \sum_j 4 \epsilon_{ij} \left[  (\frac{\sigma_{ij}}{r_{ij}})^{12} - (\frac{\sigma_{ij}}{r_{ij}})^{6}\right]+ 
  \frac{1}{4 \pi \epsilon_0} \sum_i \sum_j \frac{q_i q_j}{r_{ij}}  + \\
  &  \sum_i \sum_j  \frac{\mu_0}{4 \pi R_{ij}^3} \Big( \bf{m}_i \cdot \bf{m}_j - 3 (\bf{m}_i \cdot \hat{\bf{r}})  (\bf{m}_j \cdot \hat{\bf{r}})  \Big) \\
  & - \sum_i \bf{m}_i \cdot \bf{B}
\end{split}
\label{eq:Hmol}
\end{equation}
where $r_{ij}$ is the distance between interaction sites in a given model. On the other hand, $R_{ij}$ is the distance between the centers of masses or oxygen atoms in the water model and $\hat{\bf{r}}$ is the unit vector parallel to the line joining the centers of two magnetic dipoles. 
When the two magnetic dipoles have the same directions and their magnitudes are the same with $m_i=m_j=m_w$, one gets

\begin{equation}
    U_{dd}(R_{ij},\theta_{ij}) =  \frac{\mu_0}{4 \pi R_{ij}^3} m_w^2 \left( 1 - 3 \cos^2{\theta_{ij}}  \right)
    \label{eq:Udd}
\end{equation}
where $\theta_{ij}$ is the angle between the direction of
external magnetic field and the line joining the centers of the induced magnetic dipole moments for water molecules $i$ and $j$. 

To get the magnitudes of the magnetic dipole moments for water ($m_w$), one may assume
\begin{equation}
    m_w = \chi_w B/ (N_{\rm A} \mu_0) 
\end{equation}
where $\chi_w$ is dimensionless water magnetic susceptibility with a value of $ -9.04 \times 10^{-6}$ and $N_{\rm A}$ is the Avogadro number.  

Monte Carlo simulations were performed using a modified version of the MCCCS software program,\cite{Dinpajooh-2011} which incorporates the magnetic interactions for water molecules as described above.
The water molecule is represented by the TIP4P (transferable intermolecular potentials—four point) model,\cite{TIP4P} which uses a rigid internal structure and a single Lennard-Jones 12–6 interaction site per molecule at the oxygen position. The first-order electrostatic interactions are represented by three partial charges placed on the positions of the two
hydrogen atoms and an additional site on the bisector of the HOH angle close to the oxygen atom.

Monte Carlo simulations in the canonical ensemble were carried out for a system consisting of
$1000$ water molecules with a box length of $31.06$ \AA\ (at a density of $0.997$ gr/cm$^3$). The cutoff distance was set to $14$ \AA. Each simulation consisted of $5\times10^5$ and $10^6$ MC cycles for the equilibration
and production periods, respectively (each cycle
consists of $N$ trial moves, where $N$ is the number of molecules in the system).
Center-of-mass translations and rigid-body rotations were employed for these simulations with the probabilities of $0.5$ and $0.5$, respectively. The equal proportions of translational and rotational moves indicate that each water molecule has three translational and three rotational degrees of freedom.

The trajectories were saved every $200$ MC cycles from which the radial distribution functions and the structure factors were calculated. The oxygen-oxygen partial static structure factors were computed using

\begin{equation}
    S(\bf{Q}) = \frac{1}{N} \Big\lvert \sum_{m=1}^N e^{-i \bf{k}\cdot r_O} \Big\rvert ^2 
\end{equation}
where $N$ is the total number of water molecules and $\bf{r}_O$ is the vector position of oxygen atom $m$ in the simulation box.
The possible set of wavevectors $\bf{Q}$ is defined by
${\bf Q}=\frac{2\pi}{L}(n_x \hat{\bf{x}}+ n_y \hat{\bf{y}} + n_z \hat{\bf{z}})$, where $L$ is the side length of the simulation box; $\hat{\bf{x}}$, $\hat{\bf{y}}$, $\hat{\bf{z}}$ are the unit vectors in their respective directions; and $n_x$, $n_y$, and $n_z$ run over
all integer values.
The one-dimensional $S(Q)$ is obtained by radially averaging $S(\bf{Q})$ over each point at a given wavenumber $Q=|\bf{Q}|$.
For the TIP4P system, the position of the oxygen atom as a proxy for the position of the molecule.

The total structure factor is also calculated using 
\begin{equation}
    S_{\rm tot}(\bf{Q}) =  \frac{1}{N_{\rm tot} } \Big\lvert \sum_{m=1}^{N_{\rm tot}} e^{-i \bf{k}\cdot r_m}  \Big\rvert ^2 
\end{equation}
where $\bf{r}_m$ is the position of oxygen or hydrogen atom and $N_{\rm tot}$ is the total number of atoms in the simulation box.

\begin{figure}[tbh]
\centering
\includegraphics[width=\columnwidth]{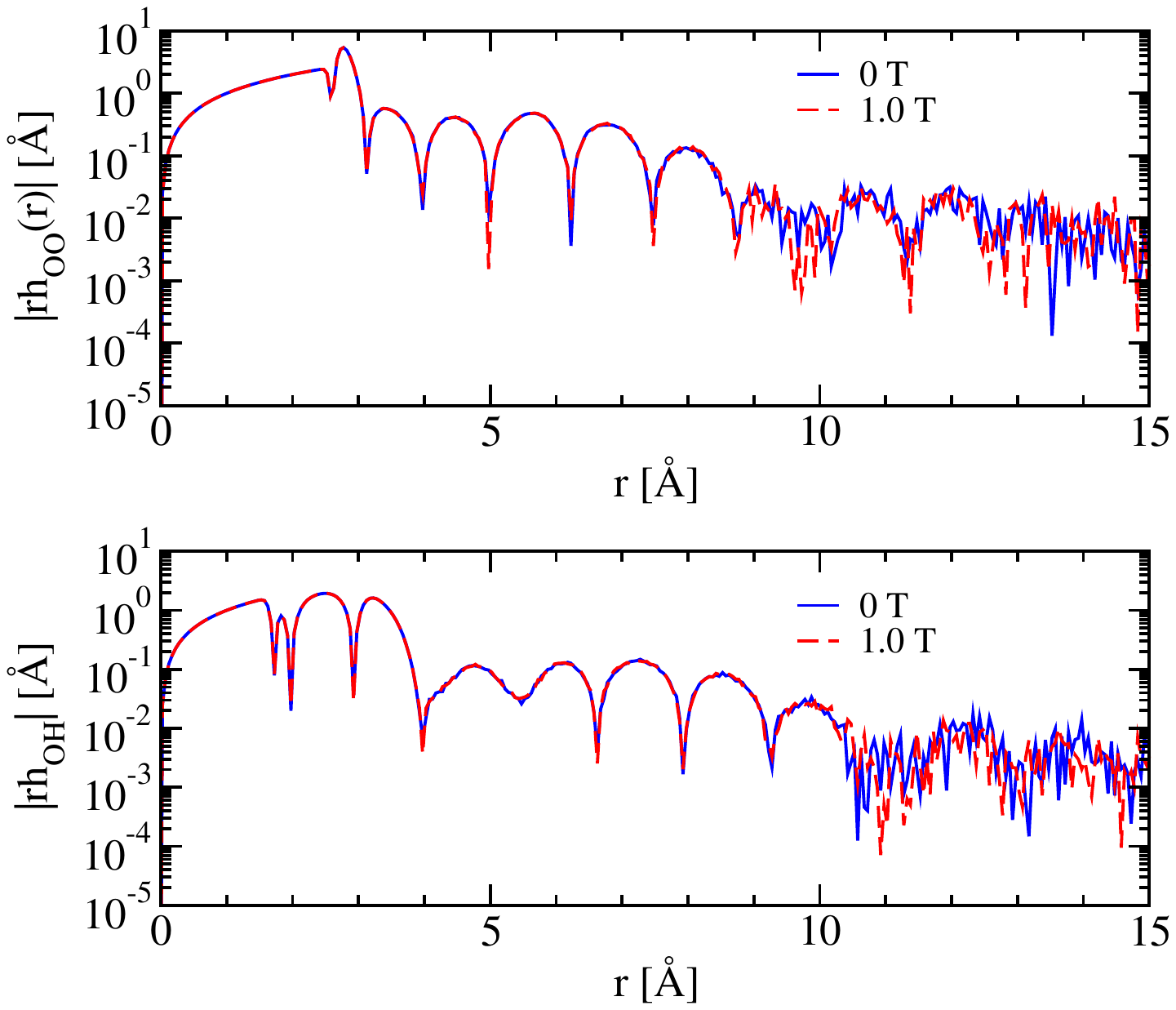}
\caption{ Comparisons of $|r h_{\rm{OO}}(r)|$ and $|r h_{\rm{OH}}(r)|$ for TIP4P water in the absence of an external magnetic field and an external magnetic field of $1$ T.
}
\label{fig:rhr}
\end{figure}

In $r$-space, one may report $|rh(r)|$ to stress the long-range correlations, where $h(r)=g(r)-1$ with $g(r)$ as the radial distribution functions.  
Figure \ref{fig:rhr} shows the resulting $|rh(r)|$ for oxygen-oxygen and oxygen-hydrogen partials indicating that the correlation functions of water remain almost constant as one puts water in the external homogeneous magnetic field.

\begin{figure}[tbh]
\centering
\includegraphics[width=\columnwidth]{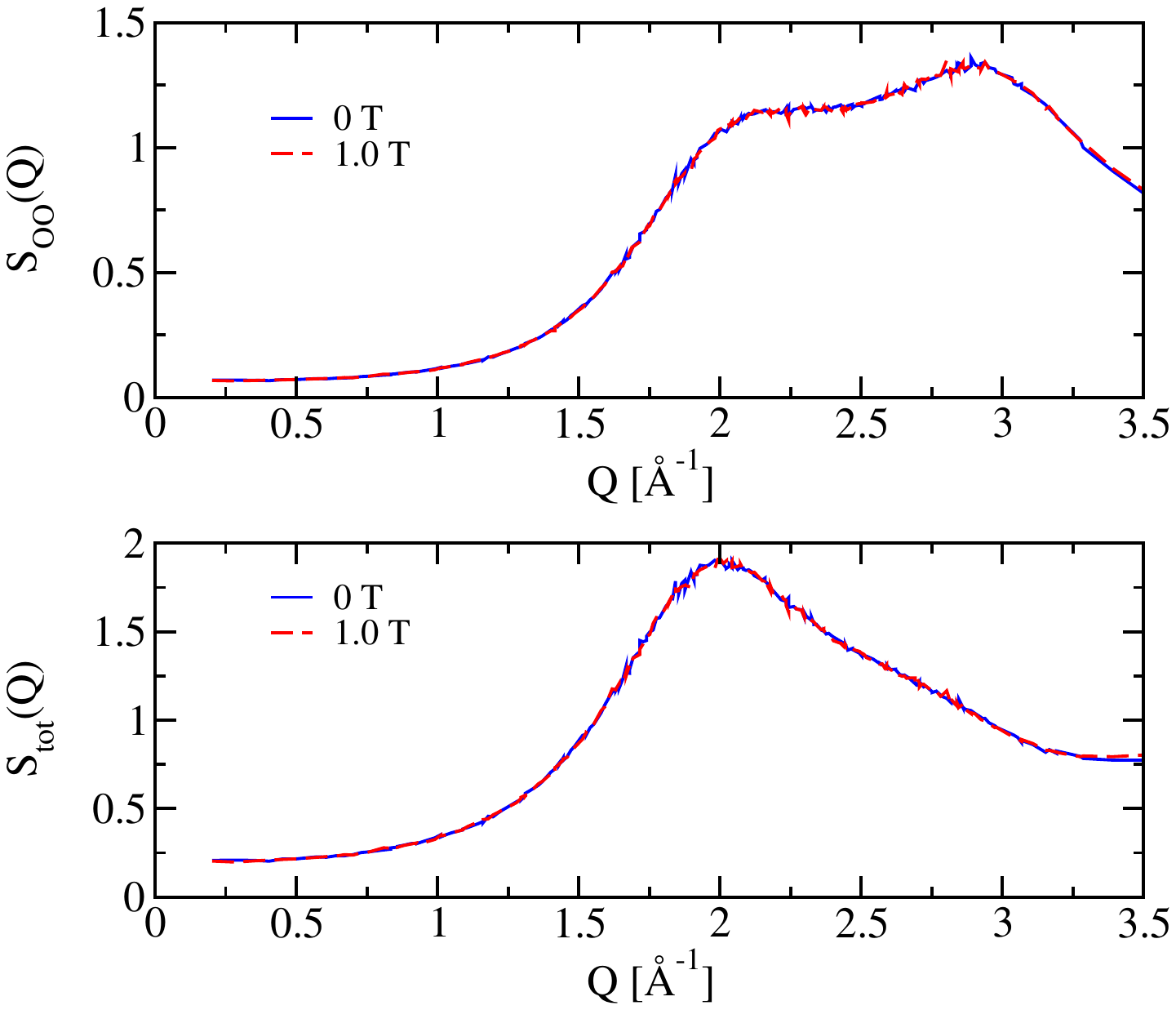}
\caption{Comparisons of structure factors for TIP4P water in the absence of an external magnetic field and an external magnetic field of $1$ T.
}
\label{fig:SQ}
\end{figure}

Figure \ref{fig:SQ} shows the oxygen-oxygen and total structure factors for water in the absence of external magnetic field and in the presence of $1$ T magnetic field. As can be seen, the structure factors of water in both cases are in close agreements confirming the assumption that the structure factor of water is almost consistent at experimental magnetic fields.

\section{Nanoparticle Interaction Potential in MD and CDFT Calculations}
\label{sec:nano_pot}

The interaction potentials for both NP-NP and NP-solvent interactions involve the standard Coulombic and Lennard-Jones (LJ) terms

\begin{equation}
    \upsilon_{nx}(r) = \upsilon_{nx}^C(r) + \upsilon_{nx}^{\mathrm{LJ}}(r)
    \label{tot_pot}
\end{equation}
where $x=n$ for NP-NP interactions and $x=s$ for NP-solvent interactions, and Coulombic interactions are given by

\begin{equation}
     \upsilon_{nx}^C(r) = q_n q_x/r
     \label{coul-term}
\end{equation}
where $q_n$ denotes the NP charge uniformly distributed over its surface. For the case $x=s$ in Eq. \ref{coul-term}, $q_s$ shows the partial charge of the solvent interaction site. In this work, the solvent is represented by a two-site model water model, \cite{Dyer2009} consisting of two interaction sites representing oxygen and hydrogen atoms and the bond length between the oxygen and hydrogen atoms is 1 \AA. See Table \ref{pot_param} for more details. 

When the NPs are solvated in a liquid characterized by $\sigma_s$ and $\epsilon_s$ for the interaction between liquid particles, the interaction between the LJ-type solvent and NP is determined by integrating the interaction between the LJ interaction sites of liquid particles and the LJ particles within a NP, which results in

\begin{equation}
    \begin{split} 
    \upsilon^{\mathrm{LJ}}_{ns} (r) = & \frac{2}{9} \frac{R^3 \sigma_{ns}^3 A_{ns}}{(R^2 - r^2)^3} \\ 
    & \left[ 1- \frac{(5 R^6 + 45 R^4 r^2 + 63 R^2 r^4 + 15 r^6)\sigma_{ns}^6 }{15(R^2-r^2)^6}  \right],
    \end{split}
    \label{Uns}
\end{equation}
where $r$ is the distance between the solvent LJ interaction site and the NP center, and the Hamaker constant $A_{ns}$ is given by

\begin{equation}
    A_{ns} = 24 \pi \epsilon_{ns} \rho_n \sigma_{ns}^3, 
    \label{Ans}
\end{equation}
where $\sigma_{ns} = (\sigma_n + \sigma_s)/2$ and $\epsilon_{ns}$ is the interaction between a LJ-type solvent particle and a particle in the nanoparticle given by  $\epsilon_{ns} = \sqrt{\epsilon_n  \epsilon_s}$. Simple estimations of magnetic $dd$ interactions between the effective magnetic dipole moment of the nanoscale domains in the NPs and the induced magnetic dipole moment of water shows that their magnitudes are several orders of magnitudes less than their electrostatics and vdW interactions. Therefore, one may neglect these magnetic $dd$ interactions in the cDFT approach.\cite{Chuev-2022}

\begin{table}[tbh]
\centering
\caption{Interaction parameters for the nanoparticle, and water. Note that for the numerical stability of cDFT calculations, the hydrogen site has $\epsilon$ and $\sigma$ values of $0.046$ kcal/mol and 0, respectively.$^a$ In molecular dynamics simulations, the hydrogen site does not have any LJ parameters.}
\begin{tabular}{cccc}
     Site & $\epsilon$[kcal mol$^{-1}$] & $\sigma$[\AA] & $q$[e]  \\
     \hline
     DyCl$_3$ & $0.0916$ & $4.775$ & varies \\
     O & $0.1554$ & $3.16$ & $-0.38$ \\
     H & $0.046^a$ & $0^a$ & $0.38$ \\
     \hline
\end{tabular}
\label{pot_param}
\end{table}

As described in the main text, it is mostly assumed that the NPs consist of neutral hydrated clusters of MCl$_3$. It is beyond the scope of this study to determine the structures of the hydrated clusters in electrolyte solutions, which is an active area of research.\cite{Kathmann-2019,Hartel-2023,Komori-2023,Safran-2023,Dinpajooh-2024,Jin-2024} Here we focus mainly on a simple model to provide insights on the significance of magnetic $dd$ interactions when compared to electrostatics and vdW interactions.  
Assuming a tetrahedral configuration for DyCl$_3$, one may estimate $\sigma_{\rm DyCl_3}=\sigma_{\rm Dy^{3+}}+\frac{\sqrt{3}}{4} \sigma_{\rm Cl^{-}}$ and $\epsilon_{\rm DyCl_3}=\sqrt{\epsilon_{\rm Dy^{3+}} \epsilon_{\rm Cl^{-}}}$, where ion-water vdW interactions inside the neutral cluster are not treated explicitly for simplicity. It is an interesting future research project to develop realistic models of hydrated-ion clusters that constitute NPs.

To treat the van der Waals (vdW) interactions between the NPs, we use a Hamaker approach in a vacuum and assume that NPs consist of a uniform distribution of atoms interacting with each other by a LJ potential with $\epsilon_n$ and $\sigma_n$ parameters. \cite{Hamaker1937,Grest2008,Cheng-2012,Everaers2003} 
For spherical NPs, the interaction between NPs can then be obtained analytically by integrating over all the interacting LJ particles within the two NPs\cite{Hamaker1937,Grest2008,Cheng2012,Everaers2003} as 

\begin{equation}
   \upsilon_{nn}^{\mathrm{LJ}}(r) = \upsilon^{\rm{A}}_{nn}(r) + \upsilon^{\rm{R}}_{nn}(r),
    \label{Unn}
\end{equation}
where $r$ is the distance between the centers of the nanoparticles, $\upsilon^{\rm{A}}_{nn}(r)$ and $\upsilon^{\rm{R}}_{nn}(r)$ are the attractive (long-range) and repulsive (short-range) parts of the interaction between nanoparticles, respectively and for nanoparticles of radii $R_a$ and $R_b$ they are given by:

\begin{equation}
  \begin{split}
     \upsilon^{\rm{A}}_{nn}(r) = & -\frac{A_{nn}}{6} \bigg[ 
     \frac{2 R_a R_b}{r^2-(R_a+R_b)^2} 
     + \frac{2R_a R_b}{r^2-(R_a-R_b)^2} \\ 
     & + \ln \left(\frac{r^2-(R_a+R_b)^2}{r^2-(R_a-R_b)^2}\right) 
     \bigg]
     \label{Uatt}
  \end{split}
\end{equation}

and 

\begin{equation}
  \begin{split}
     \upsilon^{\rm{R}}_{nn}(r) & = \frac{A_{nn}}{37800} \\
     & \frac{\sigma_n^6}{r}  \left[ \frac{r^2 - 7 r (R_a+R_b) + 6 (R_a^2 + 7R_aR_b+R_b^2) }{(r-R_a-R_b)^7} \right. \\ 
     & + \frac{r^2 + 7 r (R_a+R_b) + 6 (R_a^2 + 7R_aR_b+R_b^2) }{(r+R_a+R_b)^7} \\
     & - \frac{r^2 + 7 r (R_a-R_b) + 6 (R_a^2 - 7R_aR_b+R_b^2) }{(r+R_a-R_b)^7} \\
     & \left. - \frac{r^2 - 7 r (R_a-R_b) + 6 (R_a^2 - 7R_aR_b+R_b^2) }{(r-R_a+R_b)^7} \right] 
  \end{split}
  \label{Urep}
\end{equation}
where $A_{nn}$ is the Hamaker constant given by $A_{nn} = 4\pi^2 \epsilon_n \rho_a \rho_b \sigma_n^6$ with $\rho_a$ and $\rho_b$ as the number densities of LJ atoms within NPs of the same type $a$ and $b$, respectively.
The Hamaker constant above, $A_{nn}$, can also be calculated from a Lifshitz-Hamaker approach\cite{Hamaker1937,Evans-1999,Priye-2013,Lee-2023} , but here we use a simple approach based on the integration of LJ particles with uniform distributions in NPs.\cite{Everaers2003,Grest2008}
The densities of particles within the NPs were set to $\sigma_n^{-3}$.
When NPs are characterized by such an intermolecular potential, one can show that the van der Waals potential between two NPs has a long-range weak attractive tail.
The attractive part of the vdW interactions as presented in Eq. \ref{Uatt} can be related to the Lifshitz theory,\cite{Evans-1999} but the repulsive part of vdW interactions may be chosen as Eq. \ref{Urep} to represent the excluded volumes for the NPs similar to previous works.\cite{Henderson-1997,Rabani-2002, Everaers2003, Grest2008} Such models may be further improved by considering that the two colloidal particles/NPs with given radii can have contact with each other in a vacuum, to be consistent with the experimental observations for colloid particles.\cite{Liang2007} 
Note also that the repulsive bare interaction (Eq. \ref{Urep}) is mathematically correct but physically it is not originated from a short-range repulsion between two atoms that would indeed give an "overlapping" potential, which is not commonly used in the colloids community.
In addition, the polarization multibody effects may become important and both the attractive and repulsive potential functions need to be modified to include such effects.\cite{Batista2015} Therefore, it would be interesting to apply the cDFT approach to problems with more accurate and detailed potentials in future work.

It is noteworthy that for two NPs of equal radii $R$ whose edges are at a small distance $h$ apart ($R \gg  h$), Eqs. \ref{Uatt} and \ref{Urep} can be simplified to $-A_{nn}R/(12h)$ and  $A_{nn}R (\sigma_n/h )^6/(2520h)$, respectively, while for two NPs far apart ($h \gg R$) the interaction energies vary as $-1/h^6$ and $1/h^{12}$, respectively as for two small particles. At intermediate separations ($R \approx h$) the more complicated Eqs. \ref{Uatt} and \ref{Urep} cannot be easily simplified.

\section{Molecular Dynamics Simulation Details}
The molecular dynamics (MD) simulations were performed using the LAMMPS software program\cite{Plimpton1995} for nanoparticles (NPs) in the S2 water model at $298$ K and a density of $0.997$ g/cm$^3$.
Details about the water-NP interactions as well as NP-NP interactions are presented in Section \ref{sec:nano_pot}.
The periodic boundary conditions were applied in 3-dimensions and the MD simulations were performed in the canonical ensemble (NVT) using the Nos$\rm{\acute{e}}$-Hoover thermostat with various box lengths such as $62.15$ \AA\ consisting of $8000$ water molecules.
The standard velocity-Verlet time integrator was used with a time step of $2$ fs and the SHAKE procedure was used to conserve the intramolecular constraints.
A cutoff distance of $14$ \AA\ was used to truncate the Lennard Jones interactions for solvent-solvent interactions while a cutoff distance of $3.0-3.6\: R$ was used to truncate the solvent-NP interactions with $R$ as the radius of the NP. Finally, a cutoff distance of about $5 R$ was used to truncate NP-NP interactions due to long-range attractive tail of the NP potential. 
Due to this wide disparity in cutoff distances the multi neighbor-finding and inter-processor communication algorithms were used.\cite{Pieter-2008}
The conducting metal (tinfoil) boundary conditions were used to treat the electrostatic interactions and the particle-particle particle-mesh solver was used.
 the MD simulation results are directly obtained by solving the equations of motions at a given state point using the NP-solvent and solvent-solvent interactions for a relatively large number of solvent molecules and one NP.

Different methods are used to calculate the potential of mean forces (effective interactions) from molecular dynamics (MD) simulations.
For the NP-solvent effective interactions, the normalized site densities, $g_i(r) = \rho_i(r)/\rho_0$, can be simply used to get the effective interactions through $-k_{\mathrm{B}}T \ln(g_i(r))$, where $\rho_0$ is the bulk solvent density. Here we focus on NP-NP effective interaction.
and apply the harmonic biasing/umbrella sampling method to enforce fixed or moving restraints, where the restraints are introduced on the collective variable $\xi$, which is the distance vector between the nanoparticle centers.
A force constant of $10$ kcal/mol/\AA$^2$~ is used with the moving restraints, where the centers of the harmonic restraints are changed during MD simulations (usually from faraway distances to closer distances) in discrete stages, where each stage consists of $10000$ MD steps. The initial configuration is equilibriated for $10^5$ MD steps and a grid spacing of $0.1$ \AA\ is used to discretize the distance between the nanoparticles.
The weighted histogram analysis (WHAM) method\cite{Kumar1992,wham-code} as well as the thermodynamics integration methods that are available in the colvars package\cite{Plimpton1995,Fiorin2013,Abrams2014,Barducci2011} are used to get the free energies. In the WHAM method, the Jacobian term $2 k_{\mathrm{B}} T \ln(r)$ is added to the free energies to get the effective interactions while in the methods available in the colvars package, the Jacobian terms are already included in the outputs. In this work, the MD effective interactions between two NPs are reported based on the thermodynamics integration method with the moving restraints.
For comparison with the cDFT effective interactions, the MD effective interactions are shifted to the corresponding cDFT effective interactions at large distances. 

\section{The classical density functional theory (cDFT) Calculation Details}

The cDFT calculations were performed using a version of cdftpy software program,\cite{Chuev-2022} which utilized an equidistant radial grid with the grid spacing equal to 0.01 \AA. The grid extended to  $10000$ \AA\ for the studied NPs. 
The inputs to the cDFT calculations are the NP-solvent interaction parameters (see Table \ref{pot_param}) as well as the homogeneous structure factor of the water, $S(Q)$, at a given state point, which is obtained/estimated from MD simulations considering the low $Q$ behavior of structure factors with $S(0)=0.044$ for the water model presented in Table \ref{pot_param}. The linear response approximation as well as the HyperNetted-Chain (HNC) closure are then used to solve the self-consistent equations in the cDFT calculations.
To compare the cDFT results with the MD simulation results, it is worth noting that the MD results are directly obtained by solving the equations of motions at a given state point using the NP-solvent and solvent-solvent interactions for a relatively large number MD steps to get converged results while the cDFT calculations relies on appropriate parameters.
Finally, the forces between NPs can be obtained from the gradients of the effective interactions between NPs.

\section{Accuracy of cDFT Calculations}

\begin{figure}[tbh]
\centering
\includegraphics[width=\columnwidth]{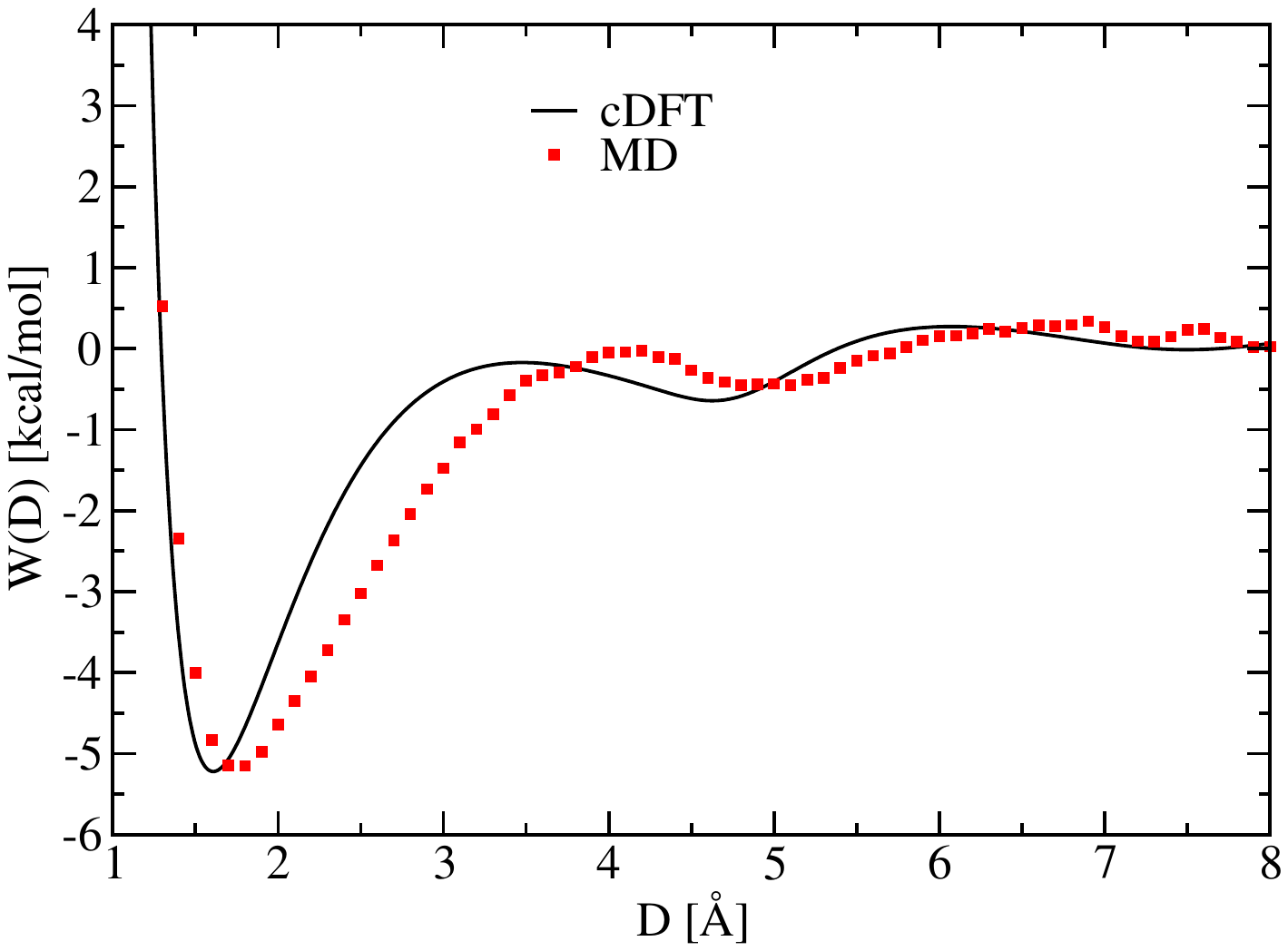}
\caption{Comparison of the NP-NP potential of mean forces (effective interactions) from the molecular dynamics simulations (filled squares) and the cDFT (solid lines) for DyCl$_3$-type NPs with radii of $5$ \AA\ in the absence of magnetic field. The NP-NP effective interactions are plotted as functions of the separation from the surface of NPs.
}
\label{fig:accuracy-np-np}
\end{figure}

The accuracy of the cDFT approach is demonstrated in Fig. \ref{fig:accuracy-np-np}, where the cDFT effective interactions for relatively small NPs in the absence of magnetic fields can be compared with the calculated ones with the MD simulations, i.e. NPs with radii of $5$ \AA. 
The general features/typologies of NP-NP MD effective interactions are well captures in cDFT NP-NP effective interactions and they involve the very short-range repulsive behavior, which becomes attractive at the first extreme point followed by their oscillatory behaviors up to distances around NP size.
The very short-range repulsive cDFT NP-NP effective interactions overlap with the MD NP-NP effective interactions. The positions and depth magnitudes of the first minima of the cDFT NP-NP effective interactions only slightly deviate from the MD NP-NP effective interactions.
Less agreement between the cDFT and MD NP-NP effective interactions is observed for the first maxima. The positions of the first maxima for the cDFT NP-NP effective interactions appear at shorter distances with the percentage absolute differences of about $15-20$ \%. Therefore, the cDFT NP-NP effective interactions have more compressed typologies than the MD NP-NP effective interactions, which can be traced back to the more compressed cDFT NP-solvent structures.\cite{Chuev-2022} However, less deviation is observed for the differences between the positions of the first maxima and second maxima of the cDFT and MD NP-NP effective interactions (about $10$ \%). This shows that the more compressed cDFT NP-NP structures at short distances are the main differences between cDFT and MD NP-NP effective interactions. Nevertheless, the long-range parts of cDFT effective interactions are probably more reliable than the MD NP-NP effective interactions given the uncertainties in MD simulations. It is worth mentioning that the cDFT calculations for these systems are about $10^6-10^7$ times faster than the corresponding MD calculations.\cite{Chuev-2022}

\section{LMNP model for relatively small NPs}

\begin{figure}[tbh]
\centering
\includegraphics[width=\columnwidth]{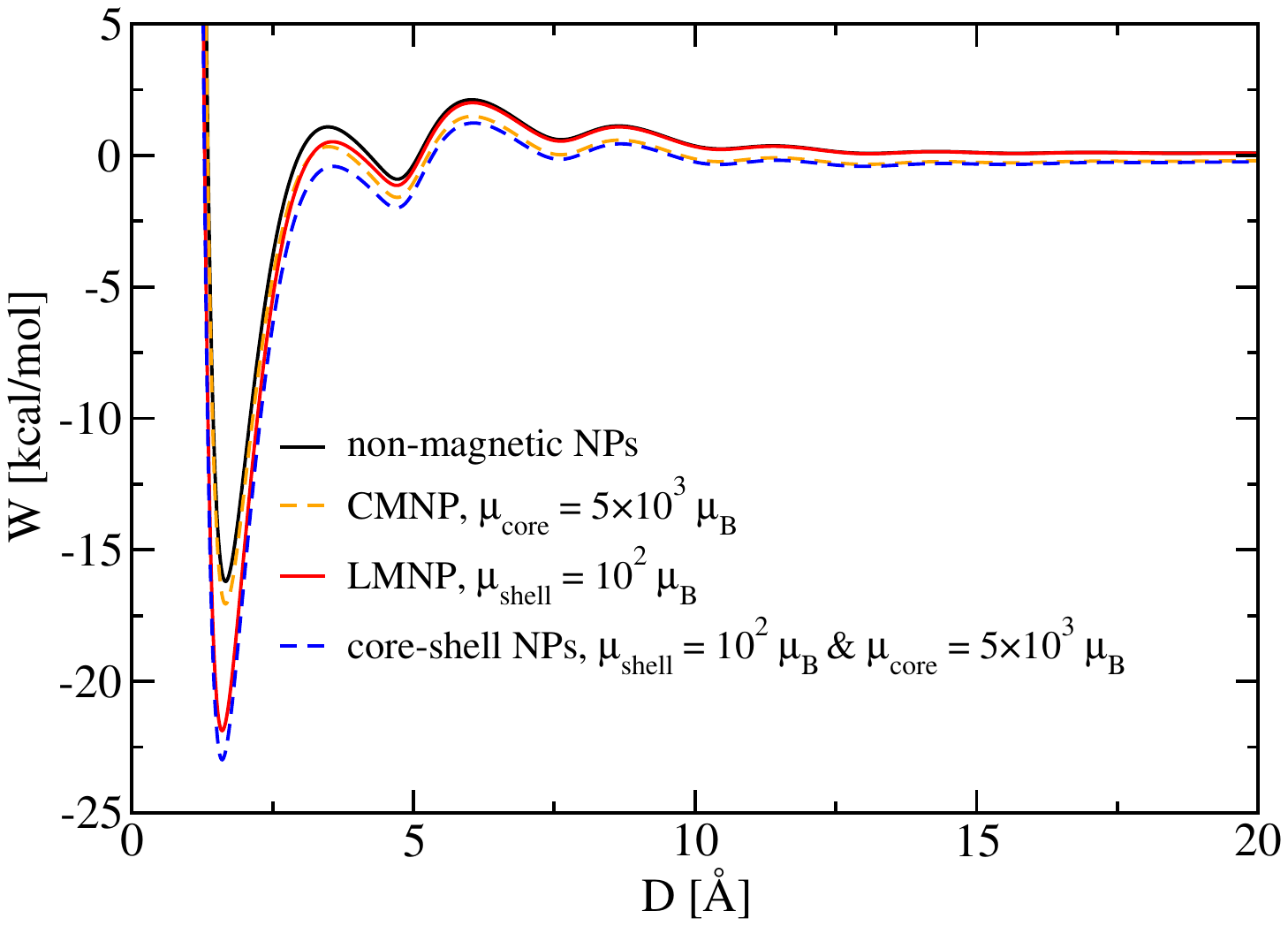}
\caption{Comparisons of effective interactions between different models of NPs with radii of $2$ nm and various effective magnetic dipole moments, $\mu_{\rm eff}$. They include non-magnetic NPs, NPs whose effective magnetic dipoles with $\mu_{\rm core}=5\times10^5\mu_{\rm B}$ are located at their centers, and the NPs whose effective magnetic interactions are located at their surfaces (the LMNP model) with magnitudes of $\mu_{\rm eff}=\mu_{\rm shell}=100\mu_{\rm B}$. All the magnetic dipoles  are aligned with the external magnetic field similar to setup shown in panel {\bf c} of Fig. $1$ in the main text.
}
\label{fig:LMNPbreak}
\end{figure}

In the main text, we show that the effective interaction potential between the LMNP model are in good agreement with the one for the core-shell model, which includes both magnetic dipoles at the surfaces and at the centers of NPs. When compared to the CMNP model, we show that for relatively small NPs with given magnitudes of magnetic dipoles, the LMNP model show relatively small deviations with respect to the core-shell model as the reference. At $D<\sim5$ \AA, Fig. \ref{fig:LMNPbreak} shows that the LMNP model are in better agreement with the core-shell NPs whose center magnetic dipole moments are $50$ times larger than the surface magnetic dipole moments. However, one can notice the values of effective interactions for the LMNP model at extreme points can deviate from the core-shell NPs by about $1.1$ kcal/mol at $D\sim 1.6$ \AA, which reduces to about $0.9$ kcal/mol at $D\sim 3.5$ \AA. On the other hand, the effective interaction between the CMNPs are in better agreement with the core-shell NPs at $D>\sim 5$ \AA . Overall the deviations of the effective interactions between the LMNPs and the core-shell NPs at $D>\sim5$ \AA\ do not exceed more than $\sim 1.5$ $k_{\rm B}T$.

\begin{thebibliography}{106}%
\makeatletter
\providecommand \@ifxundefined [1]{%
 \@ifx{#1\undefined}
}%
\providecommand \@ifnum [1]{%
 \ifnum #1\expandafter \@firstoftwo
 \else \expandafter \@secondoftwo
 \fi
}%
\providecommand \@ifx [1]{%
 \ifx #1\expandafter \@firstoftwo
 \else \expandafter \@secondoftwo
 \fi
}%
\providecommand \natexlab [1]{#1}%
\providecommand \enquote  [1]{``#1''}%
\providecommand \bibnamefont  [1]{#1}%
\providecommand \bibfnamefont [1]{#1}%
\providecommand \citenamefont [1]{#1}%
\providecommand \href@noop [0]{\@secondoftwo}%
\providecommand \href [0]{\begingroup \@sanitize@url \@href}%
\providecommand \@href[1]{\@@startlink{#1}\@@href}%
\providecommand \@@href[1]{\endgroup#1\@@endlink}%
\providecommand \@sanitize@url [0]{\catcode `\\12\catcode `\$12\catcode `\&12\catcode `\#12\catcode `\^12\catcode `\_12\catcode `\%12\relax}%
\providecommand \@@startlink[1]{}%
\providecommand \@@endlink[0]{}%
\providecommand \url  [0]{\begingroup\@sanitize@url \@url }%
\providecommand \@url [1]{\endgroup\@href {#1}{\urlprefix }}%
\providecommand \urlprefix  [0]{URL }%
\providecommand \Eprint [0]{\href }%
\providecommand \doibase [0]{http://dx.doi.org/}%
\providecommand \selectlanguage [0]{\@gobble}%
\providecommand \bibinfo  [0]{\@secondoftwo}%
\providecommand \bibfield  [0]{\@secondoftwo}%
\providecommand \translation [1]{[#1]}%
\providecommand \BibitemOpen [0]{}%
\providecommand \bibitemStop [0]{}%
\providecommand \bibitemNoStop [0]{.\EOS\space}%
\providecommand \EOS [0]{\spacefactor3000\relax}%
\providecommand \BibitemShut  [1]{\csname bibitem#1\endcsname}%
\let\auto@bib@innerbib\@empty
\bibitem [{\citenamefont {Fujiwara}\ \emph {et~al.}(2001)\citenamefont {Fujiwara}, \citenamefont {Kodoi}, \citenamefont {Duan},\ and\ \citenamefont {Tanimoto}}]{Fujiwara-2001}%
  \BibitemOpen
  \bibfield  {author} {\bibinfo {author} {\bibfnamefont {M.}~\bibnamefont {Fujiwara}}, \bibinfo {author} {\bibfnamefont {D.}~\bibnamefont {Kodoi}}, \bibinfo {author} {\bibfnamefont {W.}~\bibnamefont {Duan}}, \ and\ \bibinfo {author} {\bibfnamefont {Y.}~\bibnamefont {Tanimoto}},\ }\href {\doibase 10.1021/jp003562d} {\bibfield  {journal} {\bibinfo  {journal} {J. Phys. Chem. B}\ }\textbf {\bibinfo {volume} {105}},\ \bibinfo {pages} {3343} (\bibinfo {year} {2001})}\BibitemShut {NoStop}%
\bibitem [{\citenamefont {Fujiwara}\ \emph {et~al.}(2006)\citenamefont {Fujiwara}, \citenamefont {Mitsuda},\ and\ \citenamefont {Tanimoto}}]{Fujiwara-2006}%
  \BibitemOpen
  \bibfield  {author} {\bibinfo {author} {\bibfnamefont {M.}~\bibnamefont {Fujiwara}}, \bibinfo {author} {\bibfnamefont {K.}~\bibnamefont {Mitsuda}}, \ and\ \bibinfo {author} {\bibfnamefont {Y.}~\bibnamefont {Tanimoto}},\ }\href {\doibase 10.1021/jp061482p} {\bibfield  {journal} {\bibinfo  {journal} {J. Phys. Chem. B}\ }\textbf {\bibinfo {volume} {110}},\ \bibinfo {pages} {13965} (\bibinfo {year} {2006})}\BibitemShut {NoStop}%
\bibitem [{\citenamefont {Pulko}\ \emph {et~al.}(2014)\citenamefont {Pulko}, \citenamefont {Yang}, \citenamefont {Lei}, \citenamefont {Odenbach},\ and\ \citenamefont {Eckert}}]{Pulko-2014}%
  \BibitemOpen
  \bibfield  {author} {\bibinfo {author} {\bibfnamefont {B.}~\bibnamefont {Pulko}}, \bibinfo {author} {\bibfnamefont {X.}~\bibnamefont {Yang}}, \bibinfo {author} {\bibfnamefont {Z.}~\bibnamefont {Lei}}, \bibinfo {author} {\bibfnamefont {S.}~\bibnamefont {Odenbach}}, \ and\ \bibinfo {author} {\bibfnamefont {K.}~\bibnamefont {Eckert}},\ }\href {\doibase 10.1063/1.4903794} {\bibfield  {journal} {\bibinfo  {journal} {Appl. Phys. Lett}\ }\textbf {\bibinfo {volume} {105}},\ \bibinfo {pages} {232407} (\bibinfo {year} {2014})}\BibitemShut {NoStop}%
\bibitem [{\citenamefont {Rodrigues}\ \emph {et~al.}(2017)\citenamefont {Rodrigues}, \citenamefont {Lukina}, \citenamefont {Dehaeck}, \citenamefont {Colinet}, \citenamefont {Binnemans},\ and\ \citenamefont {Fransaer}}]{Rodrigues-2017}%
  \BibitemOpen
  \bibfield  {author} {\bibinfo {author} {\bibfnamefont {I.~R.}\ \bibnamefont {Rodrigues}}, \bibinfo {author} {\bibfnamefont {L.}~\bibnamefont {Lukina}}, \bibinfo {author} {\bibfnamefont {S.}~\bibnamefont {Dehaeck}}, \bibinfo {author} {\bibfnamefont {P.}~\bibnamefont {Colinet}}, \bibinfo {author} {\bibfnamefont {K.}~\bibnamefont {Binnemans}}, \ and\ \bibinfo {author} {\bibfnamefont {J.}~\bibnamefont {Fransaer}},\ }\href {\doibase 10.1021/acs.jpclett.7b02226} {\bibfield  {journal} {\bibinfo  {journal} {J. Phys. Chem. Lett.}\ }\textbf {\bibinfo {volume} {8}},\ \bibinfo {pages} {5301} (\bibinfo {year} {2017})}\BibitemShut {NoStop}%
\bibitem [{\citenamefont {Rodrigues}\ \emph {et~al.}(2018)\citenamefont {Rodrigues}, \citenamefont {Lukina}, \citenamefont {Dehaeck}, \citenamefont {Colinet}, \citenamefont {Binnemans},\ and\ \citenamefont {Fransaer}}]{Rodrigues-2018}%
  \BibitemOpen
  \bibfield  {author} {\bibinfo {author} {\bibfnamefont {I.~R.}\ \bibnamefont {Rodrigues}}, \bibinfo {author} {\bibfnamefont {L.}~\bibnamefont {Lukina}}, \bibinfo {author} {\bibfnamefont {S.}~\bibnamefont {Dehaeck}}, \bibinfo {author} {\bibfnamefont {P.}~\bibnamefont {Colinet}}, \bibinfo {author} {\bibfnamefont {K.}~\bibnamefont {Binnemans}}, \ and\ \bibinfo {author} {\bibfnamefont {J.}~\bibnamefont {Fransaer}},\ }\href {\doibase 10.1021/acs.jpcc.8b06471} {\bibfield  {journal} {\bibinfo  {journal} {J. Phys. Chem. C}\ }\textbf {\bibinfo {volume} {122}},\ \bibinfo {pages} {23675} (\bibinfo {year} {2018})}\BibitemShut {NoStop}%
\bibitem [{\citenamefont {Rodrigues}\ \emph {et~al.}(2019)\citenamefont {Rodrigues}, \citenamefont {Lukina}, \citenamefont {Dehaeck}, \citenamefont {Colinet}, \citenamefont {Binnemans},\ and\ \citenamefont {Fransaer}}]{Rodrigues-2019}%
  \BibitemOpen
  \bibfield  {author} {\bibinfo {author} {\bibfnamefont {I.~R.}\ \bibnamefont {Rodrigues}}, \bibinfo {author} {\bibfnamefont {L.}~\bibnamefont {Lukina}}, \bibinfo {author} {\bibfnamefont {S.}~\bibnamefont {Dehaeck}}, \bibinfo {author} {\bibfnamefont {P.}~\bibnamefont {Colinet}}, \bibinfo {author} {\bibfnamefont {K.}~\bibnamefont {Binnemans}}, \ and\ \bibinfo {author} {\bibfnamefont {J.}~\bibnamefont {Fransaer}},\ }\href {\doibase 10.1021/acs.jpcc.9b06706} {\bibfield  {journal} {\bibinfo  {journal} {J. Phys. Chem. C}\ }\textbf {\bibinfo {volume} {123}},\ \bibinfo {pages} {23131} (\bibinfo {year} {2019})}\BibitemShut {NoStop}%
\bibitem [{\citenamefont {Butcher}\ and\ \citenamefont {Coey}(2022)}]{Butcher-2023}%
  \BibitemOpen
  \bibfield  {author} {\bibinfo {author} {\bibfnamefont {T.~A.}\ \bibnamefont {Butcher}}\ and\ \bibinfo {author} {\bibfnamefont {J.~M.~D.}\ \bibnamefont {Coey}},\ }\href {\doibase 10.1088/1361-648X/aca37f} {\bibfield  {journal} {\bibinfo  {journal} {J. Phys. Condens. Matter}\ }\textbf {\bibinfo {volume} {35}},\ \bibinfo {pages} {053002} (\bibinfo {year} {2022})}\BibitemShut {NoStop}%
\bibitem [{\citenamefont {Rassolov}\ \emph {et~al.}(2024)\citenamefont {Rassolov}, \citenamefont {Ali}, \citenamefont {Siegrist}, \citenamefont {Humayun},\ and\ \citenamefont {Mohammadigoushki}}]{Rassolov-2024}%
  \BibitemOpen
  \bibfield  {author} {\bibinfo {author} {\bibfnamefont {P.}~\bibnamefont {Rassolov}}, \bibinfo {author} {\bibfnamefont {J.}~\bibnamefont {Ali}}, \bibinfo {author} {\bibfnamefont {T.}~\bibnamefont {Siegrist}}, \bibinfo {author} {\bibfnamefont {M.}~\bibnamefont {Humayun}}, \ and\ \bibinfo {author} {\bibfnamefont {H.}~\bibnamefont {Mohammadigoushki}},\ }\href {\doibase 10.1039/D3SM01607B} {\bibfield  {journal} {\bibinfo  {journal} {Soft Matter}\ }\textbf {\bibinfo {volume} {20}},\ \bibinfo {pages} {2496} (\bibinfo {year} {2024})}\BibitemShut {NoStop}%
\bibitem [{\citenamefont {Weston}\ \emph {et~al.}(2010)\citenamefont {Weston}, \citenamefont {Gerner},\ and\ \citenamefont {Fritsch}}]{Weston-2010}%
  \BibitemOpen
  \bibfield  {author} {\bibinfo {author} {\bibfnamefont {M.~C.}\ \bibnamefont {Weston}}, \bibinfo {author} {\bibfnamefont {M.~D.}\ \bibnamefont {Gerner}}, \ and\ \bibinfo {author} {\bibfnamefont {I.}~\bibnamefont {Fritsch}},\ }\href {\doibase 10.1021/ac901783n} {\bibfield  {journal} {\bibinfo  {journal} {Anal. Chem.}\ }\textbf {\bibinfo {volume} {82}},\ \bibinfo {pages} {3411} (\bibinfo {year} {2010})}\BibitemShut {NoStop}%
\bibitem [{\citenamefont {Yuan}\ \emph {et~al.}(2024)\citenamefont {Yuan}, \citenamefont {Rampal}, \citenamefont {Du}, \citenamefont {Shu}, \citenamefont {Wang}, \citenamefont {Wang}, \citenamefont {Stack}, \citenamefont {Ben~Ishai}, \citenamefont {Anovitz},\ and\ \citenamefont {Xu}}]{Anovitz-2024}%
  \BibitemOpen
  \bibfield  {author} {\bibinfo {author} {\bibfnamefont {K.}~\bibnamefont {Yuan}}, \bibinfo {author} {\bibfnamefont {N.}~\bibnamefont {Rampal}}, \bibinfo {author} {\bibfnamefont {X.}~\bibnamefont {Du}}, \bibinfo {author} {\bibfnamefont {F.}~\bibnamefont {Shu}}, \bibinfo {author} {\bibfnamefont {Y.}~\bibnamefont {Wang}}, \bibinfo {author} {\bibfnamefont {H.}~\bibnamefont {Wang}}, \bibinfo {author} {\bibfnamefont {A.~G.}\ \bibnamefont {Stack}}, \bibinfo {author} {\bibfnamefont {P.}~\bibnamefont {Ben~Ishai}}, \bibinfo {author} {\bibfnamefont {L.~M.}\ \bibnamefont {Anovitz}}, \ and\ \bibinfo {author} {\bibfnamefont {P.}~\bibnamefont {Xu}},\ }\href {\doibase 10.1039/D4CP02041C} {\bibfield  {journal} {\bibinfo  {journal} {Phys. Chem. Chem. Phys.}\ }\textbf {\bibinfo {volume} {26}},\ \bibinfo {pages} {27891} (\bibinfo {year} {2024})}\BibitemShut {NoStop}%
\bibitem [{\citenamefont {Ilgen}\ \emph {et~al.}(2024)\citenamefont {Ilgen}, \citenamefont {Borguet}, \citenamefont {Geiger}, \citenamefont {Gibbs}, \citenamefont {Grassian}, \citenamefont {Jun}, \citenamefont {Kabengi},\ and\ \citenamefont {Kubicki}}]{Gibbs-2024}%
  \BibitemOpen
  \bibfield  {author} {\bibinfo {author} {\bibfnamefont {A.~G.}\ \bibnamefont {Ilgen}}, \bibinfo {author} {\bibfnamefont {E.}~\bibnamefont {Borguet}}, \bibinfo {author} {\bibfnamefont {F.~M.}\ \bibnamefont {Geiger}}, \bibinfo {author} {\bibfnamefont {J.~M.}\ \bibnamefont {Gibbs}}, \bibinfo {author} {\bibfnamefont {V.~H.}\ \bibnamefont {Grassian}}, \bibinfo {author} {\bibfnamefont {Y.-S.}\ \bibnamefont {Jun}}, \bibinfo {author} {\bibfnamefont {N.}~\bibnamefont {Kabengi}}, \ and\ \bibinfo {author} {\bibfnamefont {J.~D.}\ \bibnamefont {Kubicki}},\ }\href {\doibase 10.1038/s41467-024-49598-y} {\bibfield  {journal} {\bibinfo  {journal} {Nat. Commun.}\ }\textbf {\bibinfo {volume} {15}},\ \bibinfo {pages} {5326} (\bibinfo {year} {2024})}\BibitemShut {NoStop}%
\bibitem [{\citenamefont {Omta}\ \emph {et~al.}(2003)\citenamefont {Omta}, \citenamefont {Kropman}, \citenamefont {Woutersen},\ and\ \citenamefont {Bakker}}]{Omta-2003}%
  \BibitemOpen
  \bibfield  {author} {\bibinfo {author} {\bibfnamefont {A.~W.}\ \bibnamefont {Omta}}, \bibinfo {author} {\bibfnamefont {M.~F.}\ \bibnamefont {Kropman}}, \bibinfo {author} {\bibfnamefont {S.}~\bibnamefont {Woutersen}}, \ and\ \bibinfo {author} {\bibfnamefont {H.~J.}\ \bibnamefont {Bakker}},\ }\href {\doibase 10.1126/science.1084801} {\bibfield  {journal} {\bibinfo  {journal} {Science}\ }\textbf {\bibinfo {volume} {301}},\ \bibinfo {pages} {347} (\bibinfo {year} {2003})}\BibitemShut {NoStop}%
\bibitem [{\citenamefont {Turton}\ \emph {et~al.}(2008)\citenamefont {Turton}, \citenamefont {Hunger}, \citenamefont {Hefter}, \citenamefont {Buchner},\ and\ \citenamefont {Wynne}}]{Turton-2008}%
  \BibitemOpen
  \bibfield  {author} {\bibinfo {author} {\bibfnamefont {D.~A.}\ \bibnamefont {Turton}}, \bibinfo {author} {\bibfnamefont {J.}~\bibnamefont {Hunger}}, \bibinfo {author} {\bibfnamefont {G.}~\bibnamefont {Hefter}}, \bibinfo {author} {\bibfnamefont {R.}~\bibnamefont {Buchner}}, \ and\ \bibinfo {author} {\bibfnamefont {K.}~\bibnamefont {Wynne}},\ }\href {\doibase 10.1063/1.2906132} {\bibfield  {journal} {\bibinfo  {journal} {J. Chem. Phys.}\ }\textbf {\bibinfo {volume} {128}},\ \bibinfo {pages} {161102} (\bibinfo {year} {2008})}\BibitemShut {NoStop}%
\bibitem [{\citenamefont {Georgalis}\ \emph {et~al.}(2000)\citenamefont {Georgalis}, \citenamefont {Kierzek},\ and\ \citenamefont {Saenger}}]{Georgalis-2000}%
  \BibitemOpen
  \bibfield  {author} {\bibinfo {author} {\bibfnamefont {Y.}~\bibnamefont {Georgalis}}, \bibinfo {author} {\bibfnamefont {A.~M.}\ \bibnamefont {Kierzek}}, \ and\ \bibinfo {author} {\bibfnamefont {W.}~\bibnamefont {Saenger}},\ }\href {\doibase 10.1021/jp000132e} {\bibfield  {journal} {\bibinfo  {journal} {J. Phys. Chem. B}\ }\textbf {\bibinfo {volume} {104}},\ \bibinfo {pages} {3405} (\bibinfo {year} {2000})}\BibitemShut {NoStop}%
\bibitem [{\citenamefont {Kim}\ \emph {et~al.}(2014)\citenamefont {Kim}, \citenamefont {Kim}, \citenamefont {Choi},\ and\ \citenamefont {Cho}}]{Kim-2014}%
  \BibitemOpen
  \bibfield  {author} {\bibinfo {author} {\bibfnamefont {S.}~\bibnamefont {Kim}}, \bibinfo {author} {\bibfnamefont {H.}~\bibnamefont {Kim}}, \bibinfo {author} {\bibfnamefont {J.-H.}\ \bibnamefont {Choi}}, \ and\ \bibinfo {author} {\bibfnamefont {M.}~\bibnamefont {Cho}},\ }\href {\doibase 10.1063/1.4896227} {\bibfield  {journal} {\bibinfo  {journal} {J. Chem. Phys.}\ }\textbf {\bibinfo {volume} {141}},\ \bibinfo {pages} {124510} (\bibinfo {year} {2014})}\BibitemShut {NoStop}%
\bibitem [{\citenamefont {Fetisov}\ \emph {et~al.}(2019)\citenamefont {Fetisov}, \citenamefont {Isley}, \citenamefont {Lumetta},\ and\ \citenamefont {Kathmann}}]{Kathmann-2019}%
  \BibitemOpen
  \bibfield  {author} {\bibinfo {author} {\bibfnamefont {E.~O.}\ \bibnamefont {Fetisov}}, \bibinfo {author} {\bibfnamefont {W.~C.~I.}\ \bibnamefont {Isley}}, \bibinfo {author} {\bibfnamefont {G.~J.}\ \bibnamefont {Lumetta}}, \ and\ \bibinfo {author} {\bibfnamefont {S.~M.}\ \bibnamefont {Kathmann}},\ }\href {\doibase 10.1021/acs.jpcc.9b02494} {\bibfield  {journal} {\bibinfo  {journal} {J. Phys. Chem. C}\ }\textbf {\bibinfo {volume} {123}},\ \bibinfo {pages} {14010} (\bibinfo {year} {2019})}\BibitemShut {NoStop}%
\bibitem [{\citenamefont {Jin}\ \emph {et~al.}(2024)\citenamefont {Jin}, \citenamefont {Chen}, \citenamefont {Pyles}, \citenamefont {Baer}, \citenamefont {Legg}, \citenamefont {Wang}, \citenamefont {Washton}, \citenamefont {Mueller}, \citenamefont {Baker}, \citenamefont {Schenter}, \citenamefont {Mundy},\ and\ \citenamefont {De~Yoreo}}]{Jin-2024}%
  \BibitemOpen
  \bibfield  {author} {\bibinfo {author} {\bibfnamefont {B.}~\bibnamefont {Jin}}, \bibinfo {author} {\bibfnamefont {Y.}~\bibnamefont {Chen}}, \bibinfo {author} {\bibfnamefont {H.}~\bibnamefont {Pyles}}, \bibinfo {author} {\bibfnamefont {M.~D.}\ \bibnamefont {Baer}}, \bibinfo {author} {\bibfnamefont {B.~A.}\ \bibnamefont {Legg}}, \bibinfo {author} {\bibfnamefont {Z.}~\bibnamefont {Wang}}, \bibinfo {author} {\bibfnamefont {N.~M.}\ \bibnamefont {Washton}}, \bibinfo {author} {\bibfnamefont {K.~T.}\ \bibnamefont {Mueller}}, \bibinfo {author} {\bibfnamefont {D.}~\bibnamefont {Baker}}, \bibinfo {author} {\bibfnamefont {G.~K.}\ \bibnamefont {Schenter}}, \bibinfo {author} {\bibfnamefont {C.~J.}\ \bibnamefont {Mundy}}, \ and\ \bibinfo {author} {\bibfnamefont {J.~J.}\ \bibnamefont {De~Yoreo}},\ }\href {\doibase 10.1038/s41563-024-02025-5} {\bibfield  {journal} {\bibinfo  {journal} {Nat. Mater}\ } (\bibinfo {year} {2024}),\ 10.1038/s41563-024-02025-5}\BibitemShut {NoStop}%
\bibitem [{\citenamefont {Dinpajooh}\ \emph {et~al.}(2024{\natexlab{a}})\citenamefont {Dinpajooh}, \citenamefont {Biasin}, \citenamefont {Nienhuis}, \citenamefont {Mergelsberg}, \citenamefont {Benmore}, \citenamefont {Schenter}, \citenamefont {Fulton}, \citenamefont {Kathmann},\ and\ \citenamefont {Mundy}}]{Dinpajooh-2024}%
  \BibitemOpen
  \bibfield  {author} {\bibinfo {author} {\bibfnamefont {M.}~\bibnamefont {Dinpajooh}}, \bibinfo {author} {\bibfnamefont {E.}~\bibnamefont {Biasin}}, \bibinfo {author} {\bibfnamefont {E.~T.}\ \bibnamefont {Nienhuis}}, \bibinfo {author} {\bibfnamefont {S.~T.}\ \bibnamefont {Mergelsberg}}, \bibinfo {author} {\bibfnamefont {C.~J.}\ \bibnamefont {Benmore}}, \bibinfo {author} {\bibfnamefont {G.~K.}\ \bibnamefont {Schenter}}, \bibinfo {author} {\bibfnamefont {J.~L.}\ \bibnamefont {Fulton}}, \bibinfo {author} {\bibfnamefont {S.~M.}\ \bibnamefont {Kathmann}}, \ and\ \bibinfo {author} {\bibfnamefont {C.~J.}\ \bibnamefont {Mundy}},\ }\href {\doibase 10.1063/5.0234518} {\bibfield  {journal} {\bibinfo  {journal} {J. Chem. Phys.}\ }\textbf {\bibinfo {volume} {161}},\ \bibinfo {pages} {151102} (\bibinfo {year} {2024}{\natexlab{a}})}\BibitemShut {NoStop}%
\bibitem [{\citenamefont {Dinpajooh}\ \emph {et~al.}(2024{\natexlab{b}})\citenamefont {Dinpajooh}, \citenamefont {Intan}, \citenamefont {Duignan}, \citenamefont {Biasin}, \citenamefont {Fulton}, \citenamefont {Kathmann}, \citenamefont {Schenter},\ and\ \citenamefont {Mundy}}]{Dinpajooh-2024-2}%
  \BibitemOpen
  \bibfield  {author} {\bibinfo {author} {\bibfnamefont {M.}~\bibnamefont {Dinpajooh}}, \bibinfo {author} {\bibfnamefont {N.~N.}\ \bibnamefont {Intan}}, \bibinfo {author} {\bibfnamefont {T.~T.}\ \bibnamefont {Duignan}}, \bibinfo {author} {\bibfnamefont {E.}~\bibnamefont {Biasin}}, \bibinfo {author} {\bibfnamefont {J.~L.}\ \bibnamefont {Fulton}}, \bibinfo {author} {\bibfnamefont {S.~M.}\ \bibnamefont {Kathmann}}, \bibinfo {author} {\bibfnamefont {G.~K.}\ \bibnamefont {Schenter}}, \ and\ \bibinfo {author} {\bibfnamefont {C.~J.}\ \bibnamefont {Mundy}},\ }\href {\doibase 10.1063/5.0238708} {\bibfield  {journal} {\bibinfo  {journal} {J. Chem. Phys.}\ }\textbf {\bibinfo {volume} {161}},\ \bibinfo {pages} {230901} (\bibinfo {year} {2024}{\natexlab{b}})}\BibitemShut {NoStop}%
\bibitem [{\citenamefont {H\"artel}\ \emph {et~al.}(2023)\citenamefont {H\"artel}, \citenamefont {B\"ultmann},\ and\ \citenamefont {Coupette}}]{Hartel-2023}%
  \BibitemOpen
  \bibfield  {author} {\bibinfo {author} {\bibfnamefont {A.}~\bibnamefont {H\"artel}}, \bibinfo {author} {\bibfnamefont {M.}~\bibnamefont {B\"ultmann}}, \ and\ \bibinfo {author} {\bibfnamefont {F.}~\bibnamefont {Coupette}},\ }\href {\doibase 10.1103/PhysRevLett.130.108202} {\bibfield  {journal} {\bibinfo  {journal} {Phys. Rev. Lett.}\ }\textbf {\bibinfo {volume} {130}},\ \bibinfo {pages} {108202} (\bibinfo {year} {2023})}\BibitemShut {NoStop}%
\bibitem [{\citenamefont {Komori}\ and\ \citenamefont {Terao}(2023)}]{Komori-2023}%
  \BibitemOpen
  \bibfield  {author} {\bibinfo {author} {\bibfnamefont {K.}~\bibnamefont {Komori}}\ and\ \bibinfo {author} {\bibfnamefont {T.}~\bibnamefont {Terao}},\ }\href {\doibase https://doi.org/10.1016/j.cplett.2023.140627} {\bibfield  {journal} {\bibinfo  {journal} {Chem. Phys. Lett.}\ }\textbf {\bibinfo {volume} {825}},\ \bibinfo {pages} {140627} (\bibinfo {year} {2023})}\BibitemShut {NoStop}%
\bibitem [{\citenamefont {Safran}\ and\ \citenamefont {Pincus}(2023)}]{Safran-2023}%
  \BibitemOpen
  \bibfield  {author} {\bibinfo {author} {\bibfnamefont {S.~A.}\ \bibnamefont {Safran}}\ and\ \bibinfo {author} {\bibfnamefont {P.~A.}\ \bibnamefont {Pincus}},\ }\href {\doibase 10.1039/D3SM01094E} {\bibfield  {journal} {\bibinfo  {journal} {Soft Matter}\ }\textbf {\bibinfo {volume} {19}},\ \bibinfo {pages} {7907} (\bibinfo {year} {2023})}\BibitemShut {NoStop}%
\bibitem [{\citenamefont {Consta}(2002)}]{Consta-2002}%
  \BibitemOpen
  \bibfield  {author} {\bibinfo {author} {\bibfnamefont {S.}~\bibnamefont {Consta}},\ }\href {\doibase https://doi.org/10.1016/S0166-1280(02)00216-6} {\bibfield  {journal} {\bibinfo  {journal} {J. Mol. Struct. THEOCHEM}\ }\textbf {\bibinfo {volume} {591}},\ \bibinfo {pages} {131} (\bibinfo {year} {2002})}\BibitemShut {NoStop}%
\bibitem [{\citenamefont {Shi}\ and\ \citenamefont {Zheng}(2012)}]{Shi-2012}%
  \BibitemOpen
  \bibfield  {author} {\bibinfo {author} {\bibfnamefont {L.}~\bibnamefont {Shi}}\ and\ \bibinfo {author} {\bibfnamefont {L.}~\bibnamefont {Zheng}},\ }\href {\doibase 10.1021/jp211338k} {\bibfield  {journal} {\bibinfo  {journal} {J. Phys. Chem. B}\ }\textbf {\bibinfo {volume} {116}},\ \bibinfo {pages} {2162} (\bibinfo {year} {2012})}\BibitemShut {NoStop}%
\bibitem [{\citenamefont {Kang}\ \emph {et~al.}(2016)\citenamefont {Kang}, \citenamefont {Ma}, \citenamefont {Zhang}, \citenamefont {Xing}, \citenamefont {Mo}, \citenamefont {Wu}, \citenamefont {Li},\ and\ \citenamefont {Han}}]{Kang-2016}%
  \BibitemOpen
  \bibfield  {author} {\bibinfo {author} {\bibfnamefont {X.}~\bibnamefont {Kang}}, \bibinfo {author} {\bibfnamefont {X.}~\bibnamefont {Ma}}, \bibinfo {author} {\bibfnamefont {J.}~\bibnamefont {Zhang}}, \bibinfo {author} {\bibfnamefont {X.}~\bibnamefont {Xing}}, \bibinfo {author} {\bibfnamefont {G.}~\bibnamefont {Mo}}, \bibinfo {author} {\bibfnamefont {Z.}~\bibnamefont {Wu}}, \bibinfo {author} {\bibfnamefont {Z.}~\bibnamefont {Li}}, \ and\ \bibinfo {author} {\bibfnamefont {B.}~\bibnamefont {Han}},\ }\href {\doibase 10.1039/C6CC08015D} {\bibfield  {journal} {\bibinfo  {journal} {Chem. Commun.}\ }\textbf {\bibinfo {volume} {52}},\ \bibinfo {pages} {14286} (\bibinfo {year} {2016})}\BibitemShut {NoStop}%
\bibitem [{\citenamefont {Dunne}(2020)}]{Dunne-2020}%
  \BibitemOpen
  \bibfield  {author} {\bibinfo {author} {\bibfnamefont {P.}~\bibnamefont {Dunne}},\ }\enquote {\bibinfo {title} {Magnetochemistry and magnetic separation},}\ in\ \href {\doibase 10.1007/978-3-030-63101-7_35-1} {\emph {\bibinfo {booktitle} {Handbook of Magnetism and Magnetic Materials}}},\ \bibinfo {editor} {edited by\ \bibinfo {editor} {\bibfnamefont {M.}~\bibnamefont {Coey}}\ and\ \bibinfo {editor} {\bibfnamefont {S.}~\bibnamefont {Parkin}}}\ (\bibinfo  {publisher} {Springer International Publishing},\ \bibinfo {address} {Cham},\ \bibinfo {year} {2020})\ pp.\ \bibinfo {pages} {1--39}\BibitemShut {NoStop}%
\bibitem [{\citenamefont {Lei}\ \emph {et~al.}(2021)\citenamefont {Lei}, \citenamefont {Fritzsche},\ and\ \citenamefont {Eckert}}]{Lei-2021}%
  \BibitemOpen
  \bibfield  {author} {\bibinfo {author} {\bibfnamefont {Z.}~\bibnamefont {Lei}}, \bibinfo {author} {\bibfnamefont {B.}~\bibnamefont {Fritzsche}}, \ and\ \bibinfo {author} {\bibfnamefont {K.}~\bibnamefont {Eckert}},\ }\href {\doibase 10.1103/PhysRevFluids.6.L021901} {\bibfield  {journal} {\bibinfo  {journal} {Phys. Rev. Fluids}\ }\textbf {\bibinfo {volume} {6}},\ \bibinfo {pages} {L021901} (\bibinfo {year} {2021})}\BibitemShut {NoStop}%
\bibitem [{\citenamefont {Panagiotopoulos}\ and\ \citenamefont {Yue}(2023)}]{Shuwen-2023}%
  \BibitemOpen
  \bibfield  {author} {\bibinfo {author} {\bibfnamefont {A.~Z.}\ \bibnamefont {Panagiotopoulos}}\ and\ \bibinfo {author} {\bibfnamefont {S.}~\bibnamefont {Yue}},\ }\href {\doibase 10.1021/acs.jpcb.2c07477} {\bibfield  {journal} {\bibinfo  {journal} {J. Phys. Chem. B}\ }\textbf {\bibinfo {volume} {127}},\ \bibinfo {pages} {430} (\bibinfo {year} {2023})}\BibitemShut {NoStop}%
\bibitem [{\citenamefont {Lee}\ \emph {et~al.}(2023)\citenamefont {Lee}, \citenamefont {Nakouzi}, \citenamefont {Heo}, \citenamefont {Legg}, \citenamefont {Schenter}, \citenamefont {Li}, \citenamefont {Park}, \citenamefont {Ma},\ and\ \citenamefont {Chun}}]{Lee-2023}%
  \BibitemOpen
  \bibfield  {author} {\bibinfo {author} {\bibfnamefont {J.}~\bibnamefont {Lee}}, \bibinfo {author} {\bibfnamefont {E.}~\bibnamefont {Nakouzi}}, \bibinfo {author} {\bibfnamefont {J.}~\bibnamefont {Heo}}, \bibinfo {author} {\bibfnamefont {B.~A.}\ \bibnamefont {Legg}}, \bibinfo {author} {\bibfnamefont {G.~K.}\ \bibnamefont {Schenter}}, \bibinfo {author} {\bibfnamefont {D.}~\bibnamefont {Li}}, \bibinfo {author} {\bibfnamefont {C.}~\bibnamefont {Park}}, \bibinfo {author} {\bibfnamefont {H.}~\bibnamefont {Ma}}, \ and\ \bibinfo {author} {\bibfnamefont {J.}~\bibnamefont {Chun}},\ }\href {\doibase https://doi.org/10.1016/j.jcis.2023.08.160} {\bibfield  {journal} {\bibinfo  {journal} {J. Colloid. Interface Sci.}\ }\textbf {\bibinfo {volume} {652}},\ \bibinfo {pages} {1974} (\bibinfo {year} {2023})}\BibitemShut {NoStop}%
\bibitem [{\citenamefont {Li}\ \emph {et~al.}(2023)\citenamefont {Li}, \citenamefont {Chen}, \citenamefont {Chun}, \citenamefont {Fichthorn}, \citenamefont {De~Yoreo},\ and\ \citenamefont {Zheng}}]{Jaehun-2023}%
  \BibitemOpen
  \bibfield  {author} {\bibinfo {author} {\bibfnamefont {D.}~\bibnamefont {Li}}, \bibinfo {author} {\bibfnamefont {Q.}~\bibnamefont {Chen}}, \bibinfo {author} {\bibfnamefont {J.}~\bibnamefont {Chun}}, \bibinfo {author} {\bibfnamefont {K.}~\bibnamefont {Fichthorn}}, \bibinfo {author} {\bibfnamefont {J.}~\bibnamefont {De~Yoreo}}, \ and\ \bibinfo {author} {\bibfnamefont {H.}~\bibnamefont {Zheng}},\ }\href {\doibase 10.1021/acs.chemrev.2c00700} {\bibfield  {journal} {\bibinfo  {journal} {Chem. Rev.}\ }\textbf {\bibinfo {volume} {123}},\ \bibinfo {pages} {3127} (\bibinfo {year} {2023})}\BibitemShut {NoStop}%
\bibitem [{\citenamefont {Butreddy}\ \emph {et~al.}(2024)\citenamefont {Butreddy}, \citenamefont {Heo}, \citenamefont {Rampal}, \citenamefont {Liu}, \citenamefont {Liu}, \citenamefont {Smith}, \citenamefont {Zhang}, \citenamefont {Prange}, \citenamefont {Legg}, \citenamefont {Schenter}, \citenamefont {De~Yoreo}, \citenamefont {Chun}, \citenamefont {Stack},\ and\ \citenamefont {Nakouzi}}]{Butreddy-2024}%
  \BibitemOpen
  \bibfield  {author} {\bibinfo {author} {\bibfnamefont {P.}~\bibnamefont {Butreddy}}, \bibinfo {author} {\bibfnamefont {J.}~\bibnamefont {Heo}}, \bibinfo {author} {\bibfnamefont {N.}~\bibnamefont {Rampal}}, \bibinfo {author} {\bibfnamefont {T.}~\bibnamefont {Liu}}, \bibinfo {author} {\bibfnamefont {L.}~\bibnamefont {Liu}}, \bibinfo {author} {\bibfnamefont {W.}~\bibnamefont {Smith}}, \bibinfo {author} {\bibfnamefont {X.}~\bibnamefont {Zhang}}, \bibinfo {author} {\bibfnamefont {M.~P.}\ \bibnamefont {Prange}}, \bibinfo {author} {\bibfnamefont {B.~A.}\ \bibnamefont {Legg}}, \bibinfo {author} {\bibfnamefont {G.~K.}\ \bibnamefont {Schenter}}, \bibinfo {author} {\bibfnamefont {J.~J.}\ \bibnamefont {De~Yoreo}}, \bibinfo {author} {\bibfnamefont {J.}~\bibnamefont {Chun}}, \bibinfo {author} {\bibfnamefont {A.~G.}\ \bibnamefont {Stack}}, \ and\ \bibinfo {author} {\bibfnamefont {E.}~\bibnamefont {Nakouzi}},\ }\href {\doibase 10.1021/acsnano.4c05563} {\bibfield  {journal} {\bibinfo  {journal} {ACS Nano}\ }\textbf
  {\bibinfo {volume} {18}},\ \bibinfo {pages} {26047} (\bibinfo {year} {2024})}\BibitemShut {NoStop}%
\bibitem [{\citenamefont {Panczyk}\ and\ \citenamefont {Camp}(2021)}]{Panczyk-2021}%
  \BibitemOpen
  \bibfield  {author} {\bibinfo {author} {\bibfnamefont {T.}~\bibnamefont {Panczyk}}\ and\ \bibinfo {author} {\bibfnamefont {P.~J.}\ \bibnamefont {Camp}},\ }\href {\doibase https://doi.org/10.1016/j.molliq.2021.115701} {\bibfield  {journal} {\bibinfo  {journal} {J. Mol. Liq.}\ }\textbf {\bibinfo {volume} {330}},\ \bibinfo {pages} {115701} (\bibinfo {year} {2021})}\BibitemShut {NoStop}%
\bibitem [{\citenamefont {Spreiter}\ and\ \citenamefont {Walter}(1999)}]{Spreiter-1999}%
  \BibitemOpen
  \bibfield  {author} {\bibinfo {author} {\bibfnamefont {Q.}~\bibnamefont {Spreiter}}\ and\ \bibinfo {author} {\bibfnamefont {M.}~\bibnamefont {Walter}},\ }\href {\doibase https://doi.org/10.1006/jcph.1999.6237} {\bibfield  {journal} {\bibinfo  {journal} {J. Comput. Phys.}\ }\textbf {\bibinfo {volume} {152}},\ \bibinfo {pages} {102} (\bibinfo {year} {1999})}\BibitemShut {NoStop}%
\bibitem [{\citenamefont {Sun}\ \emph {et~al.}(2014)\citenamefont {Sun}, \citenamefont {He}, \citenamefont {Lo}, \citenamefont {Xu}, \citenamefont {Bi}, \citenamefont {Sun}, \citenamefont {Zhang}, \citenamefont {Mao},\ and\ \citenamefont {Li}}]{Sun-2014}%
  \BibitemOpen
  \bibfield  {author} {\bibinfo {author} {\bibfnamefont {J.}~\bibnamefont {Sun}}, \bibinfo {author} {\bibfnamefont {L.}~\bibnamefont {He}}, \bibinfo {author} {\bibfnamefont {Y.-C.}\ \bibnamefont {Lo}}, \bibinfo {author} {\bibfnamefont {T.}~\bibnamefont {Xu}}, \bibinfo {author} {\bibfnamefont {H.}~\bibnamefont {Bi}}, \bibinfo {author} {\bibfnamefont {L.}~\bibnamefont {Sun}}, \bibinfo {author} {\bibfnamefont {Z.}~\bibnamefont {Zhang}}, \bibinfo {author} {\bibfnamefont {S.~X.}\ \bibnamefont {Mao}}, \ and\ \bibinfo {author} {\bibfnamefont {J.}~\bibnamefont {Li}},\ }\href {\doibase 10.1038/nmat4105} {\bibfield  {journal} {\bibinfo  {journal} {Nat. Mater.}\ }\textbf {\bibinfo {volume} {13}},\ \bibinfo {pages} {1007} (\bibinfo {year} {2014})}\BibitemShut {NoStop}%
\bibitem [{\citenamefont {Cao}\ \emph {et~al.}(2019)\citenamefont {Cao}, \citenamefont {Huang}, \citenamefont {Shi}, \citenamefont {Zheng}, \citenamefont {Wang}, \citenamefont {Gu},\ and\ \citenamefont {Bai}}]{Cao-2019}%
  \BibitemOpen
  \bibfield  {author} {\bibinfo {author} {\bibfnamefont {C.~R.}\ \bibnamefont {Cao}}, \bibinfo {author} {\bibfnamefont {K.~Q.}\ \bibnamefont {Huang}}, \bibinfo {author} {\bibfnamefont {J.~A.}\ \bibnamefont {Shi}}, \bibinfo {author} {\bibfnamefont {D.~N.}\ \bibnamefont {Zheng}}, \bibinfo {author} {\bibfnamefont {W.~H.}\ \bibnamefont {Wang}}, \bibinfo {author} {\bibfnamefont {L.}~\bibnamefont {Gu}}, \ and\ \bibinfo {author} {\bibfnamefont {H.~Y.}\ \bibnamefont {Bai}},\ }\href {\doibase 10.1038/s41467-019-09895-3} {\bibfield  {journal} {\bibinfo  {journal} {Nat. Commun.}\ }\textbf {\bibinfo {volume} {10}},\ \bibinfo {pages} {1966} (\bibinfo {year} {2019})}\BibitemShut {NoStop}%
\bibitem [{\citenamefont {Liu}\ \emph {et~al.}(2021)\citenamefont {Liu}, \citenamefont {Tian},\ and\ \citenamefont {Jiang}}]{Liu-Xubo-2021}%
  \BibitemOpen
  \bibfield  {author} {\bibinfo {author} {\bibfnamefont {X.}~\bibnamefont {Liu}}, \bibinfo {author} {\bibfnamefont {Y.}~\bibnamefont {Tian}}, \ and\ \bibinfo {author} {\bibfnamefont {L.}~\bibnamefont {Jiang}},\ }\href {\doibase 10.1021/acs.nanolett.0c04757} {\bibfield  {journal} {\bibinfo  {journal} {Nano Lett.}\ }\textbf {\bibinfo {volume} {21}},\ \bibinfo {pages} {2699} (\bibinfo {year} {2021})}\BibitemShut {NoStop}%
\bibitem [{\citenamefont {Liu}\ \emph {et~al.}(2019)\citenamefont {Liu}, \citenamefont {Kent}, \citenamefont {Ceballos}, \citenamefont {Streubel}, \citenamefont {Jiang}, \citenamefont {Chai}, \citenamefont {Kim}, \citenamefont {Forth}, \citenamefont {Hellman}, \citenamefont {Shi}, \citenamefont {Wang}, \citenamefont {Helms}, \citenamefont {Ashby}, \citenamefont {Fischer},\ and\ \citenamefont {Russell}}]{Liu-2019}%
  \BibitemOpen
  \bibfield  {author} {\bibinfo {author} {\bibfnamefont {X.}~\bibnamefont {Liu}}, \bibinfo {author} {\bibfnamefont {N.}~\bibnamefont {Kent}}, \bibinfo {author} {\bibfnamefont {A.}~\bibnamefont {Ceballos}}, \bibinfo {author} {\bibfnamefont {R.}~\bibnamefont {Streubel}}, \bibinfo {author} {\bibfnamefont {Y.}~\bibnamefont {Jiang}}, \bibinfo {author} {\bibfnamefont {Y.}~\bibnamefont {Chai}}, \bibinfo {author} {\bibfnamefont {P.~Y.}\ \bibnamefont {Kim}}, \bibinfo {author} {\bibfnamefont {J.}~\bibnamefont {Forth}}, \bibinfo {author} {\bibfnamefont {F.}~\bibnamefont {Hellman}}, \bibinfo {author} {\bibfnamefont {S.}~\bibnamefont {Shi}}, \bibinfo {author} {\bibfnamefont {D.}~\bibnamefont {Wang}}, \bibinfo {author} {\bibfnamefont {B.~A.}\ \bibnamefont {Helms}}, \bibinfo {author} {\bibfnamefont {P.~D.}\ \bibnamefont {Ashby}}, \bibinfo {author} {\bibfnamefont {P.}~\bibnamefont {Fischer}}, \ and\ \bibinfo {author} {\bibfnamefont {T.~P.}\ \bibnamefont {Russell}},\ }\href {\doibase 10.1126/science.aaw8719} {\bibfield
  {journal} {\bibinfo  {journal} {Science}\ }\textbf {\bibinfo {volume} {365}},\ \bibinfo {pages} {264} (\bibinfo {year} {2019})}\BibitemShut {NoStop}%
\bibitem [{\citenamefont {Wu}\ \emph {et~al.}(2021)\citenamefont {Wu}, \citenamefont {Streubel}, \citenamefont {Liu}, \citenamefont {Kim}, \citenamefont {Chai}, \citenamefont {Hu}, \citenamefont {Wang}, \citenamefont {Fischer},\ and\ \citenamefont {Russell}}]{Wu-2021}%
  \BibitemOpen
  \bibfield  {author} {\bibinfo {author} {\bibfnamefont {X.}~\bibnamefont {Wu}}, \bibinfo {author} {\bibfnamefont {R.}~\bibnamefont {Streubel}}, \bibinfo {author} {\bibfnamefont {X.}~\bibnamefont {Liu}}, \bibinfo {author} {\bibfnamefont {P.~Y.}\ \bibnamefont {Kim}}, \bibinfo {author} {\bibfnamefont {Y.}~\bibnamefont {Chai}}, \bibinfo {author} {\bibfnamefont {Q.}~\bibnamefont {Hu}}, \bibinfo {author} {\bibfnamefont {D.}~\bibnamefont {Wang}}, \bibinfo {author} {\bibfnamefont {P.}~\bibnamefont {Fischer}}, \ and\ \bibinfo {author} {\bibfnamefont {T.~P.}\ \bibnamefont {Russell}},\ }\href {\doibase 10.1073/pnas.2017355118} {\bibfield  {journal} {\bibinfo  {journal} {Proc. Natl. Acad. Sci. U.S. A.}\ }\textbf {\bibinfo {volume} {118}},\ \bibinfo {pages} {e2017355118} (\bibinfo {year} {2021})}\BibitemShut {NoStop}%
\bibitem [{\citenamefont {Luttinger}\ and\ \citenamefont {Tisza}(1946)}]{Luttinger-1946}%
  \BibitemOpen
  \bibfield  {author} {\bibinfo {author} {\bibfnamefont {J.~M.}\ \bibnamefont {Luttinger}}\ and\ \bibinfo {author} {\bibfnamefont {L.}~\bibnamefont {Tisza}},\ }\href {\doibase 10.1103/PhysRev.70.954} {\bibfield  {journal} {\bibinfo  {journal} {Phys. Rev.}\ }\textbf {\bibinfo {volume} {70}},\ \bibinfo {pages} {954} (\bibinfo {year} {1946})}\BibitemShut {NoStop}%
\bibitem [{\citenamefont {Bergman}\ \emph {et~al.}(2006)\citenamefont {Bergman}, \citenamefont {Nordstr\"om}, \citenamefont {Burlamaqui~Klautau}, \citenamefont {Frota-Pess\^oa},\ and\ \citenamefont {Eriksson}}]{Bergman-2006}%
  \BibitemOpen
  \bibfield  {author} {\bibinfo {author} {\bibfnamefont {A.}~\bibnamefont {Bergman}}, \bibinfo {author} {\bibfnamefont {L.}~\bibnamefont {Nordstr\"om}}, \bibinfo {author} {\bibfnamefont {A.}~\bibnamefont {Burlamaqui~Klautau}}, \bibinfo {author} {\bibfnamefont {S.}~\bibnamefont {Frota-Pess\^oa}}, \ and\ \bibinfo {author} {\bibfnamefont {O.}~\bibnamefont {Eriksson}},\ }\href {\doibase 10.1103/PhysRevB.73.174434} {\bibfield  {journal} {\bibinfo  {journal} {Phys. Rev. B}\ }\textbf {\bibinfo {volume} {73}},\ \bibinfo {pages} {174434} (\bibinfo {year} {2006})}\BibitemShut {NoStop}%
\bibitem [{\citenamefont {Majetich}\ and\ \citenamefont {Sachan}(2006)}]{Majetich-2006}%
  \BibitemOpen
  \bibfield  {author} {\bibinfo {author} {\bibfnamefont {S.~A.}\ \bibnamefont {Majetich}}\ and\ \bibinfo {author} {\bibfnamefont {M.}~\bibnamefont {Sachan}},\ }\href {\doibase 10.1088/0022-3727/39/21/R02} {\bibfield  {journal} {\bibinfo  {journal} {J. Phys. D Appl. Phys.}\ }\textbf {\bibinfo {volume} {39}},\ \bibinfo {pages} {R407} (\bibinfo {year} {2006})}\BibitemShut {NoStop}%
\bibitem [{\citenamefont {Haghgooie}\ and\ \citenamefont {Doyle}(2007)}]{Doyle-2007}%
  \BibitemOpen
  \bibfield  {author} {\bibinfo {author} {\bibfnamefont {R.}~\bibnamefont {Haghgooie}}\ and\ \bibinfo {author} {\bibfnamefont {P.~S.}\ \bibnamefont {Doyle}},\ }\href {\doibase 10.1103/PhysRevE.75.061406} {\bibfield  {journal} {\bibinfo  {journal} {Phys. Rev. E}\ }\textbf {\bibinfo {volume} {75}},\ \bibinfo {pages} {061406} (\bibinfo {year} {2007})}\BibitemShut {NoStop}%
\bibitem [{\citenamefont {Schaller}\ \emph {et~al.}(2010)\citenamefont {Schaller}, \citenamefont {Wahnström}, \citenamefont {Sanz-Velasco}, \citenamefont {Enoksson},\ and\ \citenamefont {Johansson}}]{Schaller-2010}%
  \BibitemOpen
  \bibfield  {author} {\bibinfo {author} {\bibfnamefont {V.}~\bibnamefont {Schaller}}, \bibinfo {author} {\bibfnamefont {G.}~\bibnamefont {Wahnström}}, \bibinfo {author} {\bibfnamefont {A.}~\bibnamefont {Sanz-Velasco}}, \bibinfo {author} {\bibfnamefont {P.}~\bibnamefont {Enoksson}}, \ and\ \bibinfo {author} {\bibfnamefont {C.}~\bibnamefont {Johansson}},\ }\href {\doibase 10.1088/1742-6596/200/7/072085} {\bibfield  {journal} {\bibinfo  {journal} {J. Phys. Conf. Ser.}\ }\textbf {\bibinfo {volume} {200}},\ \bibinfo {pages} {072085} (\bibinfo {year} {2010})}\BibitemShut {NoStop}%
\bibitem [{\citenamefont {Muratov}(2002)}]{Muratov-2002}%
  \BibitemOpen
  \bibfield  {author} {\bibinfo {author} {\bibfnamefont {C.~B.}\ \bibnamefont {Muratov}},\ }\href {\doibase 10.1103/PhysRevE.66.066108} {\bibfield  {journal} {\bibinfo  {journal} {Phys. Rev. E}\ }\textbf {\bibinfo {volume} {66}},\ \bibinfo {pages} {066108} (\bibinfo {year} {2002})}\BibitemShut {NoStop}%
\bibitem [{\citenamefont {Furrer}\ and\ \citenamefont {Waldmann}(2013)}]{Furrer-2013}%
  \BibitemOpen
  \bibfield  {author} {\bibinfo {author} {\bibfnamefont {A.}~\bibnamefont {Furrer}}\ and\ \bibinfo {author} {\bibfnamefont {O.}~\bibnamefont {Waldmann}},\ }\href {\doibase 10.1103/RevModPhys.85.367} {\bibfield  {journal} {\bibinfo  {journal} {Rev. Mod. Phys.}\ }\textbf {\bibinfo {volume} {85}},\ \bibinfo {pages} {367} (\bibinfo {year} {2013})}\BibitemShut {NoStop}%
\bibitem [{\citenamefont {Hamaker}(1937)}]{Hamaker1937}%
  \BibitemOpen
  \bibfield  {author} {\bibinfo {author} {\bibfnamefont {H.~C.}\ \bibnamefont {Hamaker}},\ }\href@noop {} {\bibfield  {journal} {\bibinfo  {journal} {Physica}\ }\textbf {\bibinfo {volume} {4}},\ \bibinfo {pages} {1058} (\bibinfo {year} {1937})}\BibitemShut {NoStop}%
\bibitem [{\citenamefont {Evans}\ and\ \citenamefont {Wennerström}(1999)}]{Evans-1999}%
  \BibitemOpen
  \bibfield  {author} {\bibinfo {author} {\bibfnamefont {D.~F.}\ \bibnamefont {Evans}}\ and\ \bibinfo {author} {\bibfnamefont {H.}~\bibnamefont {Wennerström}},\ }\href@noop {} {\emph {\bibinfo {title} {The Colloidal Domain: Where Physics, Chemistry and Biology Meet}}}\ (\bibinfo  {publisher} {Wiley},\ \bibinfo {year} {1999})\BibitemShut {NoStop}%
\bibitem [{\citenamefont {Priye}\ and\ \citenamefont {Marlow}(2013)}]{Priye-2013}%
  \BibitemOpen
  \bibfield  {author} {\bibinfo {author} {\bibfnamefont {A.}~\bibnamefont {Priye}}\ and\ \bibinfo {author} {\bibfnamefont {W.~H.}\ \bibnamefont {Marlow}},\ }\href {\doibase 10.1088/0022-3727/46/42/425306} {\bibfield  {journal} {\bibinfo  {journal} {J. Phys. D: Appl. Phys.}\ }\textbf {\bibinfo {volume} {46}},\ \bibinfo {pages} {425306} (\bibinfo {year} {2013})}\BibitemShut {NoStop}%
\bibitem [{\citenamefont {Henderson}\ \emph {et~al.}(1997)\citenamefont {Henderson}, \citenamefont {Duh}, \citenamefont {Chu},\ and\ \citenamefont {Wasan}}]{Henderson-1997}%
  \BibitemOpen
  \bibfield  {author} {\bibinfo {author} {\bibfnamefont {D.}~\bibnamefont {Henderson}}, \bibinfo {author} {\bibfnamefont {D.-M.}\ \bibnamefont {Duh}}, \bibinfo {author} {\bibfnamefont {X.}~\bibnamefont {Chu}}, \ and\ \bibinfo {author} {\bibfnamefont {D.}~\bibnamefont {Wasan}},\ }\href {\doibase https://doi.org/10.1006/jcis.1996.4600} {\bibfield  {journal} {\bibinfo  {journal} {J. Colloid Interface Sci.}\ }\textbf {\bibinfo {volume} {185}},\ \bibinfo {pages} {265} (\bibinfo {year} {1997})}\BibitemShut {NoStop}%
\bibitem [{\citenamefont {Rabani}\ and\ \citenamefont {Egorov}(2002)}]{Rabani-2002}%
  \BibitemOpen
  \bibfield  {author} {\bibinfo {author} {\bibfnamefont {E.}~\bibnamefont {Rabani}}\ and\ \bibinfo {author} {\bibfnamefont {S.}~\bibnamefont {Egorov}},\ }\href {\doibase 10.1021/jp025693f} {\bibfield  {journal} {\bibinfo  {journal} {J. Phys. Chem. B}\ }\textbf {\bibinfo {volume} {106}},\ \bibinfo {pages} {6771} (\bibinfo {year} {2002})}\BibitemShut {NoStop}%
\bibitem [{\citenamefont {Everaers}\ and\ \citenamefont {Ejtehadi}(2003)}]{Everaers2003}%
  \BibitemOpen
  \bibfield  {author} {\bibinfo {author} {\bibfnamefont {R.}~\bibnamefont {Everaers}}\ and\ \bibinfo {author} {\bibfnamefont {M.~R.}\ \bibnamefont {Ejtehadi}},\ }\href {\doibase 10.1103/PhysRevE.67.041710} {\bibfield  {journal} {\bibinfo  {journal} {Phys. Rev. E}\ }\textbf {\bibinfo {volume} {67}},\ \bibinfo {pages} {041710} (\bibinfo {year} {2003})}\BibitemShut {NoStop}%
\bibitem [{\citenamefont {Grest}\ \emph {et~al.}(2008)\citenamefont {Grest}, \citenamefont {in~'t Veld},\ and\ \citenamefont {Lechman}}]{Grest2008}%
  \BibitemOpen
  \bibfield  {author} {\bibinfo {author} {\bibfnamefont {G.~S.}\ \bibnamefont {Grest}}, \bibinfo {author} {\bibfnamefont {P.~J.}\ \bibnamefont {in~'t Veld}}, \ and\ \bibinfo {author} {\bibfnamefont {J.~B.}\ \bibnamefont {Lechman}},\ }\href {\doibase 10.1063/1.2897804} {\bibfield  {journal} {\bibinfo  {journal} {AIP Conf. Proc.}\ }\textbf {\bibinfo {volume} {982}},\ \bibinfo {pages} {304} (\bibinfo {year} {2008})}\BibitemShut {NoStop}%
\bibitem [{\citenamefont {Cheng}\ and\ \citenamefont {Grest}(2012)}]{Cheng-2012}%
  \BibitemOpen
  \bibfield  {author} {\bibinfo {author} {\bibfnamefont {S.}~\bibnamefont {Cheng}}\ and\ \bibinfo {author} {\bibfnamefont {G.~S.}\ \bibnamefont {Grest}},\ }\href {\doibase 10.1063/1.4725543} {\bibfield  {journal} {\bibinfo  {journal} {J. Chem. Phys.}\ }\textbf {\bibinfo {volume} {136}},\ \bibinfo {pages} {214702} (\bibinfo {year} {2012})}\BibitemShut {NoStop}%
\bibitem [{\citenamefont {Chuev}\ \emph {et~al.}(2022)\citenamefont {Chuev}, \citenamefont {Dinpajooh},\ and\ \citenamefont {Valiev}}]{Chuev-2022}%
  \BibitemOpen
  \bibfield  {author} {\bibinfo {author} {\bibfnamefont {G.}~\bibnamefont {Chuev}}, \bibinfo {author} {\bibfnamefont {M.}~\bibnamefont {Dinpajooh}}, \ and\ \bibinfo {author} {\bibfnamefont {M.}~\bibnamefont {Valiev}},\ }\href {\doibase 10.1063/5.0128817} {\bibfield  {journal} {\bibinfo  {journal} {J. Chem. Phys.}\ }\textbf {\bibinfo {volume} {157}},\ \bibinfo {pages} {184505} (\bibinfo {year} {2022})}\BibitemShut {NoStop}%
\bibitem [{\citenamefont {Nakouzi}\ \emph {et~al.}(2023)\citenamefont {Nakouzi}, \citenamefont {Kerisit}, \citenamefont {Legg}, \citenamefont {Yadav}, \citenamefont {Li}, \citenamefont {Stack}, \citenamefont {Mundy}, \citenamefont {Chun}, \citenamefont {Schenter},\ and\ \citenamefont {De~Yoreo}}]{Nakouzi-2023}%
  \BibitemOpen
  \bibfield  {author} {\bibinfo {author} {\bibfnamefont {E.}~\bibnamefont {Nakouzi}}, \bibinfo {author} {\bibfnamefont {S.}~\bibnamefont {Kerisit}}, \bibinfo {author} {\bibfnamefont {B.~A.}\ \bibnamefont {Legg}}, \bibinfo {author} {\bibfnamefont {S.}~\bibnamefont {Yadav}}, \bibinfo {author} {\bibfnamefont {D.}~\bibnamefont {Li}}, \bibinfo {author} {\bibfnamefont {A.~G.}\ \bibnamefont {Stack}}, \bibinfo {author} {\bibfnamefont {C.~J.}\ \bibnamefont {Mundy}}, \bibinfo {author} {\bibfnamefont {J.}~\bibnamefont {Chun}}, \bibinfo {author} {\bibfnamefont {G.~K.}\ \bibnamefont {Schenter}}, \ and\ \bibinfo {author} {\bibfnamefont {J.~J.}\ \bibnamefont {De~Yoreo}},\ }\href {\doibase 10.1021/acs.jpcc.2c09120} {\bibfield  {journal} {\bibinfo  {journal} {J. Phys. Chem. C.}\ }\textbf {\bibinfo {volume} {127}},\ \bibinfo {pages} {2741} (\bibinfo {year} {2023})}\BibitemShut {NoStop}%
\bibitem [{\citenamefont {Israelachvili}(1987)}]{Israelachvili-1987}%
  \BibitemOpen
  \bibfield  {author} {\bibinfo {author} {\bibfnamefont {J.}~\bibnamefont {Israelachvili}},\ }\href@noop {} {\bibfield  {journal} {\bibinfo  {journal} {Proc. Natl. Acad. Sci. U.S.A.}\ }\textbf {\bibinfo {volume} {84}},\ \bibinfo {pages} {4722} (\bibinfo {year} {1987})}\BibitemShut {NoStop}%
\bibitem [{\citenamefont {Israelachvili}\ and\ \citenamefont {McGuiggan}(1988)}]{Israelachvili-1988}%
  \BibitemOpen
  \bibfield  {author} {\bibinfo {author} {\bibfnamefont {J.~N.}\ \bibnamefont {Israelachvili}}\ and\ \bibinfo {author} {\bibfnamefont {P.~M.}\ \bibnamefont {McGuiggan}},\ }\href {\doibase 10.1126/science.241.4867.795} {\bibfield  {journal} {\bibinfo  {journal} {Science}\ }\textbf {\bibinfo {volume} {241}},\ \bibinfo {pages} {795} (\bibinfo {year} {1988})}\BibitemShut {NoStop}%
\bibitem [{\citenamefont {Israelachvili}\ and\ \citenamefont {Wennerstr{\"o}m}(1996)}]{Israelachvili-1996}%
  \BibitemOpen
  \bibfield  {author} {\bibinfo {author} {\bibfnamefont {J.}~\bibnamefont {Israelachvili}}\ and\ \bibinfo {author} {\bibfnamefont {H.}~\bibnamefont {Wennerstr{\"o}m}},\ }\href {\doibase 10.1038/379219a0} {\bibfield  {journal} {\bibinfo  {journal} {Nature}\ }\textbf {\bibinfo {volume} {379}},\ \bibinfo {pages} {219} (\bibinfo {year} {1996})}\BibitemShut {NoStop}%
\bibitem [{\citenamefont {Israelachvili}(2011)}]{Israelachvili2011}%
  \BibitemOpen
  \bibfield  {author} {\bibinfo {author} {\bibfnamefont {J.~N.}\ \bibnamefont {Israelachvili}},\ }\href@noop {} {\emph {\bibinfo {title} {Intermolecular and Surface Forces}}}\ (\bibinfo  {publisher} {Academic Press},\ \bibinfo {year} {2011})\BibitemShut {NoStop}%
\bibitem [{\citenamefont {Dormann}\ \emph {et~al.}(1988)\citenamefont {Dormann}, \citenamefont {Bessais},\ and\ \citenamefont {Fiorani}}]{Dormann-1988}%
  \BibitemOpen
  \bibfield  {author} {\bibinfo {author} {\bibfnamefont {J.~L.}\ \bibnamefont {Dormann}}, \bibinfo {author} {\bibfnamefont {L.}~\bibnamefont {Bessais}}, \ and\ \bibinfo {author} {\bibfnamefont {D.}~\bibnamefont {Fiorani}},\ }\href {\doibase 10.1088/0022-3719/21/10/019} {\bibfield  {journal} {\bibinfo  {journal} {J. Phys. C: Solid State Phys}\ }\textbf {\bibinfo {volume} {21}},\ \bibinfo {pages} {2015} (\bibinfo {year} {1988})}\BibitemShut {NoStop}%
\bibitem [{\citenamefont {Hansen}\ and\ \citenamefont {Mørup}(1998)}]{Hansen-1998}%
  \BibitemOpen
  \bibfield  {author} {\bibinfo {author} {\bibfnamefont {M.}~\bibnamefont {Hansen}}\ and\ \bibinfo {author} {\bibfnamefont {S.}~\bibnamefont {Mørup}},\ }\href {\doibase https://doi.org/10.1016/S0304-8853(97)01165-7} {\bibfield  {journal} {\bibinfo  {journal} {J. Magn. Magn. Mater.}\ }\textbf {\bibinfo {volume} {184}},\ \bibinfo {pages} {L262} (\bibinfo {year} {1998})}\BibitemShut {NoStop}%
\bibitem [{\citenamefont {Ahrentorp}\ \emph {et~al.}(2015)\citenamefont {Ahrentorp}, \citenamefont {Astalan}, \citenamefont {Blomgren}, \citenamefont {Jonasson}, \citenamefont {Wetterskog}, \citenamefont {Svedlindh}, \citenamefont {Lak}, \citenamefont {Ludwig}, \citenamefont {{van IJzendoorn}}, \citenamefont {Westphal}, \citenamefont {Grüttner}, \citenamefont {Gehrke}, \citenamefont {Gustafsson}, \citenamefont {Olsson},\ and\ \citenamefont {Johansson}}]{Ahrentorp-2015}%
  \BibitemOpen
  \bibfield  {author} {\bibinfo {author} {\bibfnamefont {F.}~\bibnamefont {Ahrentorp}}, \bibinfo {author} {\bibfnamefont {A.}~\bibnamefont {Astalan}}, \bibinfo {author} {\bibfnamefont {J.}~\bibnamefont {Blomgren}}, \bibinfo {author} {\bibfnamefont {C.}~\bibnamefont {Jonasson}}, \bibinfo {author} {\bibfnamefont {E.}~\bibnamefont {Wetterskog}}, \bibinfo {author} {\bibfnamefont {P.}~\bibnamefont {Svedlindh}}, \bibinfo {author} {\bibfnamefont {A.}~\bibnamefont {Lak}}, \bibinfo {author} {\bibfnamefont {F.}~\bibnamefont {Ludwig}}, \bibinfo {author} {\bibfnamefont {L.~J.}\ \bibnamefont {{van IJzendoorn}}}, \bibinfo {author} {\bibfnamefont {F.}~\bibnamefont {Westphal}}, \bibinfo {author} {\bibfnamefont {C.}~\bibnamefont {Grüttner}}, \bibinfo {author} {\bibfnamefont {N.}~\bibnamefont {Gehrke}}, \bibinfo {author} {\bibfnamefont {S.}~\bibnamefont {Gustafsson}}, \bibinfo {author} {\bibfnamefont {E.}~\bibnamefont {Olsson}}, \ and\ \bibinfo {author} {\bibfnamefont {C.}~\bibnamefont {Johansson}},\ }\href {\doibase
  https://doi.org/10.1016/j.jmmm.2014.09.070} {\bibfield  {journal} {\bibinfo  {journal} {J. Magn. Magn. Mater.}\ }\textbf {\bibinfo {volume} {380}},\ \bibinfo {pages} {221} (\bibinfo {year} {2015})}\BibitemShut {NoStop}%
\bibitem [{\citenamefont {Bender}\ \emph {et~al.}(2017)\citenamefont {Bender}, \citenamefont {Bogart}, \citenamefont {Posth}, \citenamefont {Szczerba}, \citenamefont {Rogers}, \citenamefont {Castro}, \citenamefont {Nilsson}, \citenamefont {Zeng}, \citenamefont {Sugunan}, \citenamefont {Sommertune}, \citenamefont {Fornara}, \citenamefont {Gonz{\'a}lez-Alonso}, \citenamefont {Barqu{\'i}n},\ and\ \citenamefont {Johansson}}]{Bender-2017}%
  \BibitemOpen
  \bibfield  {author} {\bibinfo {author} {\bibfnamefont {P.}~\bibnamefont {Bender}}, \bibinfo {author} {\bibfnamefont {L.~K.}\ \bibnamefont {Bogart}}, \bibinfo {author} {\bibfnamefont {O.}~\bibnamefont {Posth}}, \bibinfo {author} {\bibfnamefont {W.}~\bibnamefont {Szczerba}}, \bibinfo {author} {\bibfnamefont {S.~E.}\ \bibnamefont {Rogers}}, \bibinfo {author} {\bibfnamefont {A.}~\bibnamefont {Castro}}, \bibinfo {author} {\bibfnamefont {L.}~\bibnamefont {Nilsson}}, \bibinfo {author} {\bibfnamefont {L.~J.}\ \bibnamefont {Zeng}}, \bibinfo {author} {\bibfnamefont {A.}~\bibnamefont {Sugunan}}, \bibinfo {author} {\bibfnamefont {J.}~\bibnamefont {Sommertune}}, \bibinfo {author} {\bibfnamefont {A.}~\bibnamefont {Fornara}}, \bibinfo {author} {\bibfnamefont {D.}~\bibnamefont {Gonz{\'a}lez-Alonso}}, \bibinfo {author} {\bibfnamefont {L.~F.}\ \bibnamefont {Barqu{\'i}n}}, \ and\ \bibinfo {author} {\bibfnamefont {C.}~\bibnamefont {Johansson}},\ }\href {\doibase 10.1038/srep45990} {\bibfield  {journal} {\bibinfo  {journal}
  {Sci. Rep.}\ }\textbf {\bibinfo {volume} {7}},\ \bibinfo {pages} {45990} (\bibinfo {year} {2017})}\BibitemShut {NoStop}%
\bibitem [{\citenamefont {De~Biasi}\ \emph {et~al.}(2002)\citenamefont {De~Biasi}, \citenamefont {Ramos}, \citenamefont {Zysler},\ and\ \citenamefont {Romero}}]{Biasi-2002}%
  \BibitemOpen
  \bibfield  {author} {\bibinfo {author} {\bibfnamefont {E.}~\bibnamefont {De~Biasi}}, \bibinfo {author} {\bibfnamefont {C.~A.}\ \bibnamefont {Ramos}}, \bibinfo {author} {\bibfnamefont {R.~D.}\ \bibnamefont {Zysler}}, \ and\ \bibinfo {author} {\bibfnamefont {H.}~\bibnamefont {Romero}},\ }\href {\doibase 10.1103/PhysRevB.65.144416} {\bibfield  {journal} {\bibinfo  {journal} {Phys. Rev. B}\ }\textbf {\bibinfo {volume} {65}},\ \bibinfo {pages} {144416} (\bibinfo {year} {2002})}\BibitemShut {NoStop}%
\bibitem [{\citenamefont {Derjaguin}(1934)}]{Derjaguin1934}%
  \BibitemOpen
  \bibfield  {author} {\bibinfo {author} {\bibfnamefont {B.}~\bibnamefont {Derjaguin}},\ }\href {\doibase 10.1007/BF01433225} {\bibfield  {journal} {\bibinfo  {journal} {Kolloid-Zeitschrift}\ }\textbf {\bibinfo {volume} {69}},\ \bibinfo {pages} {155} (\bibinfo {year} {1934})}\BibitemShut {NoStop}%
\bibitem [{\citenamefont {Wennerström}\ and\ \citenamefont {Stenhammar}(2020)}]{Wennerstroem2020}%
  \BibitemOpen
  \bibfield  {author} {\bibinfo {author} {\bibfnamefont {H.}~\bibnamefont {Wennerström}}\ and\ \bibinfo {author} {\bibfnamefont {J.}~\bibnamefont {Stenhammar}},\ }\href {\doibase 10.1063/5.0011446} {\bibfield  {journal} {\bibinfo  {journal} {J. Chem. Phys.}\ }\textbf {\bibinfo {volume} {152}},\ \bibinfo {pages} {234704} (\bibinfo {year} {2020})}\BibitemShut {NoStop}%
\bibitem [{\citenamefont {Remsing}\ \emph {et~al.}(2014)\citenamefont {Remsing}, \citenamefont {Baer}, \citenamefont {Schenter}, \citenamefont {Mundy},\ and\ \citenamefont {Weeks}}]{Remsing2014}%
  \BibitemOpen
  \bibfield  {author} {\bibinfo {author} {\bibfnamefont {R.~C.}\ \bibnamefont {Remsing}}, \bibinfo {author} {\bibfnamefont {M.~D.}\ \bibnamefont {Baer}}, \bibinfo {author} {\bibfnamefont {G.~K.}\ \bibnamefont {Schenter}}, \bibinfo {author} {\bibfnamefont {C.~J.}\ \bibnamefont {Mundy}}, \ and\ \bibinfo {author} {\bibfnamefont {J.~D.}\ \bibnamefont {Weeks}},\ }\href {\doibase 10.1021/jz501067w} {\bibfield  {journal} {\bibinfo  {journal} {J. Phys. Chem. Lett.}\ }\textbf {\bibinfo {volume} {5}},\ \bibinfo {pages} {2767} (\bibinfo {year} {2014})}\BibitemShut {NoStop}%
\bibitem [{\citenamefont {Dinpajooh}\ and\ \citenamefont {Matyushov}(2015)}]{Dinpajooh2015-born}%
  \BibitemOpen
  \bibfield  {author} {\bibinfo {author} {\bibfnamefont {M.}~\bibnamefont {Dinpajooh}}\ and\ \bibinfo {author} {\bibfnamefont {D.~V.}\ \bibnamefont {Matyushov}},\ }\href {\doibase 10.1063/1.4927570} {\bibfield  {journal} {\bibinfo  {journal} {J. Chem. Phys.}\ }\textbf {\bibinfo {volume} {143}},\ \bibinfo {pages} {044511} (\bibinfo {year} {2015})}\BibitemShut {NoStop}%
\bibitem [{\citenamefont {Wittmann}\ \emph {et~al.}(2025)\citenamefont {Wittmann}, \citenamefont {Krucker-Velasquez}, \citenamefont {Schaupp}, \citenamefont {Westphal}, \citenamefont {Swan}, \citenamefont {Alexander-Katz}, \citenamefont {Bazant}, \citenamefont {Schwaminger},\ and\ \citenamefont {Berensmeier}}]{Bazant-2024}%
  \BibitemOpen
  \bibfield  {author} {\bibinfo {author} {\bibfnamefont {L.}~\bibnamefont {Wittmann}}, \bibinfo {author} {\bibfnamefont {E.}~\bibnamefont {Krucker-Velasquez}}, \bibinfo {author} {\bibfnamefont {J.}~\bibnamefont {Schaupp}}, \bibinfo {author} {\bibfnamefont {L.}~\bibnamefont {Westphal}}, \bibinfo {author} {\bibfnamefont {J.~W.}\ \bibnamefont {Swan}}, \bibinfo {author} {\bibfnamefont {A.}~\bibnamefont {Alexander-Katz}}, \bibinfo {author} {\bibfnamefont {M.~Z.}\ \bibnamefont {Bazant}}, \bibinfo {author} {\bibfnamefont {S.~P.}\ \bibnamefont {Schwaminger}}, \ and\ \bibinfo {author} {\bibfnamefont {S.}~\bibnamefont {Berensmeier}},\ }\href {\doibase 10.1039/D4NR02225D} {\bibfield  {journal} {\bibinfo  {journal} {Nanoscale}\ ,\ } (\bibinfo {year} {2025})}\BibitemShut {NoStop}%
\bibitem [{\citenamefont {Guo}\ \emph {et~al.}(2024)\citenamefont {Guo}, \citenamefont {Du}, \citenamefont {Ling}, \citenamefont {L{\"u}}, \citenamefont {He},\ and\ \citenamefont {Luo}}]{Guo-2024}%
  \BibitemOpen
  \bibfield  {author} {\bibinfo {author} {\bibfnamefont {K.}~\bibnamefont {Guo}}, \bibinfo {author} {\bibfnamefont {L.}~\bibnamefont {Du}}, \bibinfo {author} {\bibfnamefont {X.}~\bibnamefont {Ling}}, \bibinfo {author} {\bibfnamefont {Y.}~\bibnamefont {L{\"u}}}, \bibinfo {author} {\bibfnamefont {L.}~\bibnamefont {He}}, \ and\ \bibinfo {author} {\bibfnamefont {X.}~\bibnamefont {Luo}},\ }\href {\doibase 10.1021/acs.langmuir.4c01913} {\bibfield  {journal} {\bibinfo  {journal} {Langmuir}\ }\textbf {\bibinfo {volume} {40}},\ \bibinfo {pages} {15926} (\bibinfo {year} {2024})}\BibitemShut {NoStop}%
\bibitem [{\citenamefont {Swan}\ \emph {et~al.}(2014)\citenamefont {Swan}, \citenamefont {Bauer}, \citenamefont {Liu},\ and\ \citenamefont {Furst}}]{Swan-2014}%
  \BibitemOpen
  \bibfield  {author} {\bibinfo {author} {\bibfnamefont {J.~W.}\ \bibnamefont {Swan}}, \bibinfo {author} {\bibfnamefont {J.~L.}\ \bibnamefont {Bauer}}, \bibinfo {author} {\bibfnamefont {Y.}~\bibnamefont {Liu}}, \ and\ \bibinfo {author} {\bibfnamefont {E.~M.}\ \bibnamefont {Furst}},\ }\href {\doibase 10.1039/C3SM52663A} {\bibfield  {journal} {\bibinfo  {journal} {Soft Matter}\ }\textbf {\bibinfo {volume} {10}},\ \bibinfo {pages} {1102} (\bibinfo {year} {2014})}\BibitemShut {NoStop}%
\bibitem [{\citenamefont {Sherman}\ and\ \citenamefont {Swan}(2016)}]{Sherman-2016}%
  \BibitemOpen
  \bibfield  {author} {\bibinfo {author} {\bibfnamefont {Z.~M.}\ \bibnamefont {Sherman}}\ and\ \bibinfo {author} {\bibfnamefont {J.~W.}\ \bibnamefont {Swan}},\ }\href {\doibase 10.1021/acsnano.6b01050} {\bibfield  {journal} {\bibinfo  {journal} {ACS Nano}\ }\textbf {\bibinfo {volume} {10}},\ \bibinfo {pages} {5260} (\bibinfo {year} {2016})}\BibitemShut {NoStop}%
\bibitem [{\citenamefont {Kachman}\ \emph {et~al.}(2017)\citenamefont {Kachman}, \citenamefont {Owen},\ and\ \citenamefont {England}}]{Kachman-2017}%
  \BibitemOpen
  \bibfield  {author} {\bibinfo {author} {\bibfnamefont {T.}~\bibnamefont {Kachman}}, \bibinfo {author} {\bibfnamefont {J.~A.}\ \bibnamefont {Owen}}, \ and\ \bibinfo {author} {\bibfnamefont {J.~L.}\ \bibnamefont {England}},\ }\href {\doibase 10.1103/PhysRevLett.119.038001} {\bibfield  {journal} {\bibinfo  {journal} {Phys. Rev. Lett.}\ }\textbf {\bibinfo {volume} {119}},\ \bibinfo {pages} {038001} (\bibinfo {year} {2017})}\BibitemShut {NoStop}%
\bibitem [{\citenamefont {Higashitani}\ and\ \citenamefont {Oshitani}(1997)}]{Higashitani-1997}%
  \BibitemOpen
  \bibfield  {author} {\bibinfo {author} {\bibfnamefont {K.}~\bibnamefont {Higashitani}}\ and\ \bibinfo {author} {\bibfnamefont {J.}~\bibnamefont {Oshitani}},\ }\href {\doibase https://doi.org/10.1205/095758297528887} {\bibfield  {journal} {\bibinfo  {journal} {Process Saf. Environ. Prot.}\ }\textbf {\bibinfo {volume} {75}},\ \bibinfo {pages} {115} (\bibinfo {year} {1997})}\BibitemShut {NoStop}%
\bibitem [{\citenamefont {Koga}\ and\ \citenamefont {Zeng}(1999)}]{Koga1999}%
  \BibitemOpen
  \bibfield  {author} {\bibinfo {author} {\bibfnamefont {K.}~\bibnamefont {Koga}}\ and\ \bibinfo {author} {\bibfnamefont {X.~C.}\ \bibnamefont {Zeng}},\ }\href {\doibase 10.1103/PhysRevB.60.14328} {\bibfield  {journal} {\bibinfo  {journal} {Phys. Rev. B}\ }\textbf {\bibinfo {volume} {60}},\ \bibinfo {pages} {14328} (\bibinfo {year} {1999})}\BibitemShut {NoStop}%
\bibitem [{\citenamefont {Tikhomirov}\ \emph {et~al.}(2008)\citenamefont {Tikhomirov}, \citenamefont {Yamazaki}, \citenamefont {Kovalenko},\ and\ \citenamefont {Fenniri}}]{Tikhomirov2008}%
  \BibitemOpen
  \bibfield  {author} {\bibinfo {author} {\bibfnamefont {G.}~\bibnamefont {Tikhomirov}}, \bibinfo {author} {\bibfnamefont {T.}~\bibnamefont {Yamazaki}}, \bibinfo {author} {\bibfnamefont {A.}~\bibnamefont {Kovalenko}}, \ and\ \bibinfo {author} {\bibfnamefont {H.}~\bibnamefont {Fenniri}},\ }\href {\doibase 10.1021/la8001114} {\bibfield  {journal} {\bibinfo  {journal} {Langmuir}\ }\textbf {\bibinfo {volume} {24}},\ \bibinfo {pages} {4447} (\bibinfo {year} {2008})}\BibitemShut {NoStop}%
\bibitem [{\citenamefont {Harada}\ and\ \citenamefont {Tsukada}(2010)}]{Harada2010}%
  \BibitemOpen
  \bibfield  {author} {\bibinfo {author} {\bibfnamefont {M.}~\bibnamefont {Harada}}\ and\ \bibinfo {author} {\bibfnamefont {M.}~\bibnamefont {Tsukada}},\ }\href {\doibase 10.1103/PhysRevB.82.035414} {\bibfield  {journal} {\bibinfo  {journal} {Phys. Rev. B}\ }\textbf {\bibinfo {volume} {82}},\ \bibinfo {pages} {035414} (\bibinfo {year} {2010})}\BibitemShut {NoStop}%
\bibitem [{\citenamefont {Ballone}\ \emph {et~al.}(2012)\citenamefont {Ballone}, \citenamefont {Del~Popolo}, \citenamefont {Bovio}, \citenamefont {Podesta}, \citenamefont {Milani},\ and\ \citenamefont {Manini}}]{Ballone2012}%
  \BibitemOpen
  \bibfield  {author} {\bibinfo {author} {\bibfnamefont {P.}~\bibnamefont {Ballone}}, \bibinfo {author} {\bibfnamefont {M.~G.}\ \bibnamefont {Del~Popolo}}, \bibinfo {author} {\bibfnamefont {S.}~\bibnamefont {Bovio}}, \bibinfo {author} {\bibfnamefont {A.}~\bibnamefont {Podesta}}, \bibinfo {author} {\bibfnamefont {P.}~\bibnamefont {Milani}}, \ and\ \bibinfo {author} {\bibfnamefont {N.}~\bibnamefont {Manini}},\ }\href {\doibase 10.1039/C2CP23459A} {\bibfield  {journal} {\bibinfo  {journal} {Phys. Chem. Chem. Phys.}\ }\textbf {\bibinfo {volume} {14}},\ \bibinfo {pages} {2475} (\bibinfo {year} {2012})}\BibitemShut {NoStop}%
\bibitem [{\citenamefont {Li}\ \emph {et~al.}(2020{\natexlab{a}})\citenamefont {Li}, \citenamefont {Steinmetz}, \citenamefont {Eppell},\ and\ \citenamefont {Zypman}}]{Li2020}%
  \BibitemOpen
  \bibfield  {author} {\bibinfo {author} {\bibfnamefont {L.}~\bibnamefont {Li}}, \bibinfo {author} {\bibfnamefont {N.~F.}\ \bibnamefont {Steinmetz}}, \bibinfo {author} {\bibfnamefont {S.~J.}\ \bibnamefont {Eppell}}, \ and\ \bibinfo {author} {\bibfnamefont {F.~R.}\ \bibnamefont {Zypman}},\ }\href {\doibase 10.1021/acs.langmuir.0c02455} {\bibfield  {journal} {\bibinfo  {journal} {Langmuir}\ }\textbf {\bibinfo {volume} {36}},\ \bibinfo {pages} {13621} (\bibinfo {year} {2020}{\natexlab{a}})}\BibitemShut {NoStop}%
\bibitem [{\citenamefont {Li}\ \emph {et~al.}(2020{\natexlab{b}})\citenamefont {Li}, \citenamefont {Eppell},\ and\ \citenamefont {Zypman}}]{Li-2020-2}%
  \BibitemOpen
  \bibfield  {author} {\bibinfo {author} {\bibfnamefont {L.}~\bibnamefont {Li}}, \bibinfo {author} {\bibfnamefont {S.~J.}\ \bibnamefont {Eppell}}, \ and\ \bibinfo {author} {\bibfnamefont {F.~R.}\ \bibnamefont {Zypman}},\ }\href {\doibase 10.1021/acs.langmuir.9b03602} {\bibfield  {journal} {\bibinfo  {journal} {Langmuir}\ }\textbf {\bibinfo {volume} {36}},\ \bibinfo {pages} {4123} (\bibinfo {year} {2020}{\natexlab{b}})}\BibitemShut {NoStop}%
\bibitem [{\citenamefont {Nakouzi}\ \emph {et~al.}(2021)\citenamefont {Nakouzi}, \citenamefont {Stack}, \citenamefont {Kerisit}, \citenamefont {Legg}, \citenamefont {Mundy}, \citenamefont {Schenter}, \citenamefont {Chun},\ and\ \citenamefont {De~Yoreo}}]{Nakouzi-2021}%
  \BibitemOpen
  \bibfield  {author} {\bibinfo {author} {\bibfnamefont {E.}~\bibnamefont {Nakouzi}}, \bibinfo {author} {\bibfnamefont {A.~G.}\ \bibnamefont {Stack}}, \bibinfo {author} {\bibfnamefont {S.}~\bibnamefont {Kerisit}}, \bibinfo {author} {\bibfnamefont {B.~A.}\ \bibnamefont {Legg}}, \bibinfo {author} {\bibfnamefont {C.~J.}\ \bibnamefont {Mundy}}, \bibinfo {author} {\bibfnamefont {G.~K.}\ \bibnamefont {Schenter}}, \bibinfo {author} {\bibfnamefont {J.}~\bibnamefont {Chun}}, \ and\ \bibinfo {author} {\bibfnamefont {J.~J.}\ \bibnamefont {De~Yoreo}},\ }\href {\doibase 10.1021/acs.jpcc.0c07901} {\bibfield  {journal} {\bibinfo  {journal} {J. Phys. Chem. C}\ }\textbf {\bibinfo {volume} {125}},\ \bibinfo {pages} {1282} (\bibinfo {year} {2021})}\BibitemShut {NoStop}%
\bibitem [{\citenamefont {Liu}\ \emph {et~al.}(2024)\citenamefont {Liu}, \citenamefont {Yadav~Schmid}, \citenamefont {Feng}, \citenamefont {Li}, \citenamefont {Droubay}, \citenamefont {Pauzauskie}, \citenamefont {Schenter}, \citenamefont {De~Yoreo}, \citenamefont {Chun},\ and\ \citenamefont {Nakouzi}}]{Liu-2024}%
  \BibitemOpen
  \bibfield  {author} {\bibinfo {author} {\bibfnamefont {L.}~\bibnamefont {Liu}}, \bibinfo {author} {\bibfnamefont {S.}~\bibnamefont {Yadav~Schmid}}, \bibinfo {author} {\bibfnamefont {Z.}~\bibnamefont {Feng}}, \bibinfo {author} {\bibfnamefont {D.}~\bibnamefont {Li}}, \bibinfo {author} {\bibfnamefont {T.~C.}\ \bibnamefont {Droubay}}, \bibinfo {author} {\bibfnamefont {P.~J.}\ \bibnamefont {Pauzauskie}}, \bibinfo {author} {\bibfnamefont {G.~K.}\ \bibnamefont {Schenter}}, \bibinfo {author} {\bibfnamefont {J.~J.}\ \bibnamefont {De~Yoreo}}, \bibinfo {author} {\bibfnamefont {J.}~\bibnamefont {Chun}}, \ and\ \bibinfo {author} {\bibfnamefont {E.}~\bibnamefont {Nakouzi}},\ }\href {\doibase 10.1021/acsnano.4c01797} {\bibfield  {journal} {\bibinfo  {journal} {ACS Nano}\ }\textbf {\bibinfo {volume} {18}},\ \bibinfo {pages} {16743} (\bibinfo {year} {2024})}\BibitemShut {NoStop}%
\bibitem [{\citenamefont {Yuan}\ \emph {et~al.}(2013)\citenamefont {Yuan}, \citenamefont {Li}, \citenamefont {Wang},\ and\ \citenamefont {Li}}]{Yuan-2013}%
  \BibitemOpen
  \bibfield  {author} {\bibinfo {author} {\bibfnamefont {B.}~\bibnamefont {Yuan}}, \bibinfo {author} {\bibfnamefont {W.}~\bibnamefont {Li}}, \bibinfo {author} {\bibfnamefont {C.}~\bibnamefont {Wang}}, \ and\ \bibinfo {author} {\bibfnamefont {L.}~\bibnamefont {Li}},\ }\href {\doibase https://doi.org/10.1016/j.snb.2012.10.097} {\bibfield  {journal} {\bibinfo  {journal} {Sens Actuators B Chem}\ }\textbf {\bibinfo {volume} {176}},\ \bibinfo {pages} {509} (\bibinfo {year} {2013})}\BibitemShut {NoStop}%
\bibitem [{\citenamefont {Lei}\ \emph {et~al.}(2017)\citenamefont {Lei}, \citenamefont {Fritzsche},\ and\ \citenamefont {Eckert}}]{Lei-2017}%
  \BibitemOpen
  \bibfield  {author} {\bibinfo {author} {\bibfnamefont {Z.}~\bibnamefont {Lei}}, \bibinfo {author} {\bibfnamefont {B.}~\bibnamefont {Fritzsche}}, \ and\ \bibinfo {author} {\bibfnamefont {K.}~\bibnamefont {Eckert}},\ }\href {\doibase 10.1021/acs.jpcc.7b07344} {\bibfield  {journal} {\bibinfo  {journal} {J. Phys. Chem. C}\ }\textbf {\bibinfo {volume} {121}},\ \bibinfo {pages} {24576} (\bibinfo {year} {2017})}\BibitemShut {NoStop}%
\bibitem [{\citenamefont {Huang}\ \emph {et~al.}(2019)\citenamefont {Huang}, \citenamefont {Marinaro}, \citenamefont {Yang}, \citenamefont {Fritzsche}, \citenamefont {Lei}, \citenamefont {Uhlemann}, \citenamefont {Eckert},\ and\ \citenamefont {Mutschke}}]{Huang-2019}%
  \BibitemOpen
  \bibfield  {author} {\bibinfo {author} {\bibfnamefont {M.}~\bibnamefont {Huang}}, \bibinfo {author} {\bibfnamefont {G.}~\bibnamefont {Marinaro}}, \bibinfo {author} {\bibfnamefont {X.}~\bibnamefont {Yang}}, \bibinfo {author} {\bibfnamefont {B.}~\bibnamefont {Fritzsche}}, \bibinfo {author} {\bibfnamefont {Z.}~\bibnamefont {Lei}}, \bibinfo {author} {\bibfnamefont {M.}~\bibnamefont {Uhlemann}}, \bibinfo {author} {\bibfnamefont {K.}~\bibnamefont {Eckert}}, \ and\ \bibinfo {author} {\bibfnamefont {G.}~\bibnamefont {Mutschke}},\ }\href {\doibase https://doi.org/10.1016/j.jelechem.2019.04.043} {\bibfield  {journal} {\bibinfo  {journal} {J. Electroanal. Chem.}\ }\textbf {\bibinfo {volume} {842}},\ \bibinfo {pages} {203} (\bibinfo {year} {2019})}\BibitemShut {NoStop}%
\bibitem [{\citenamefont {Ricchiuti}\ \emph {et~al.}(2022)\citenamefont {Ricchiuti}, \citenamefont {Dabrowska}, \citenamefont {Pinto}, \citenamefont {Ramer},\ and\ \citenamefont {Lendl}}]{Ricchiuti-2022}%
  \BibitemOpen
  \bibfield  {author} {\bibinfo {author} {\bibfnamefont {G.}~\bibnamefont {Ricchiuti}}, \bibinfo {author} {\bibfnamefont {A.}~\bibnamefont {Dabrowska}}, \bibinfo {author} {\bibfnamefont {D.}~\bibnamefont {Pinto}}, \bibinfo {author} {\bibfnamefont {G.}~\bibnamefont {Ramer}}, \ and\ \bibinfo {author} {\bibfnamefont {B.}~\bibnamefont {Lendl}},\ }\href {\doibase 10.1021/acs.analchem.2c03303} {\bibfield  {journal} {\bibinfo  {journal} {Analytical Chemistry}\ }\textbf {\bibinfo {volume} {94}},\ \bibinfo {pages} {16353} (\bibinfo {year} {2022})}\BibitemShut {NoStop}%
\bibitem [{\citenamefont {Ayansiji}\ \emph {et~al.}(2020)\citenamefont {Ayansiji}, \citenamefont {Dighe}, \citenamefont {Linninger},\ and\ \citenamefont {Singh}}]{Ayansiji-2020}%
  \BibitemOpen
  \bibfield  {author} {\bibinfo {author} {\bibfnamefont {A.~O.}\ \bibnamefont {Ayansiji}}, \bibinfo {author} {\bibfnamefont {A.~V.}\ \bibnamefont {Dighe}}, \bibinfo {author} {\bibfnamefont {A.~A.}\ \bibnamefont {Linninger}}, \ and\ \bibinfo {author} {\bibfnamefont {M.~R.}\ \bibnamefont {Singh}},\ }\href {\doibase 10.1073/pnas.2018568117} {\bibfield  {journal} {\bibinfo  {journal} {Proc. Natl. Acad. Sci.}\ }\textbf {\bibinfo {volume} {117}},\ \bibinfo {pages} {30208} (\bibinfo {year} {2020})}\BibitemShut {NoStop}%
\bibitem [{\citenamefont {Leong}\ \emph {et~al.}(2020)\citenamefont {Leong}, \citenamefont {Ahmad}, \citenamefont {Low}, \citenamefont {Camacho}, \citenamefont {Faraudo},\ and\ \citenamefont {Lim}}]{Leong-2020}%
  \BibitemOpen
  \bibfield  {author} {\bibinfo {author} {\bibfnamefont {S.~S.}\ \bibnamefont {Leong}}, \bibinfo {author} {\bibfnamefont {Z.}~\bibnamefont {Ahmad}}, \bibinfo {author} {\bibfnamefont {S.~C.}\ \bibnamefont {Low}}, \bibinfo {author} {\bibfnamefont {J.}~\bibnamefont {Camacho}}, \bibinfo {author} {\bibfnamefont {J.}~\bibnamefont {Faraudo}}, \ and\ \bibinfo {author} {\bibfnamefont {J.}~\bibnamefont {Lim}},\ }\href {\doibase 10.1021/acs.langmuir.0c00839} {\bibfield  {journal} {\bibinfo  {journal} {Langmuir}\ }\textbf {\bibinfo {volume} {36}},\ \bibinfo {pages} {8033} (\bibinfo {year} {2020})}\BibitemShut {NoStop}%
\bibitem [{\citenamefont {Franczak}\ \emph {et~al.}(2016)\citenamefont {Franczak}, \citenamefont {Binnemans},\ and\ \citenamefont {Fransaer}}]{Franczak-2016}%
  \BibitemOpen
  \bibfield  {author} {\bibinfo {author} {\bibfnamefont {A.}~\bibnamefont {Franczak}}, \bibinfo {author} {\bibfnamefont {K.}~\bibnamefont {Binnemans}}, \ and\ \bibinfo {author} {\bibfnamefont {J.}~\bibnamefont {Fransaer}},\ }\href {\doibase 10.1039/C6CP02575G} {\bibfield  {journal} {\bibinfo  {journal} {Phys. Chem. Chem. Phys.}\ }\textbf {\bibinfo {volume} {18}},\ \bibinfo {pages} {27342} (\bibinfo {year} {2016})}\BibitemShut {NoStop}%
\bibitem [{\citenamefont {te~Vrugt}\ and\ \citenamefont {Wittkowski}(2022)}]{Vrugt-2023}%
  \BibitemOpen
  \bibfield  {author} {\bibinfo {author} {\bibfnamefont {M.}~\bibnamefont {te~Vrugt}}\ and\ \bibinfo {author} {\bibfnamefont {R.}~\bibnamefont {Wittkowski}},\ }\href {\doibase 10.1088/1361-648X/ac8633} {\bibfield  {journal} {\bibinfo  {journal} {J. Phys. Condens. Matter}\ }\textbf {\bibinfo {volume} {35}},\ \bibinfo {pages} {041501} (\bibinfo {year} {2022})}\BibitemShut {NoStop}%
\bibitem [{\citenamefont {Tschopp}\ and\ \citenamefont {Brader}(2024)}]{Brader-2024}%
  \BibitemOpen
  \bibfield  {author} {\bibinfo {author} {\bibfnamefont {S.~M.}\ \bibnamefont {Tschopp}}\ and\ \bibinfo {author} {\bibfnamefont {J.~M.}\ \bibnamefont {Brader}},\ }\href {\doibase 10.1063/5.0211198} {\bibfield  {journal} {\bibinfo  {journal} {J. Chem. Phys.}\ }\textbf {\bibinfo {volume} {160}},\ \bibinfo {pages} {214124} (\bibinfo {year} {2024})}\BibitemShut {NoStop}%
\bibitem [{\citenamefont {Kuchel}\ \emph {et~al.}(2003)\citenamefont {Kuchel}, \citenamefont {Chapman}, \citenamefont {Bubb}, \citenamefont {Hansen}, \citenamefont {Durrant},\ and\ \citenamefont {Hertzberg}}]{Kuchel-2003}%
  \BibitemOpen
  \bibfield  {author} {\bibinfo {author} {\bibfnamefont {P.}~\bibnamefont {Kuchel}}, \bibinfo {author} {\bibfnamefont {B.}~\bibnamefont {Chapman}}, \bibinfo {author} {\bibfnamefont {W.}~\bibnamefont {Bubb}}, \bibinfo {author} {\bibfnamefont {P.}~\bibnamefont {Hansen}}, \bibinfo {author} {\bibfnamefont {C.}~\bibnamefont {Durrant}}, \ and\ \bibinfo {author} {\bibfnamefont {M.}~\bibnamefont {Hertzberg}},\ }\href {\doibase https://doi.org/10.1002/cmr.a.10066} {\bibfield  {journal} {\bibinfo  {journal} {Concept Magn. Reson. A}\ }\textbf {\bibinfo {volume} {18A}},\ \bibinfo {pages} {56} (\bibinfo {year} {2003})}\BibitemShut {NoStop}%
\bibitem [{\citenamefont {Ikeda}\ and\ \citenamefont {Yoshioka}(1968)}]{Ikeda-1968}%
  \BibitemOpen
  \bibfield  {author} {\bibinfo {author} {\bibfnamefont {T.}~\bibnamefont {Ikeda}}\ and\ \bibinfo {author} {\bibfnamefont {H.}~\bibnamefont {Yoshioka}},\ }\href {\doibase 10.1021/j100859a007} {\bibfield  {journal} {\bibinfo  {journal} {J. Phys. Chem.}\ }\textbf {\bibinfo {volume} {72}},\ \bibinfo {pages} {4392} (\bibinfo {year} {1968})}\BibitemShut {NoStop}%
\bibitem [{\citenamefont {Ikeda}\ and\ \citenamefont {Sueoka}(1970)}]{Ikeda-1970}%
  \BibitemOpen
  \bibfield  {author} {\bibinfo {author} {\bibfnamefont {T.}~\bibnamefont {Ikeda}}\ and\ \bibinfo {author} {\bibfnamefont {K.}~\bibnamefont {Sueoka}},\ }\href {\doibase 10.1246/bcsj.43.1273} {\bibfield  {journal} {\bibinfo  {journal} {Bull. Chem. Soc. Jpn.}\ }\textbf {\bibinfo {volume} {43}},\ \bibinfo {pages} {1273} (\bibinfo {year} {1970})}\BibitemShut {NoStop}%
\bibitem [{\citenamefont {Dinpajooh}\ \emph {et~al.}(2011)\citenamefont {Dinpajooh}, \citenamefont {Keasler}, \citenamefont {Truhlar},\ and\ \citenamefont {Siepmann}}]{Dinpajooh-2011}%
  \BibitemOpen
  \bibfield  {author} {\bibinfo {author} {\bibfnamefont {M.}~\bibnamefont {Dinpajooh}}, \bibinfo {author} {\bibfnamefont {S.~J.}\ \bibnamefont {Keasler}}, \bibinfo {author} {\bibfnamefont {D.~G.}\ \bibnamefont {Truhlar}}, \ and\ \bibinfo {author} {\bibfnamefont {J.~I.}\ \bibnamefont {Siepmann}},\ }\href {\doibase 10.1007/s00214-011-0973-1} {\bibfield  {journal} {\bibinfo  {journal} {Theor. Chem. Acc.}\ }\textbf {\bibinfo {volume} {130}},\ \bibinfo {pages} {83} (\bibinfo {year} {2011})}\BibitemShut {NoStop}%
\bibitem [{\citenamefont {Jorgensen}\ \emph {et~al.}(1983)\citenamefont {Jorgensen}, \citenamefont {Chandrasekhar}, \citenamefont {Madura}, \citenamefont {Impey},\ and\ \citenamefont {Klein}}]{TIP4P}%
  \BibitemOpen
  \bibfield  {author} {\bibinfo {author} {\bibfnamefont {W.~L.}\ \bibnamefont {Jorgensen}}, \bibinfo {author} {\bibfnamefont {J.}~\bibnamefont {Chandrasekhar}}, \bibinfo {author} {\bibfnamefont {J.~D.}\ \bibnamefont {Madura}}, \bibinfo {author} {\bibfnamefont {R.~W.}\ \bibnamefont {Impey}}, \ and\ \bibinfo {author} {\bibfnamefont {M.~L.}\ \bibnamefont {Klein}},\ }\href {\doibase 10.1063/1.445869} {\bibfield  {journal} {\bibinfo  {journal} {J. Chem. Phys.}\ }\textbf {\bibinfo {volume} {79}},\ \bibinfo {pages} {926} (\bibinfo {year} {1983})}\BibitemShut {NoStop}%
\bibitem [{\citenamefont {Dyer}\ \emph {et~al.}(2009)\citenamefont {Dyer}, \citenamefont {Perkyns}, \citenamefont {Stell},\ and\ \citenamefont {Pettitt}}]{Dyer2009}%
  \BibitemOpen
  \bibfield  {author} {\bibinfo {author} {\bibfnamefont {K.~M.}\ \bibnamefont {Dyer}}, \bibinfo {author} {\bibfnamefont {J.~S.}\ \bibnamefont {Perkyns}}, \bibinfo {author} {\bibfnamefont {G.}~\bibnamefont {Stell}}, \ and\ \bibinfo {author} {\bibfnamefont {B.~M.}\ \bibnamefont {Pettitt}},\ }\href {\doibase 10.1080/00268970902845313} {\bibfield  {journal} {\bibinfo  {journal} {Mol. Phys.}\ }\textbf {\bibinfo {volume} {107}},\ \bibinfo {pages} {423} (\bibinfo {year} {2009})}\BibitemShut {NoStop}%
\bibitem [{\citenamefont {Liang}\ \emph {et~al.}(2007)\citenamefont {Liang}, \citenamefont {Hilal}, \citenamefont {Langston},\ and\ \citenamefont {Starov}}]{Liang2007}%
  \BibitemOpen
  \bibfield  {author} {\bibinfo {author} {\bibfnamefont {Y.}~\bibnamefont {Liang}}, \bibinfo {author} {\bibfnamefont {N.}~\bibnamefont {Hilal}}, \bibinfo {author} {\bibfnamefont {P.}~\bibnamefont {Langston}}, \ and\ \bibinfo {author} {\bibfnamefont {V.}~\bibnamefont {Starov}},\ }\href {\doibase https://doi.org/10.1016/j.cis.2007.04.003} {\bibfield  {journal} {\bibinfo  {journal} {Adv. Colloid Interface Sci.}\ }\textbf {\bibinfo {volume} {134-135}},\ \bibinfo {pages} {151} (\bibinfo {year} {2007})}\BibitemShut {NoStop}%
\bibitem [{\citenamefont {Batista}\ \emph {et~al.}(2015)\citenamefont {Batista}, \citenamefont {Larson},\ and\ \citenamefont {Kotov}}]{Batista2015}%
  \BibitemOpen
  \bibfield  {author} {\bibinfo {author} {\bibfnamefont {C.~A.}\ \bibnamefont {Batista}}, \bibinfo {author} {\bibfnamefont {R.~G.}\ \bibnamefont {Larson}}, \ and\ \bibinfo {author} {\bibfnamefont {N.~A.}\ \bibnamefont {Kotov}},\ }\href {\doibase 10.1126/science.1242477} {\bibfield  {journal} {\bibinfo  {journal} {Science}\ }\textbf {\bibinfo {volume} {350}},\ \bibinfo {pages} {1242477} (\bibinfo {year} {2015})}\BibitemShut {NoStop}%
\bibitem [{\citenamefont {Plimpton}(1995)}]{Plimpton1995}%
  \BibitemOpen
  \bibfield  {author} {\bibinfo {author} {\bibfnamefont {S.}~\bibnamefont {Plimpton}},\ }\href {\doibase 10.1006/jcph.1995.1039} {\bibfield  {journal} {\bibinfo  {journal} {J. Comput. Phys.}\ }\textbf {\bibinfo {volume} {117}},\ \bibinfo {pages} {1 } (\bibinfo {year} {1995})}\BibitemShut {NoStop}%
\bibitem [{\citenamefont {in~'t Veld}\ \emph {et~al.}(2008)\citenamefont {in~'t Veld}, \citenamefont {Plimpton},\ and\ \citenamefont {Grest}}]{Pieter-2008}%
  \BibitemOpen
  \bibfield  {author} {\bibinfo {author} {\bibfnamefont {P.~J.}\ \bibnamefont {in~'t Veld}}, \bibinfo {author} {\bibfnamefont {S.~J.}\ \bibnamefont {Plimpton}}, \ and\ \bibinfo {author} {\bibfnamefont {G.~S.}\ \bibnamefont {Grest}},\ }\href {\doibase 10.1016/j.cpc.2008.03.005} {\bibfield  {journal} {\bibinfo  {journal} {Comput. Phys. Commun.}\ }\textbf {\bibinfo {volume} {179}},\ \bibinfo {pages} {320} (\bibinfo {year} {2008})}\BibitemShut {NoStop}%
\bibitem [{\citenamefont {Kumar}\ \emph {et~al.}(1992)\citenamefont {Kumar}, \citenamefont {Rosenberg}, \citenamefont {Bouzida}, \citenamefont {Swendsen},\ and\ \citenamefont {Kollman}}]{Kumar1992}%
  \BibitemOpen
  \bibfield  {author} {\bibinfo {author} {\bibfnamefont {S.}~\bibnamefont {Kumar}}, \bibinfo {author} {\bibfnamefont {J.~M.}\ \bibnamefont {Rosenberg}}, \bibinfo {author} {\bibfnamefont {D.}~\bibnamefont {Bouzida}}, \bibinfo {author} {\bibfnamefont {R.~H.}\ \bibnamefont {Swendsen}}, \ and\ \bibinfo {author} {\bibfnamefont {P.~A.}\ \bibnamefont {Kollman}},\ }\href {\doibase 10.1002/jcc.540130812} {\bibfield  {journal} {\bibinfo  {journal} {J. Comput. Chem.}\ }\textbf {\bibinfo {volume} {13}},\ \bibinfo {pages} {1011} (\bibinfo {year} {1992})}\BibitemShut {NoStop}%
\bibitem [{\citenamefont {Grossfield}(2022)}]{wham-code}%
  \BibitemOpen
  \bibfield  {author} {\bibinfo {author} {\bibfnamefont {A.}~\bibnamefont {Grossfield}},\ }\href@noop {} {\enquote {\bibinfo {title} {{WHAM: the weighted histogram analysis method, version 2.0.11}},}\ }\bibinfo {howpublished} {\url{http://membrane.urmc.rochester.edu/wordpress/content/wham}} (\bibinfo {year} {2022})\BibitemShut {NoStop}%
\bibitem [{\citenamefont {Fiorin}\ \emph {et~al.}(2013)\citenamefont {Fiorin}, \citenamefont {Klein},\ and\ \citenamefont {Hénin}}]{Fiorin2013}%
  \BibitemOpen
  \bibfield  {author} {\bibinfo {author} {\bibfnamefont {G.}~\bibnamefont {Fiorin}}, \bibinfo {author} {\bibfnamefont {M.~L.}\ \bibnamefont {Klein}}, \ and\ \bibinfo {author} {\bibfnamefont {J.}~\bibnamefont {Hénin}},\ }\href {\doibase 10.1080/00268976.2013.813594} {\bibfield  {journal} {\bibinfo  {journal} {Mol. Phys.}\ }\textbf {\bibinfo {volume} {111}},\ \bibinfo {pages} {3345} (\bibinfo {year} {2013})}\BibitemShut {NoStop}%
\bibitem [{\citenamefont {Abrams}\ and\ \citenamefont {Bussi}(2014)}]{Abrams2014}%
  \BibitemOpen
  \bibfield  {author} {\bibinfo {author} {\bibfnamefont {C.}~\bibnamefont {Abrams}}\ and\ \bibinfo {author} {\bibfnamefont {G.}~\bibnamefont {Bussi}},\ }\href {\doibase 10.3390/e16010163} {\bibfield  {journal} {\bibinfo  {journal} {Entropy}\ }\textbf {\bibinfo {volume} {16}},\ \bibinfo {pages} {163} (\bibinfo {year} {2014})}\BibitemShut {NoStop}%
\bibitem [{\citenamefont {Barducci}\ \emph {et~al.}(2011)\citenamefont {Barducci}, \citenamefont {Bonomi},\ and\ \citenamefont {Parrinello}}]{Barducci2011}%
  \BibitemOpen
  \bibfield  {author} {\bibinfo {author} {\bibfnamefont {A.}~\bibnamefont {Barducci}}, \bibinfo {author} {\bibfnamefont {M.}~\bibnamefont {Bonomi}}, \ and\ \bibinfo {author} {\bibfnamefont {M.}~\bibnamefont {Parrinello}},\ }\href {\doibase 10.1002/wcms.31} {\bibfield  {journal} {\bibinfo  {journal} {Wiley Interdiscip. Rev. Comput. Mol. Sci.}\ }\textbf {\bibinfo {volume} {1}},\ \bibinfo {pages} {826} (\bibinfo {year} {2011})}\BibitemShut {NoStop}%
\end{thebibliography}
\end{document}